\documentclass[fleqn,usenatbib]{mnras}

\usepackage{newtxtext,newtxmath}
\usepackage[T1]{fontenc}
\usepackage{ae,aecompl}

\usepackage{graphicx}	% Including figure files
\usepackage{amsmath}	% Advanced maths commands
\usepackage{gensymb}
\usepackage{multirow}
\usepackage[dvipsnames]{xcolor}%temporary, to add color text
\usepackage{lscape}
\usepackage{longtable}
\usepackage{booktabs}
\usepackage[flushleft]{threeparttable}
\usepackage{rotating}
\title[MATLAS dwarfs structure and morphology]{Structure and morphology of the MATLAS dwarf galaxies and their central nuclei}
\author[M. Poulain et al.]{
M{\'e}lina Poulain$^{1}$,
Francine R. Marleau$^{1}$,
Rebecca Habas$^{2}$,
Pierre-Alain Duc$^{2}$,
\newauthor Rub{\'e}n S{\'a}nchez-Janssen$^{3}$,
Patrick R. Durrell$^{4}$,
Sanjaya Paudel$^{5}$,
\newauthor Syeda Lammim Ahad$^{6}$,
Abhishek Chougule$^{7,8}$,
Oliver M{\"u}ller$^{2}$,
Sungsoon Lim$^{9}$,
\newauthor Michal B{\'i}lek$^{10}$,
J{\'e}r{\'e}my Fensch$^{11}$
\\
$^{1}$Institute f{\"u}r  Astro- und Teilchenphysik, Universit{\"a}t Innsbruck, Technikerstra{\ss}e 25/8, Innsbruck, A-6020, Austria\\
e-mail: melina.poulain45@gmail.com, Melina.Poulain@student.uibk.ac.at\\
$^{2}$Observatoire Astronomique, Universit{\'e} de Strasbourg, CNRS, 11, rue de l'Universit{\'e}. F-67000 Strasbourg, France\\
$^{3}$UK Astronomy Technology Centre, Royal Observatory Edinburgh, Blackford Hill, Edinburgh EH9 3HJ, UK\\
$^{4}$Dept. of Physics, Astronomy, Geology, and Environmental Sciences, Youngstown State University, Youngstown, OH 44555 USA\\
$^{5}$Department of Astronomy and centre for Galaxy Evolution Research, Yonsei University, Seoul 03722\\
$^{6}$Leiden Observatory, Leiden University, P.O. Box 9513, 2300 RA Leiden, The Netherlands\\
$^{7}$Instituto de Astrofísica e Ciências do Espaço, Universidade do Porto, CAUP, Rua das Estrelas, 4150-762 Porto, Portugal\\
$^{8}$Departamento de Física e Astronomia, Faculdade de Ciências, Universidade do Porto, Rua do Campo Alegre 687, 4169-007 Porto, Portugal\\
$^{9}$Department of Astronomy, Yonsei University, 50 Yonsei-ro Seodaemun-gu, Seoul, 03722, Republic of Korea\\
$^{10}$Nicolaus Copernicus Astronomical centre, Polish Academy of Sciences, Bartycka 18, 00-716 Warsaw, Poland\\
$^{11}$Univ. Lyon, ENS de Lyon, Univ. Lyon 1, CNRS, Centre de Recherche Astrophysique de Lyon, UMR5574, 69007 Lyon, France\\
}

\date{Accepted XXX. Received YYY; in original form ZZZ}

\pubyear{2021}

\begin{document}
\label{firstpage}
\pagerange{\pageref{firstpage}--\pageref{lastpage}}
\maketitle

\begin{abstract}
We present a photometric study of the dwarf galaxy population in the low to moderate density environments of the MATLAS (Mass Assembly of early-Type gaLAxies with their fine Structures) deep imaging survey. The sample consists of 2210 dwarfs, including 508 nucleated. We define a nucleus as a compact source that is close to the galaxy photocentre (within 0.5 $R_e$) which is also the brightest such source within the galaxy's effective radius. The morphological analysis is performed using a 2D surface brightness profile modelling on the g-band images of both the galaxies and nuclei. Our study reveals that, for similar luminosities, the MATLAS dwarfs show ranges in the distribution of structural properties comparable to cluster (Virgo and Fornax) dwarfs and a range of sizes comparable to the Local Group and Local Volume dwarfs. Colour measurements using the r- and i-band images indicate that the dwarfs in low and moderate density environments are as red as cluster dwarfs on average. The observed similarities between dwarf ellipticals in vastly different environments imply that dEs are not uniquely the product of morphological transformation due to ram-pressure stripping and galaxy harassment in high density environments. We measure that the dwarf nuclei are located predominantly in massive, bright and round dwarfs and observe fewer nuclei in dwarfs with a faint centre and a small size. The colour of the galaxy nucleus shows no clear relation to the colour of the dwarf, in agreement with the migration and wet migration nucleus formation scenarios. The catalogues of the MATLAS dwarfs photometric and structural properties are provided.
\end{abstract}


%%%%%%%%%%%%%%%%%%%%%%%%%%%%%%%%%%%%%%%%%%%%%%%%%%

%%%%%%%%%%%%%%%%% BODY OF PAPER %%%%%%%%%%%%%%%%%%

\section{Introduction}

Galaxy structure and morphology is one of the most fundamental 
ways used to characterize galaxies and infer their formation scenario.
These structural properties are found to be correlated with the physical 
properties of the galaxy, such as their star formation rate, merging history 
and overall scale. When observed over a range of cosmic time, they can 
also be used as a tool for measuring galaxy evolution. Although the 
morphological nomenclature across galaxies of all luminosity (or mass)
have similarities, there is a clear distinction between the so-called 
"massive" and "dwarf" galaxies. While the morphology and structure 
of massive galaxies have been studied in equal measure over a wide range 
of environments, this is not the case for dwarf galaxies. Indeed, 
the characterization of their structure and morphology have relied until 
very recently almost entirely, due to their low surface brightness, on
those found in the Local Group (LG; \citealt{Mcconachie2012}) and nearby clusters 
(Virgo: \citealt{Ferrarese2012}; Fornax: \citealt{Eigenthaler2018,Venhola2019}).

The morphological types of dwarf galaxies have been found to
belong to two main groups: elliptical/spheroidal (dE/dSph), the most commonly
observed type \citep{Ferguson1994}, and irregular (dI; \citealt{vandenBergh1960}).
The dIs, typically located in the outskirts of groups and clusters, are gas rich galaxies with ongoing star formation. They are physically similar to irregular galaxies and thus can be defined as their low mass end \citep{Grebel2004}. The dEs/dSphs are gas poor galaxies and, until recently, generally observed in dense environment like groups and clusters \citep{Ferguson1991}. The dSphs are usually defined as the low luminosity end of the elliptical dwarfs and have so far mostly been observed in the LG \citep{Grebel2001}.\\

In the last decades, many discussions concerning the formation scenarios of dEs have been carried out, with an emphasis on two distinct processes. The first process involves environmental effects like the harassment/ram-pressure stripping of late-type galaxies, either bright spiral galaxies or dIs. Some studies have shown the tendency of bright spiral galaxies to change morphologically and become more compact when going through harassment inside clusters \citep{Moore1996,Mastropietro2005,Aguerri2009}. Other models describe the evolution of dIs through external processes such as ram-pressure stripping \citep{Boselli2008} or tidal interactions \citep{Mayer2001,Pasetto2003}. 
The second scenario suggests a formation caused by internal processes \citep{Dekel1986}. It is supported by the results from simulations of isolated dwarf galaxies \citep{Valcke2008,Sawala2010} as well as the relations observed between dEs and bright ellipticals structural properties in clusters and the LG \citep{Adami2006,Misgeld2008}.

A significant fraction of the dwarfs, called nucleated dwarfs, contains a central compact nucleus. The origin of these nuclei is still unclear. Possible formation scenario suggests that the nuclear star cluster (NSC) forms from accreted gas at the centre of the galaxy (in-situ scenario, \citealt{Loose1982}). Another proposes that an NSC is formed by globular clusters (GCs) that migrate toward the centre and merge due to dynamical frictions (migration scenario, \citealt{Tremaine1975}). Simulations combined to observations show that a combination of both processes is necessary to reconstruct the observed NSCs in dwarfs and more massive galaxies \citep{Bekki2007,Antonini2012,Ordenes2018,Sills2019}. However, the level of contribution from each process seems to vary with the mass of the host galaxy, with the infalling GCs scenario dominating in galaxies with a stellar mass below $10^9 M_{\odot}$ due to the shorter dynamical friction timescales \citep{Turner2012,Janssen2019,Fahrion2020,Neumayer2020}.
A third formation model of the nucleus has been proposed by \citet{Guillard2016} for dwarfs galaxies. Called the wet migration scenario, it combines the in-situ and migration scenarios by considering an off-centred formation and growth of a massive cluster which then migrates to the centre while retaining part of its gas. This migration can then be followed by a merger with another cluster.
Dwarf galaxies are also known to host active galactic nuclei (AGNs; e.g, \citealt{Marleau2013,Marleau2017,Sugata2019,Mezcua2020,Molina2021}) or both NSC and AGNs \citep{Cote2006,denBrock2015}.

Due to their shallow gravitational potential wells, dwarf galaxies are expected to be more affected by the environment than more massive galaxies. The possible effects of the environment on the structural properties as well as on the formation scenarios have been studied during the last decades. Comparison of the scaling relations for dwarfs in clusters and in the LG support a picture where the dwarfs are shaped more by internal than external processes \citep{Weisz2011,Young2014,Dunn2015}.
Some formation scenarios, such as the transformation of late type galaxies to dEs via harassment, have been put forward to explain the morphology-density relation of dwarf galaxies \citep{Ferguson1991}.
However, these works mainly focused on the dwarf galaxies in clusters and in the LG. Some studies have looked for dwarf galaxies in the field and nearby groups environments in the Local Volume (LV; \citealt{Sharina2008,Mueller2017,Mueller2018,Carlsten2020}). But without a large sample of dwarfs in low density environments, the role of the environment cannot be properly estimated.

The study by \citet{Habas2020} of 2210 dwarf galaxies, beyond the LV, in the low to medium-density environments of the MATLAS (Mass Assembly of early-Type gaLAxies with their fine Structures) survey \citep{Duc2014} started to address this problem. In this paper, we present the results of a detailed structural and morphological analysis of this large sample of field dwarf galaxies. The paper is organized as follows. In Section 2, we present the MATLAS survey and the dwarfs sample. In Section 3, we detail the galaxy modelling method. The structural and photometric properties of the MATLAS dwarfs are presented in Section 4. In Section 5, we describe and show the results of the study of the central compact nuclei. In Section 6 we discuss the results, and conclude in Section 7.

\section{The MATLAS dwarf galaxy sample}
\label{section:dwarfsample}

The MATLAS survey is associated to the ATLAS$^{3D}$ project \citep{Capellari2011}. This project aims to study the kinematic and structural properties of a complete sample of 260 nearby early type galaxies (ETGs). The galaxies were selected so they have an absolute magnitude $M_K < -21.5 $ and are situated at $D \lesssim 45$ Mpc,  $|\delta - 29\degree| < 35\degree$, and $|b| > 15\degree$ where $D$ is the distance, $\delta$ is the declination and b the Galactic latitude. The ATLAS$^{3D}$ observations have been performed in radio with the Westerbork Radio Synthesis Telescope (WRST), in millimetre with the IRAM 30m telescope and the Combined Array for Research in Millimeter-wave Astronomy (CARMA), and in optical with the Canada-France-Hawaii telescope (CFHT). They are combined to spectroscopic observations from the Spectroscopic Areal Unit for Research on Optical Nebulae (SAURON) integral-field unit of the William Herschel Telescope (WHT), numerical simulations and semi-analytic models of galaxy formation.\\

The MATLAS survey, coupled to the Next Generation Virgo Cluster Survey (NGVS; \citealt{Ferrarese2012}), focused on the deep optical observations from the CFHT. The NGVS ETGs, situated in the Virgo cluster, were observed from 2009 to 2013 over an area of 104 deg$^2$. The dwarfs detected in the NGVS are being studied in a series of independent papers (e.g.,\citealt{Sanchez-Janssen2016,Janssen2019,Ferrarese2020}). The ETGs outside the Virgo cluster, located in low to moderate density environments, were observed for the MATLAS survey from 2012 to 2015.

The MATLAS survey aims to study the low surface brightness structures of the 202 ETGs situated outside the Virgo cluster. Rejecting the galaxies close to bright stars, the observations were carried out over 150 fields containing 180 ETGs and 55 late type galaxies (LTGs) using the MegaCam camera. The data is composed of 1\degree $\,\times$ 1\degree\,images for 150, 148 and 78 fields in the g,r and i bands respectively as well as additional u-band observations for the 12 closest targets ($D$ < 20 Mpc). The images have a resolution of 0.19 arcsec/pixel and, in the g-band, a seeing between $0.5-1.61$\arcsec\ and a depth in surface brightness of ~28.5 -- 29 mag/arcsec$^{2}$ \citep{Duc2014}. As it has the largest number of fields observed, the g-band observations have been primarily used for our analysis. The complete list of observed bands for each MATLAS target is available in \citet{Habas2020}.\\

\begin{figure}
\centering
\includegraphics[width=\linewidth]{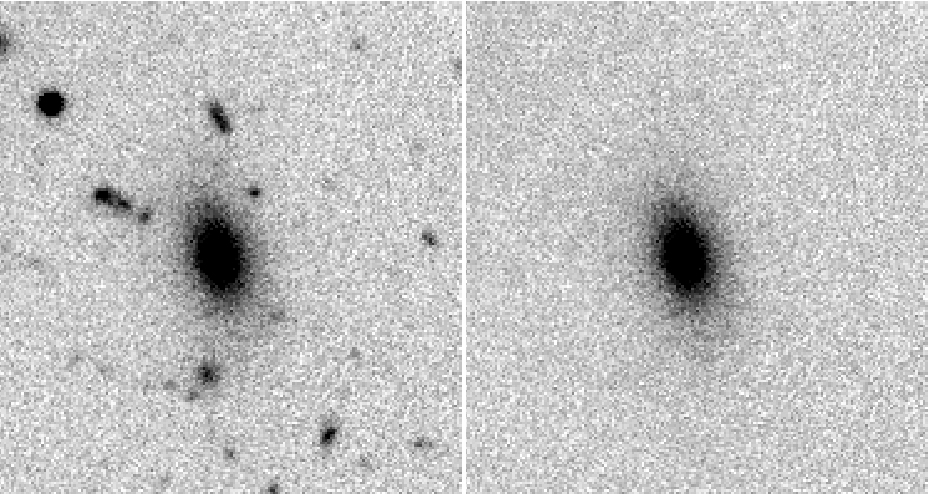}
\caption{Example of the cleaning of foreground stars and background galaxies for a dE. The cutouts are 37\arcsec\ $\times$ 37\arcsec\ with North up and East left.}
\label{fig:cleaning}
\end{figure}

As the MATLAS survey allows us to study low surface brightness structures thanks to its deep imaging, it has provided a unique data set to search and identify a statistically significant number of dwarf galaxies in low-density environments. Two methods were used to identify the dwarf candidates: a visual inspection coupled to a semi-automated catalogue of candidates. The visual inspection was performed over the 150 fields available in the g band. Each field was studied by at least one of the team members and galaxies with extended central light concentration or spiral structures was rejected to avoid contamination by background lenticulars or spirals. A catalogue of 1349 candidates was obtained after this first step. To create the semi-automated catalog, the software \textsc{source extractor} \citep{Bertin1996} was run on the g-band images as well as on \textsc{RMEDIAN} filtered images \citep{Secker1995} which emphasize the extended low surface brightness objects and removes small bright objects. Only sources detected in both images were selected. A cut in average surface brightness, apparent magnitude and size was then applied. The photometric selection criteria for the automated selection algorithm were defined based on the visual catalog, while the size was set such that dwarf candidates were reliably differentiated from background galaxies. The initial visual catalogue was cross-matched against the semi-automated catalog, and any candidates not detected by \textsc{source extractor} were appended to the semi-automated catalog, generating a sample of 25522 galaxies. The final list of dwarf candidates was created after two visual inspections of the data; in the first, potential dwarf galaxies were identified by a minimum of three team members, which were then inspected a second time by five team members to produce a final clean sample. All the details on the selection method can be found in \citet{Habas2020}.

This produced a final catalogue of 2210 dwarf galaxies with 73.4\% of ellipticals (dEs) and 26.6\% of irregulars (dIs). In the entire catalog, 23.2\% of the dwarfs are nucleated. We will now refer to this sample as the MATLAS dwarfs.
We note that the dI sample is likely not complete due to the possible confusion between dwarf irregular galaxies and a background galaxies. As explained in \citet{Habas2020}, a small number of galaxies with hints of spiral structure were rejected from the dwarf sample as likely background galaxies, but were later found to have absolute magnitudes M$_g$ indicating they are dwarf galaxies, based on pre-existing distance estimates. These galaxies were not added back into the sample.

An important parameter, not obtainable from the CFHT images, is the distance to the dwarfs. It allows us, for example, to confirm the dwarf nature of our galaxies and to compute physical sizes and luminosities. Using spectroscopic redshifts and HI measurements, distances were measured for 13.5\% of the catalogue (the HI properties of the dwarf galaxies will be discussed in an upcoming paper, Poulain et al. in preparation). Considering a cut in M$_g$ of $-18$, the dwarf nature of the candidates was confirmed for 99\% of the dwarfs for which a distance could be computed. Distances from available surveys allowed to measure relative velocities suggesting that $\sim$90\% of the 13.5\% form as satellite population around the nearest massive galaxies. 
For this subsample of dwarfs with distances, it was also shown that assuming the dwarfs are satellites of the targeted massive ETG located near the centre of each field is nearly as accurate ($\sim$80\%) as matching them with the nearest massive galaxy in 3D space; thus, for the full sample, we have assumed that the MATLAS dwarfs are satellites located at the same distance as the targeted massive ETG. And therefore, unless a distance measurement is available, we will use that distance for each dwarf. However, some fields contain several ETGs with different distances, causing uncertainties on the host of the dwarf and its distance. The dwarfs for which there is an ambiguity in the distance have been flagged in Table \ref{tab:catalog_nonnucleated}.

\begin{figure}
  \centering
  \includegraphics[width=\linewidth]{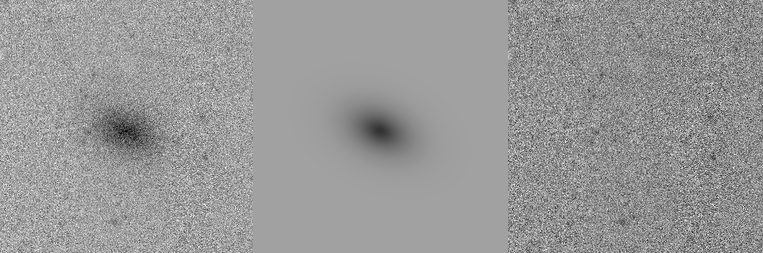}
  \includegraphics[width=\linewidth]{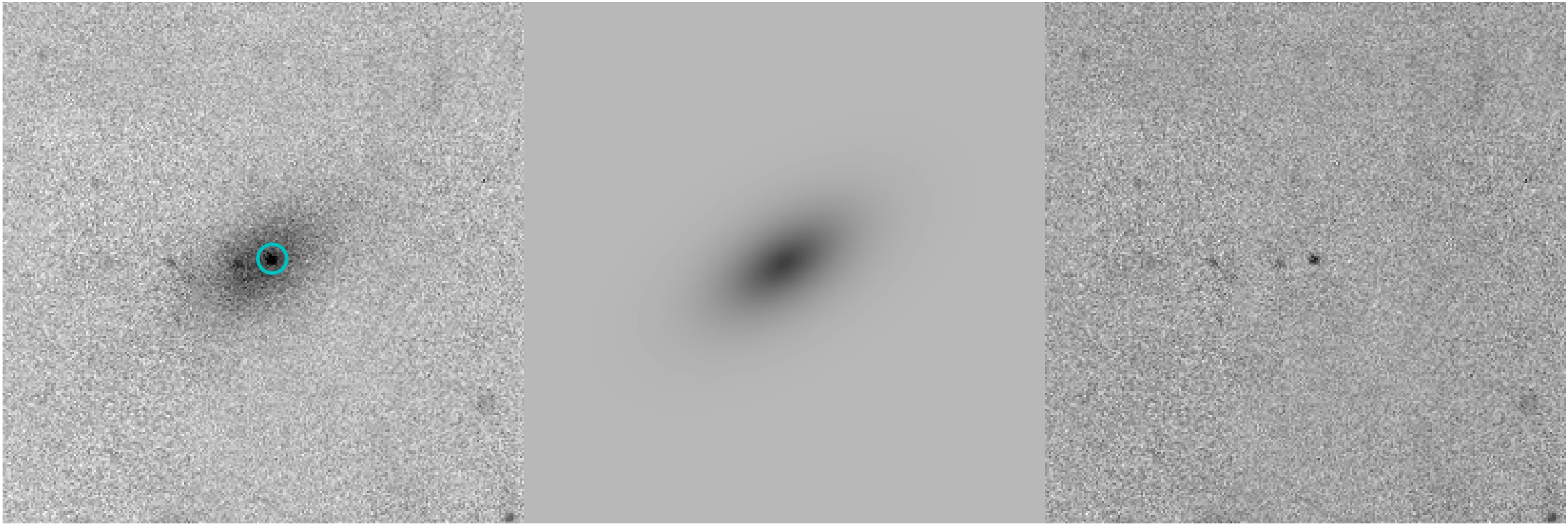}
  \caption{Top: S\'ersic modelling of a dE. Bottom: Masked S\'ersic modelling of a dE,N. The mask (cyan) is visible on the cutout (left) image. Left: Cleaned cutout. Middle: Model. Right: Residual image. The cutouts are 48\arcsec\ $\times$ 48\arcsec\ with North up and East left.}
  \label{fig:Sersic}
\end{figure}

\onecolumn
\begin{landscape}
\begin{table*}
	\small%\addtolength{\tabcolsep}{-2.5pt}
   \caption{\label{tab:catalog_nonnucleated}Structural and photometric properties of the MATLAS dwarfs.}
   \begin{center}
   \begin{tabular}{ccccccccccccccccc}
    \toprule
    ID & Host ETG & \multicolumn{3}{c}{Distance} & RA & Dec & m$_g$ & $(g-i)_0$ & $(g-r)_0$ & n & R$_e$ & b/a & PA & $\mu_{0,g}$ & $\langle\mu_{e,g}\rangle$ & Morph \\
     &  & \multicolumn{3}{c}{(Mpc)} & (deg) & (deg) &  &  &  &  & (arcsec) &  &  & (mag/arcsec$^2$) & (mag/arcsec$^2$) & \\
    (1) & (2) & (3) & (4) & (5) & (6) & (7) & (8) & (9) & (10) & (11) & (12) & (13) & (14) & (15) & (16) & (17)\\
    \toprule
MATLAS-1 & NGC0448 & 29.5 & 1 & -- & 18.2998 & -2.0792 & -- & -- & -- & -- & -- & -- & -- & -- & -- & dI\\
MATLAS-2 & NGC0448 & 29.5 & 1 & -- & 18.3236 & -1.1708 & 19.90 & 0.71 & -- & 1.21   & 3.81 & 0.45 & 1.35 & 23.26 & 24.05 & dE\\
MATLAS-3 & NGC0448 & 29.5 & 1 & -- & 18.3396 & -1.8231 & 20.44 & -- & -- & 0.61** & 4.54 & 0.52 & 85.42 & 24.18 & 24.96 & dI\\
MATLAS-4 & NGC0448 & 29.5 & 1 & -- & 18.4198 & -1.9169 & -- & -- & -- & -- & -- & -- & -- & -- & -- & dI\\
MATLAS-5 & NGC0448 & 29.5 & 1 & -- & 18.6778 & -1.0972 & 20.55 & 0.72 & 0.50 & 1.33   & 3.52 & 0.75 & 38.52 & 23.74 & 24.52 & dE\\
MATLAS-6 & NGC0448 & 29.5 & 1 & -- & 18.6875 & -1.2415 & 19.95 & 0.58 & -- & 0.69   & 3.88 & 0.75 & 46.88 & 23.35 & 24.14 & dE\\
MATLAS-7 & NGC0448 & 29.5 & 1 & 28.96a & 18.7193 & -1.0953 & -- & -- & -- & -- & -- & -- & -- & -- & -- & dE\\
MATLAS-8 & NGC0448 & 29.5 & 1 & -- & 18.7567 & -1.4806 & 19.63 & 1.26 & 0.75 & 0.57   & 8.75 & 0.78 & 48.37 & 24.79 & 25.58 & dEN\\
MATLAS-9 & NGC0448 & 29.5 & 1 & -- & 18.7782 & -1.2741 & -- & -- & -- & -- & -- & -- & -- & -- & -- & dIN\\
MATLAS-10 & NGC0448 & 29.5 & 1 & -- & 18.7949 & -1.4718 & 17.27 & -- & -- & 1.09   & 4.45 & 0.82 & 31.35 & 20.98 & 21.76 & dEN\\
...& ...& ... & ... & ... & ... & ... & ... & ... & ... & ... & ... & ... & ... \\
    \bottomrule
		\end{tabular}
   \end{center}
   \begin{tablenotes}
      \small
      \item \textbf{Notes.} The full table is available as supplementary material as well as at CDS. Columns meanings: (1) Dwarf ID; (2) Assumed host ETG ; (3) Distance; (4) Distance flag: 1 = one single massive Atlas$^{3D}$ galaxy in the field with distance; 2 = several massive Atlas3D galaxies in the field all located at roughly the same distance; 3 = several massive Atlas$^{3D}$ galaxies in the field with a majority at roughly the same distance, but some discrepant one(s); 4 = several massive Atlas3D galaxies in the field with a range of distances; (5) Distance measurement from $^{(a)}$SDSS DR13 database, $^{(b)}$Poulain et al., in preparation, $^{(c)}$\citet{Ann2015}, $^{(d)}$\citet{Karachentsev2013}, $^{(e)}$\citet{Mueller2021}; (6) and (7) Right ascension and declination of the dwarf; (8) Apparent magnitude in the g-band; (9) $g-i$ colour corrected for Galactic extinction; (10) g$-$r colour corrected for Galactic extinction; (11) Sérsic index (12) Effective radius; (13) Axis-ratio; (14) Position angle; (15) Central surface brightness in the g-band; (16) Average surface brightness within R$_e$ in the g-band; (17) Dwarf morphology.
      \item dE* corresponds to the dwarf whose nucleus matches with a star in the Gaia DR2 catalog.
      \item dE** corresponds to the dwarf for which one of the central bright GCs is possibly an NSC \citep{Forbes2019}.
      \item ** corresponds to a Sérsic index derived on a binned image.
    \end{tablenotes}
\end{table*}
\end{landscape}
\twocolumn

\begin{figure}
  \centering
    \includegraphics[scale=0.4]{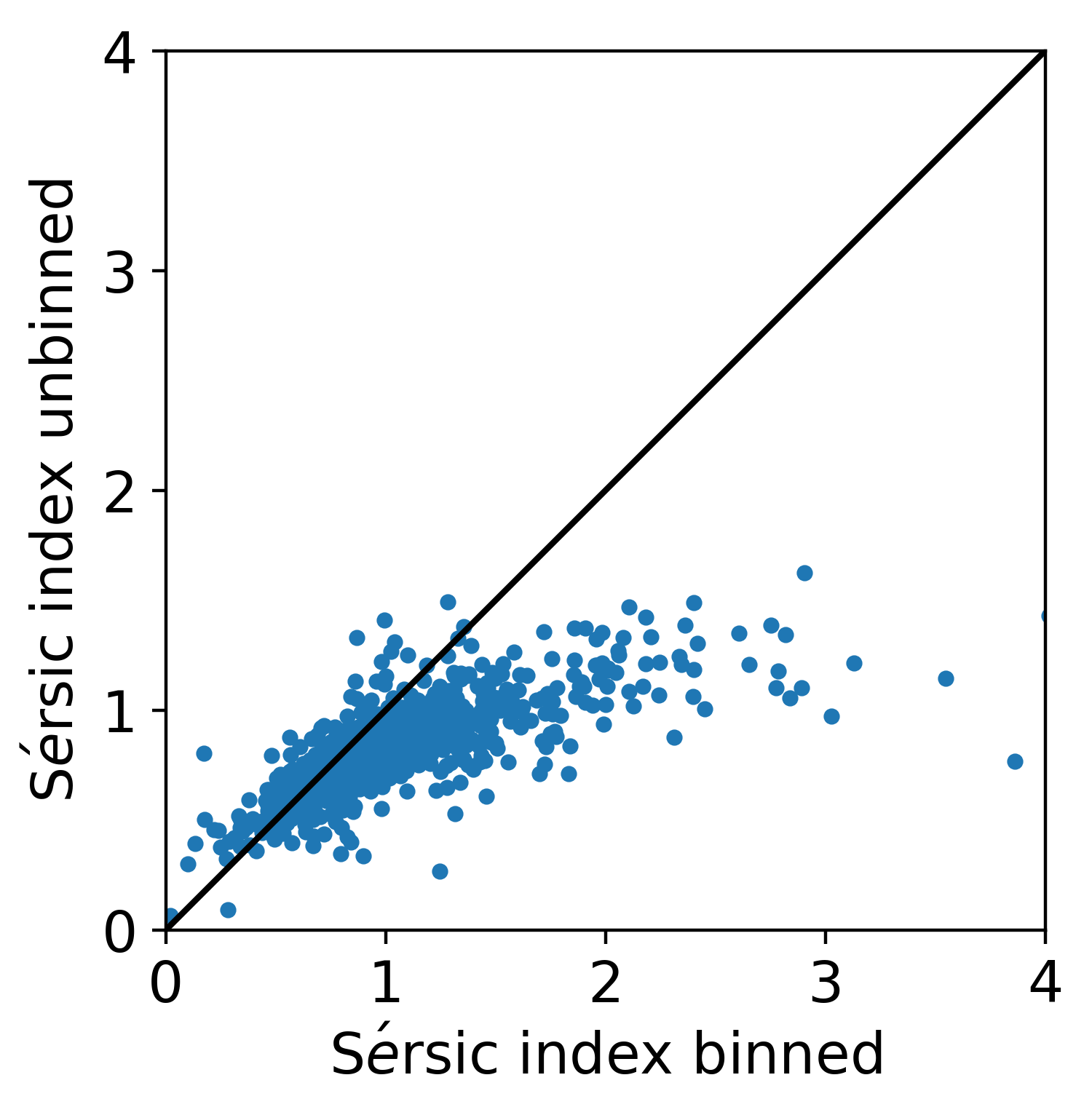}
    \includegraphics[scale=0.4]{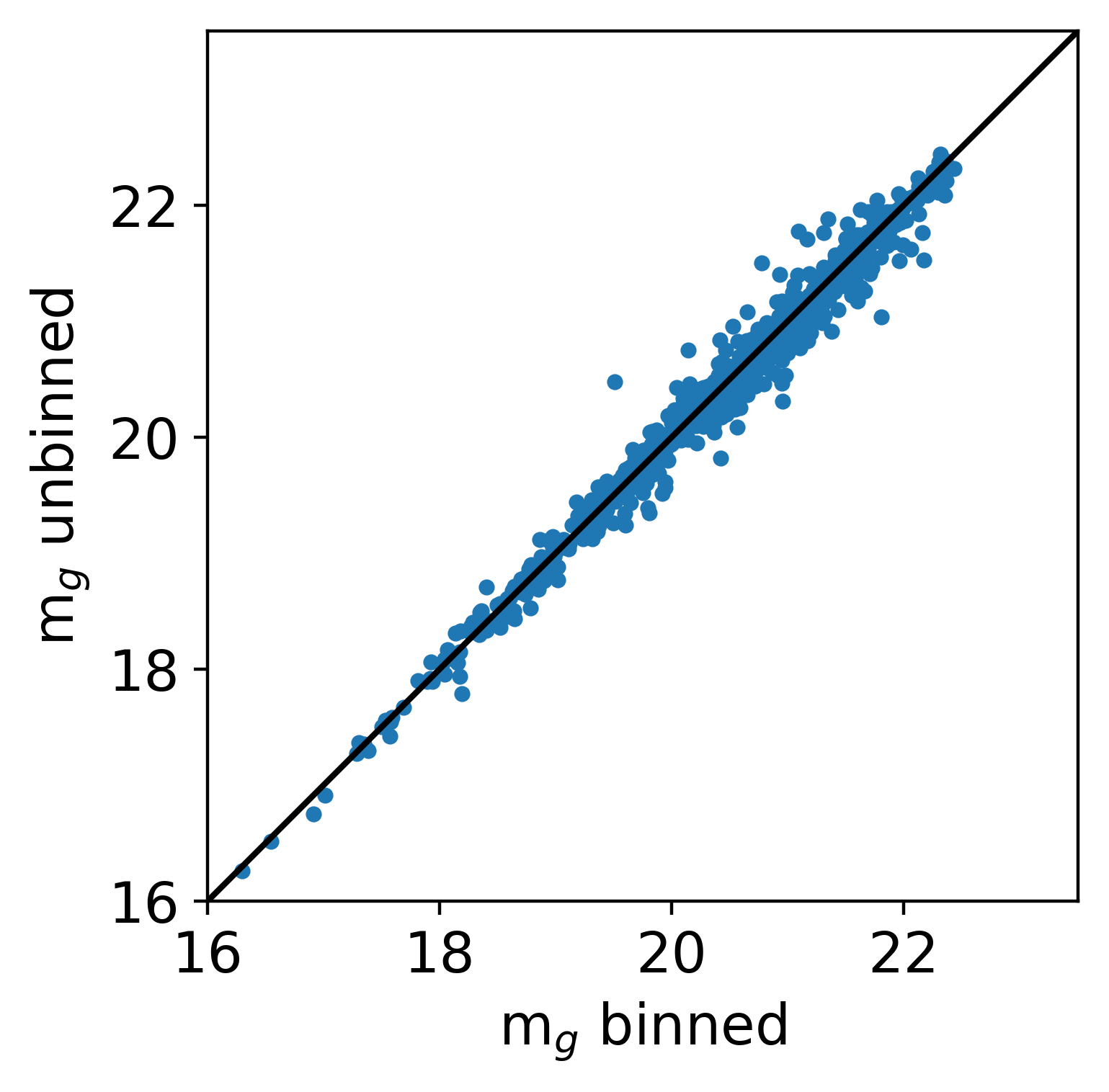}
    \includegraphics[scale=0.4]{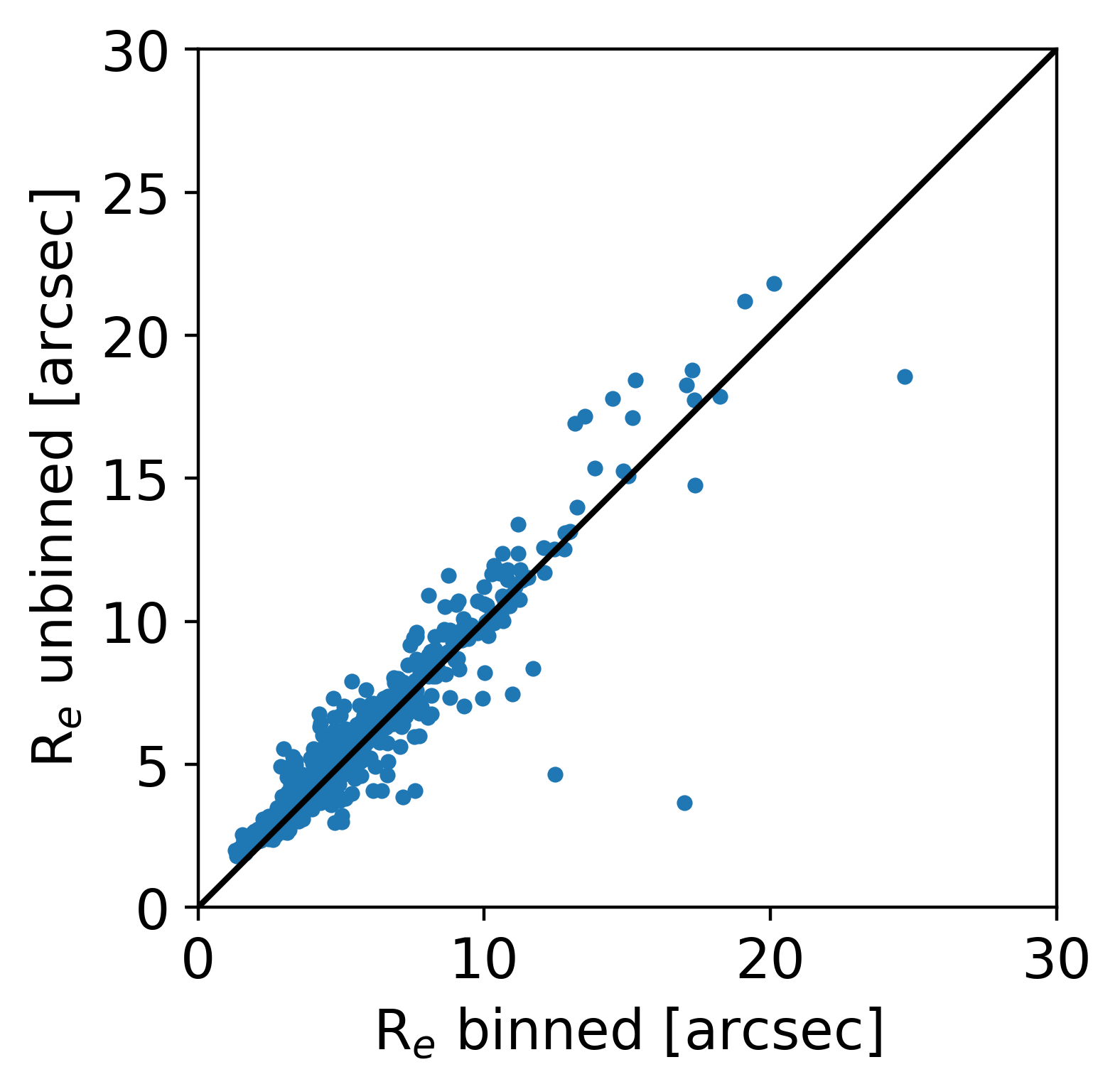}
    \includegraphics[scale=0.4]{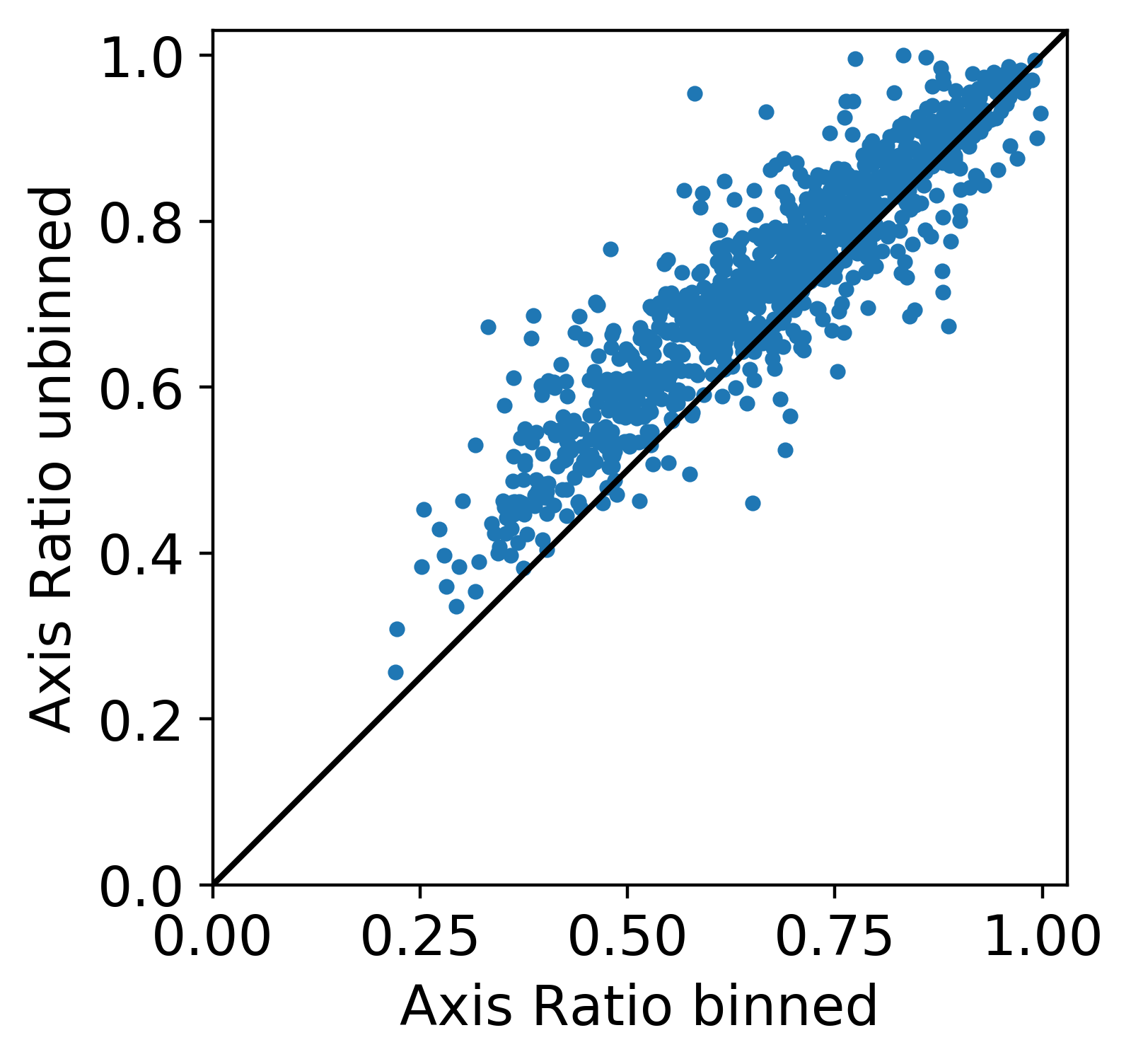}
  \caption{Comparison of the obtained S\'ersic index, m$_g$, R$_e$ and axis ratio} with image resolution used for the model fit. Black line: both axis are equal. The S\'ersic index shows a strong dependence on the resolution of the image.
  \label{fig:unbvsbin_param}
\end{figure}

\section{Galaxy modelling}

The galaxy modelling was performed on the g band images using the software \textsc{galfit} \citep{Peng2010}. We applied a 2D S\'{e}rsic model \citep{Sersic1963} to estimate the surface brightness profiles of the dwarfs.
For each dwarf, we have produced a cutout image using as a side length, a multiple of the effective radius (R$_e$) of the galaxy as estimated by \textsc{source extractor}. We looked for a size of the cutout that would provide the best sky background estimate, i.e. that would allow us to keep sufficient background without including many surrounding bright sources. A sample of 16 dwarf galaxies belonging to the MegaCam field was used for this analysis. For each dwarf, we produced several cutouts with sizes ranging from 5 to 15 R$_e$ and estimated the sky background value for each of them. We found that these values converged at 9 times the effective radius and therefore postage stamps were cut
at 9 R$_e$. On average, this corresponds to an image of approximately 1\arcmin on each side. We then extracted the PSF image of each field using \textsc{PSFEx} \citep{Bertin2011} and performed a first S\'{e}rsic fit of our candidates with \textsc{source extractor} in order to get a first estimate of the structural parameters. These were then used as input parameters in a S\'{e}rsic fit with \textsc{galfit}.

\subsection{Image cleaning}

\textsc{galfit} is not designed to separate individual sources, and therefore light from nearby sources (either point-like or extended) can impact the quality of the modelling. This is especially problematic for the faint dwarfs in our sample. To prevent such contamination, we applied patches to remove sources from the cutouts used for the \textsc{galfit} modelling. To do this, we created a noise-model image, which is our base for the patching, as we will use part of it to mask the undesired sources. This image is composed of the addition of the first S\'{e}rsic model from \textsc{galfit} to a grid of background noise estimated from the cutout.

Our cleaning consists of replacing the bright sources by the corresponding area of pixels from the noise-model image. To detect the bright sources that fall either on top of the galaxy or close to it and may therefore affect the fit, we first ran the \textsc{daophot} algorithm \citep{Stetson1987} above one effective radius of the dwarf. We applied a first cleaning using circular regions of radius 1\arcsec\ as patches. Then, we ran \textsc{source extractor} to detect all remaining bright sources on the cutout image and patched the extended sources, essentially composed of stars with halos and background galaxies, more distant than 2 R$_e$ using the \textsc{source extractor} segmentation image.

We obtained a good result for about two thirds of the galaxies, of which an example can be seen in Figure \ref{fig:cleaning}. For the remaining dwarfs, the cleaning was not sufficient due to a bad first estimate of the model by \textsc{galfit} or a poor estimate of the size of the detected sources by \textsc{source extractor}. To fix this problem, we had to proceed with a manual cleaning using the image patching tool of the software \textsc{gaia} (\textsc{graphical astronomy and image analysis tool}, \citealt{Draper2014}).

After this cleaning, we performed a second S\'{e}rsic fit, of which an example is visible in Figure \ref{fig:Sersic}. For our sample of nucleated dwarfs, to obtain the properties of the diffuse part of the galaxy, we executed a masked S\'ersic fit as well as a double profile (a S\'{e}rsic coupled with a PSF or King profile) fit. For the masked S\'ersic fit, we created a pixel mask at the centre of the galaxies with a circle of area 50 pixels$^2$, which corresponds to a radius $\sim$1.3\arcsec. This size was a good compromise to mask the nucleus without masking a too large part of the dwarf. We gave this mask as input to \textsc{galfit} to perform the S\'ersic model. The double profile fit will be discussed later is Section \ref{section:nuclei}. Combining the results from both types of fit, we obtained a \textsc{galfit} model of the diffuse part for 84\% of our nucleated sample. For most of the nucleated dwarfs without a \textsc{galfit} model, the bad fits are caused by similar features as the non-nucleated dwarfs (see Section \ref{section:binned}).

 \begin{table}
\begin{center}
\begin{tabular}{ccccc} 
 \hline
 Morphology & Number of & Dwarfs with & Unbinned & Binned \\ 
  & dwarfs & model & model & model \\
 \hline
 dE & 1181 & 1022 & 892 & 130 \\ 
 dI & 522 & 142 & 116 & 26 \\ 
 dE,N & 453 & 415 & 415 & - \\ 
 dI,N & 54 & 10 & 10 & - \\ 
 \hline
\end{tabular}
\end{center}
\caption{\label{tab:modeling}Results of the modelling of the MATLAS dwarfs for each morphology and according to the resolution of the images used.}
\end{table}

\begin{figure}
  \centering
    \includegraphics[width=\linewidth]{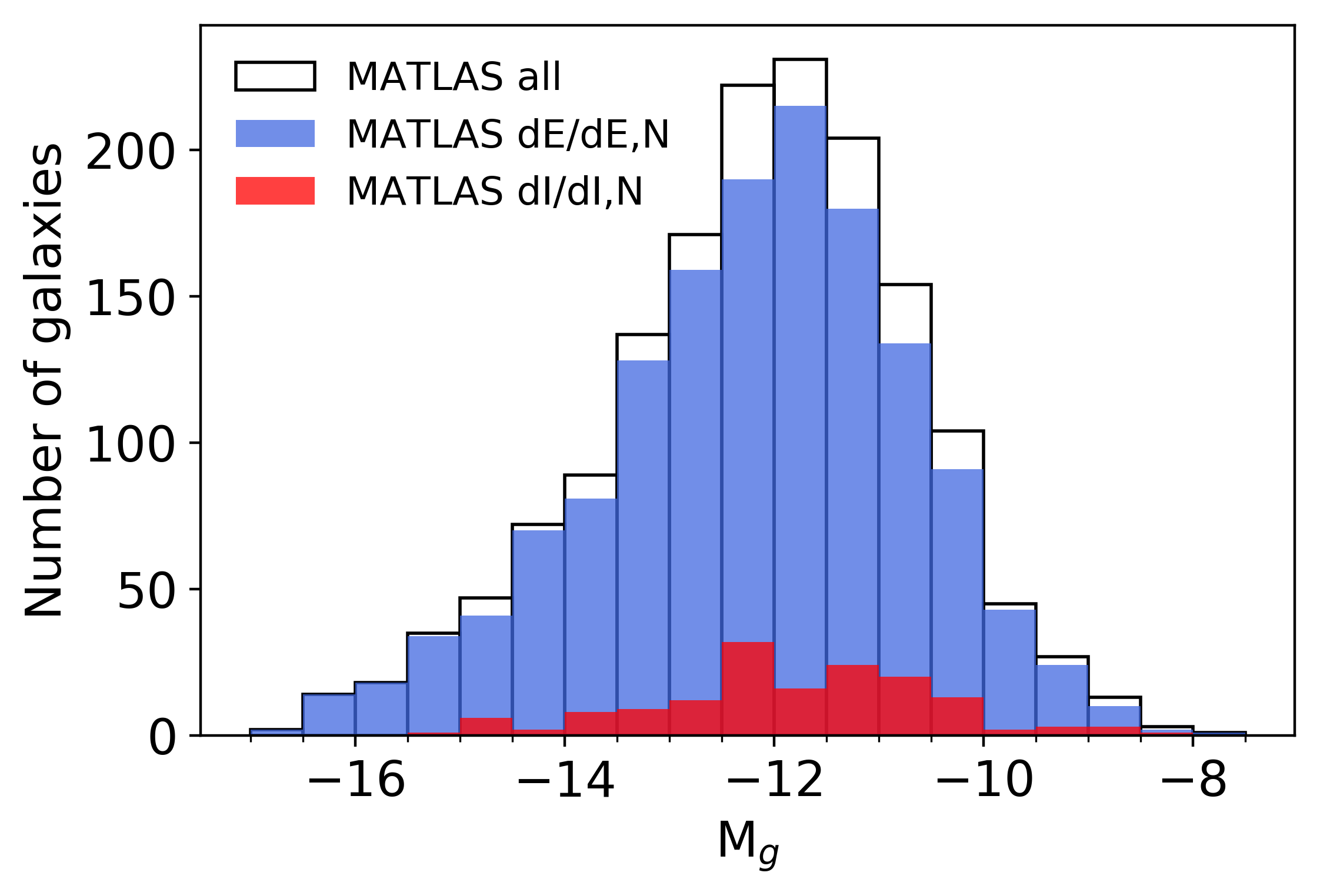}
    \includegraphics[width=\linewidth]{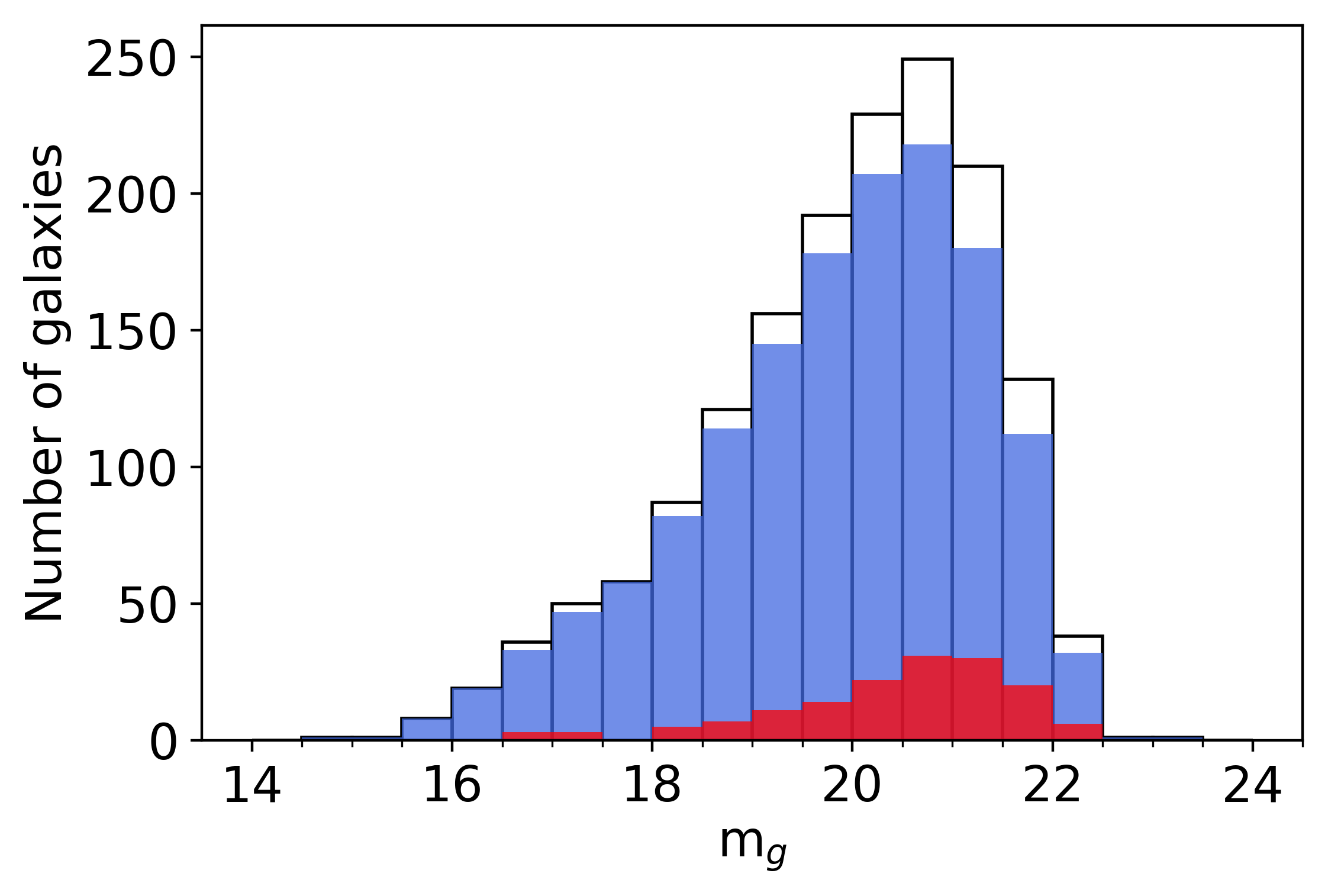}
  \caption{Distribution of M$_g$ and m$_g$ of the modelled MATLAS dwarfs for the full sample (empty bars), the dEs (blue) and the dIs (red).}
  \label{fig:magvs}
\end{figure}

 \begin{figure*}
  \centering
    \includegraphics[width=0.48\linewidth]{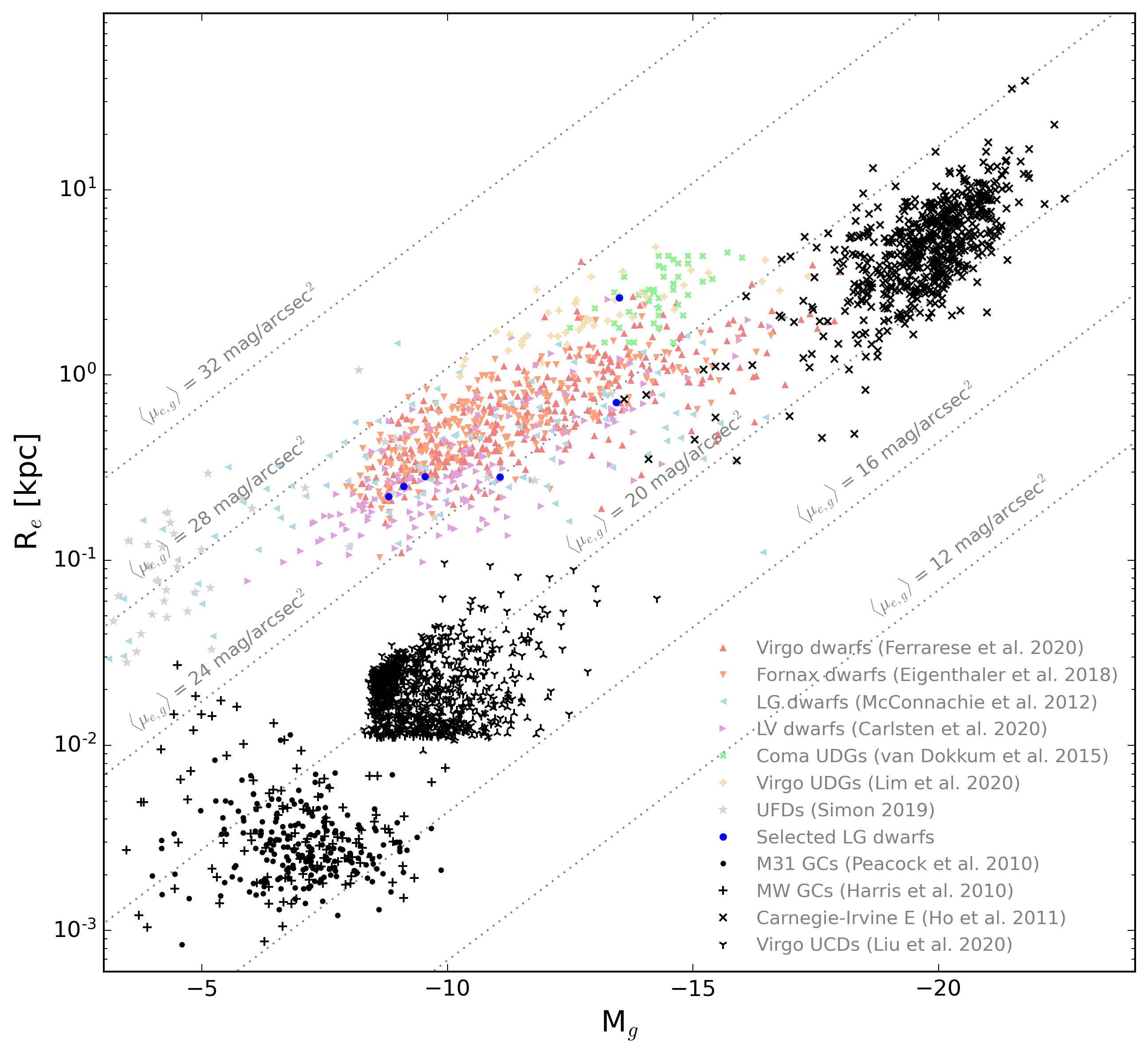}
    \includegraphics[width=0.48\linewidth]{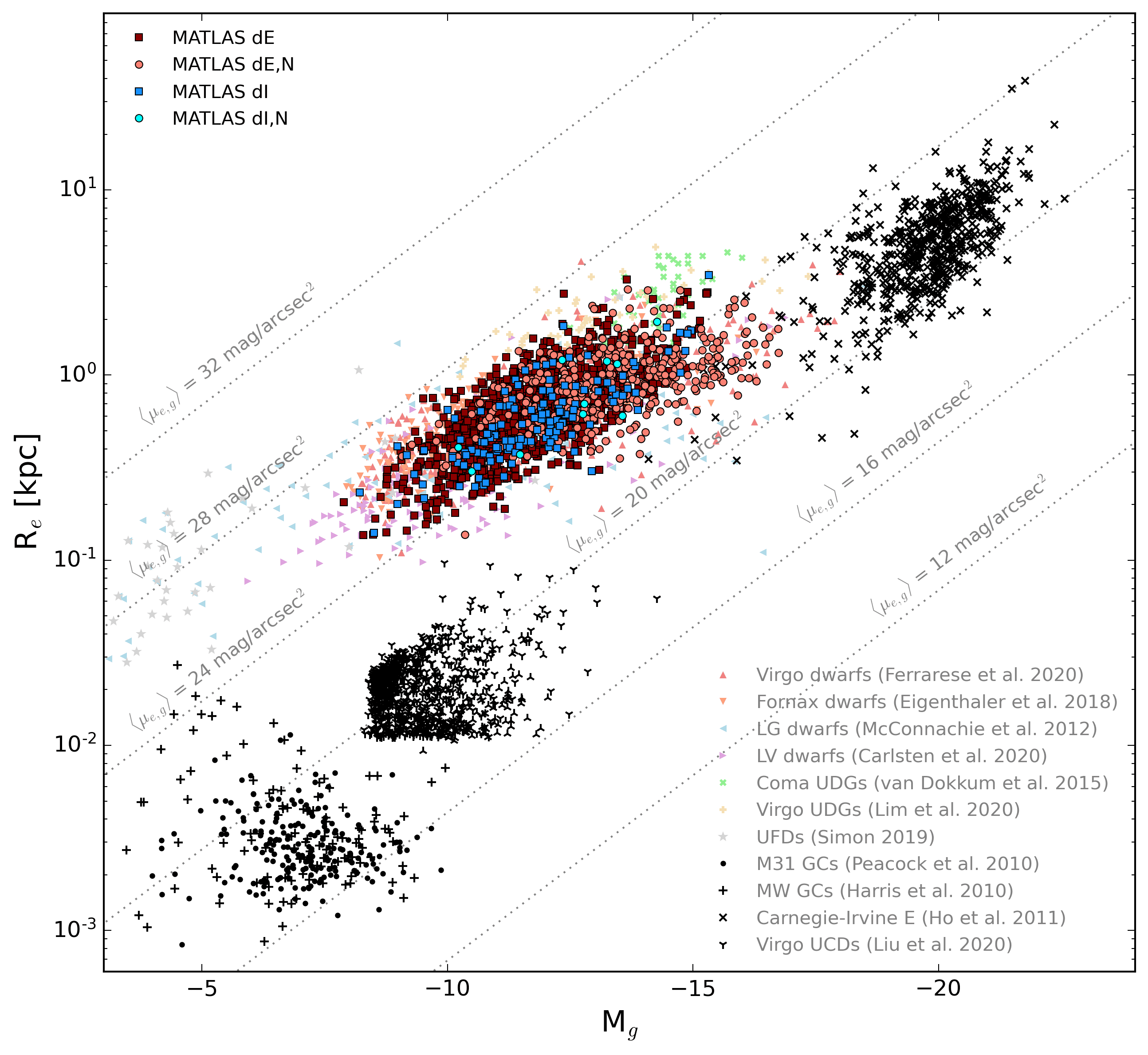}
  \caption{Scaling relation comparing the MATLAS dwarfs with the ones from clusters (NGVS, NGFS), the LG and the LV. We show the dwarfs together with ultra-diffuse galaxies \citep{vanDokkum2015,Lim2020}, ultra-faint dwarfs \citep{Simon2019}, globular clusters \citep{Peacock2010,Harris2010}, massive elliptical galaxies \citep{Ho2011} and ultra-compact dwarfs \citep{Liu2020} for comparison. Left: scaling relation without the MATLAS dwarfs. Some well-known dwarfs from the LG are highlighted with blue dots. Right: Scaling relation with the MATLAS dwarfs. The region defined by the MATLAS dwarfs is precisely overlapping the clusters, LG and LV dwarfs.}
  \label{fig:scaling_relation}
\end{figure*}

\subsection{Image resolution}
\label{section:binned}

Spatially binning the images can enhance the visibility of the dwarfs. For this reason, 3 $\times$ 3 binned images (0.56 arcsec/pixel) were used for the dwarf selection process (Section \ref{section:dwarfsample}). For the sample of nucleated dwarfs, as we wanted the best resolution for the nuclei, we have used the unbinned images (0.19 arcsec/pixel). To determine which image resolution is the best to produce a good S\'{e}rsic model of the non-nucleated dwarfs, we have run \textsc{galfit} on both resolution images. We could obtain a model for 1164 and 1005 galaxies using the binned and unbinned images, respectively. Models could only be obtained in both images for 978 dwarfs.

We have compared the structural parameters (S\'ersic index, apparent magnitude, effective radius and axis ratio) of the dwarfs having a model from both image resolution (see Figure \ref{fig:unbvsbin_param}). All parameters but one are independent of the resolution of the modelled image. The S\'ersic index has a strong dependence to the type of image, as the strength of the effect of binning seems to increase with the S\'ersic index value. 
This is likely due to the fact that the resulting smoothing enhances the central surface brightness of the dwarfs with a bright centre.
This effect explains in part why we could obtain some models using binned images only, as on the unbinned image the galaxy would appear too faint to be correctly fitted by Galfit.

Due to a better modelling of the inner regions, our final sample of dwarf \textsc{galfit} properties is based on the unbinned images, unless the fits failed in which case we used the successful \textsc{galfit} results from the binned images (156 galaxies). As the S\'ersic indices were not accurately recovered for the binned images, we excluded them from future analysis.
In total, for the non nucleated dwarf galaxies, we could model 1164 (68\%) objects in the sample. The bad modelling is mainly caused by the presence of structures like star-forming regions, especially inside the irregular galaxies which, coupled with the irregular shape, result in only 27\% of the dIs having a \textsc{galfit} model. Other reasons, pertaining mainly to the dwarf ellipticals, are the presence of a bright centre or the too faint luminosity of the galaxy. As a consequence, 87\% of the dEs were modeled. A summary of the results of the modelling for each morphology and according to the resolution of the images used is available in Table \ref{tab:modeling}.

\section{The structural and photometric properties of the MATLAS dwarfs}

We present here the structural properties of 1022 dEs, 142 dIs, 415 dE,N and 10 dI,N through scaling relations and parameter distributions. The distributions of absolute and apparent magnitude of the modelled MATLAS dwarfs are presented in Figure \ref{fig:magvs}. The structural and photometric properties of the dwarfs can be found in Table \ref{tab:catalog_nonnucleated}. We compare them to the LG, LV and cluster dwarf galaxies. 

\subsection{Catalogues of dwarfs}

We obtained catalogues of dwarfs from the LG, the LV, and select clusters from the literature to use as comparison samples. For the LG, we used the catalogue from \citet{Mcconachie2012} which summarizes the properties of the LG dwarfs. For the LV, we obtained the catalogue from \citet{Carlsten2020}, a study of the properties of 155 dwarfs located around 10 LV hosts based on data from the CFHT.
To represent the cluster environment, we utilized dwarfs samples from the two nearby Virgo and Fornax clusters. The Virgo cluster dwarfs catalogue comes from the NGVS; introduced in Section \ref{section:dwarfsample}, and focuses on the core (inner one virial radius, R$_{vir}$) of the cluster. We chose two surveys to characterize the Fornax cluster dwarf population: the Next Generation Fornax survey (NGFS; \citealt{Eigenthaler2018}) and the Fornax Deep survey (FDS; \citealt{Venhola2018}). The first makes use of observations from the Dark Energy Camera of the 4-meter Blanco telescope in the u, g, i bands and the second is based on data from the VLT Survey telescope (VST) in the u, g, r, i bands. Both telescopes show similar seeing to the CFHT. These surveys are complementary, as the NGFS focuses on the inner part of the Fornax cluster (within $\sim$ 0.25 R$_{vir}$) while the FDS studies the entire cluster, with galaxies located within and beyond the cluster virial radius. As a consequence, the first includes only dEs, while the second is composed of both dEs and dIs.

\subsection{Comparison with dwarf galaxies from different environments}

Three scaling relations were presented in \citet{Habas2020} (Figure 11) for a sample of 1470 MATLAS dwarfs\footnote{With the use of unbinned and binned images for the modelling of the nucleated and non-nucleated population, respectively.}. These plots show that the MATLAS dwarfs are similar to dwarf galaxies in clusters and the LG in term of size (effective radius), luminosity (absolute magnitude) and surface brightness. We derived the g-band surface brightnesses (central $\mu_{0,g}$; at R$_e$ $\mu_{e,g}$; and within R$_e$ $\langle \mu_{e,g} \rangle$) of the modelled dwarfs based on the equations from \citet{Graham2005}, based on the total magnitude and effective radius returned by \textsc{galfit}. We show in Figure \ref{fig:scaling_relation} the relation between M$_g$ and R$_e$ for a sightly larger sample, including parameters of 119 additional galaxies (mainly dE,N) that we compare with an additional sample of dwarfs from the LV \citep{Carlsten2020}. 
As in \citet{Habas2020}, the region defined by the MATLAS dwarfs is precisely overlapping the clusters and LG. Moreover, it also overlaps the LV dwarfs. This result means that the environment does not dramatically alter the scaling relations, at least in a statistical sense.
Focusing on the morphologies of the MATLAS dwarfs, we can see that both dEs and dIs have the same range of luminosities and sizes. However, as is visible in Figure \ref{fig:magvs}, only faint dIs have been successfully modelled, and thus may introduce a bias regarding the results.\\

\begin{figure*}
  \centering
    \includegraphics[scale=0.42]{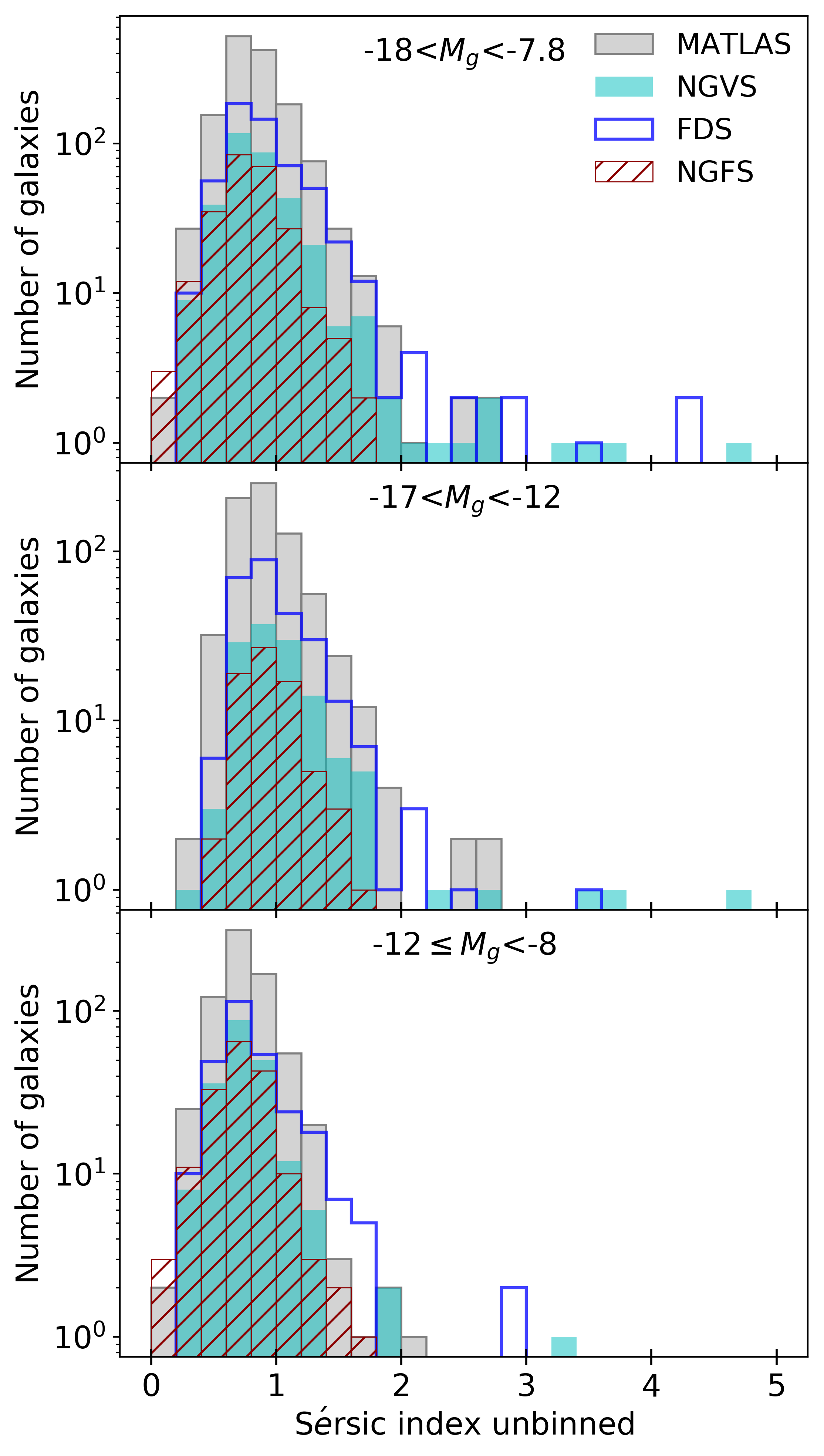}
    \includegraphics[scale=0.42]{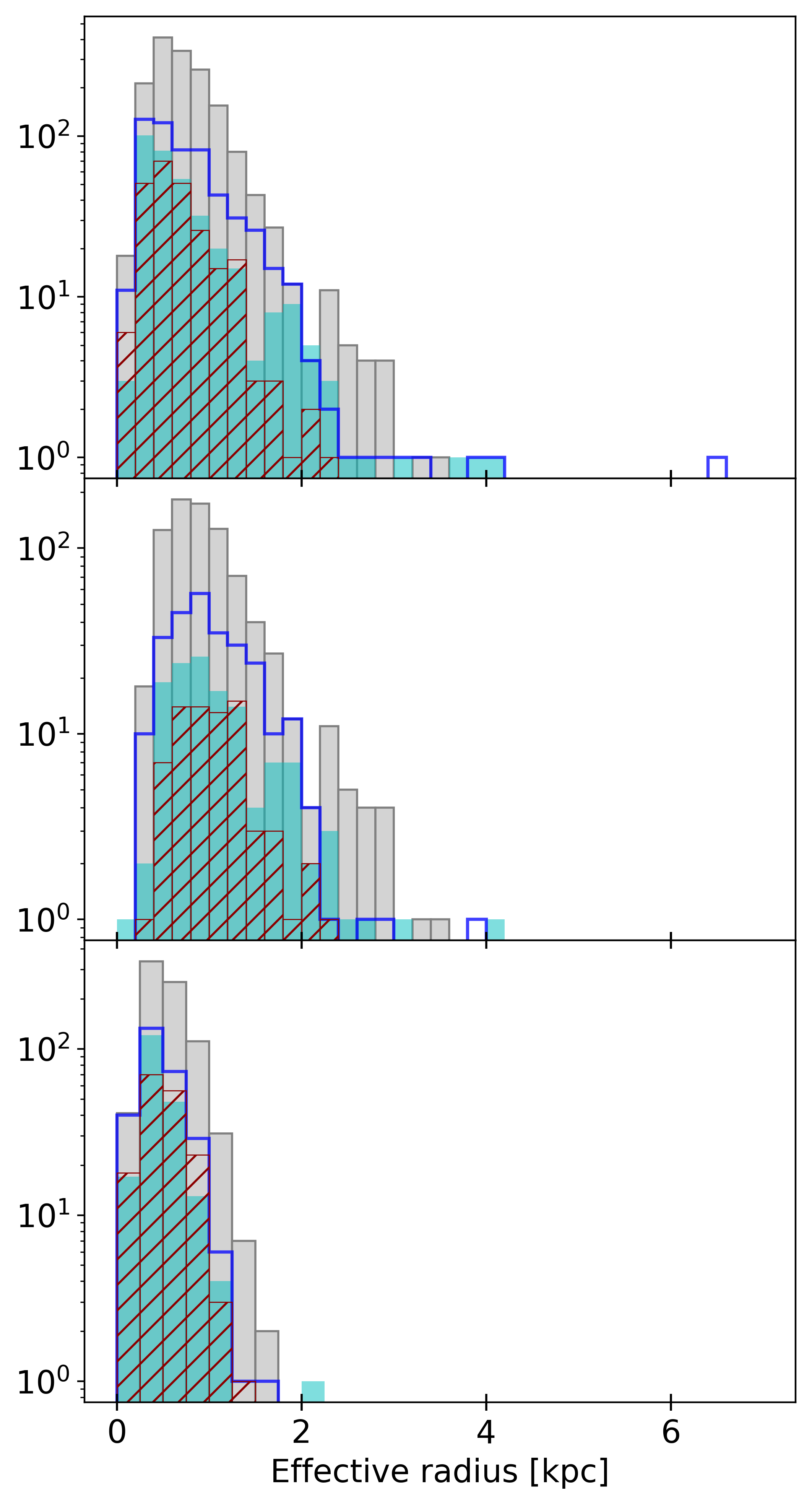}
    \includegraphics[scale=0.42]{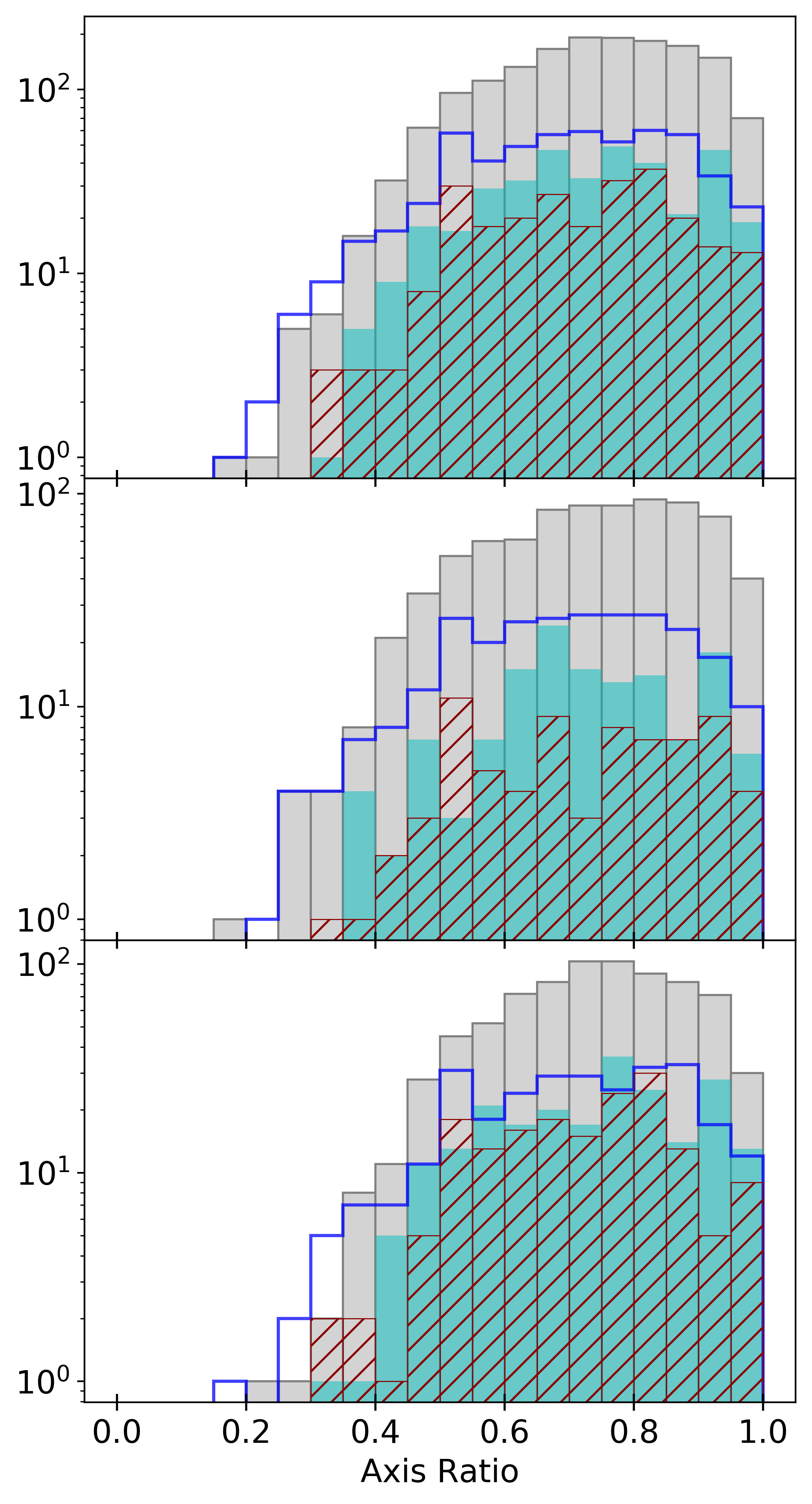}
  \caption{Comparison of distribution of the structural properties of the modelled MATLAS dwarfs with the ones from NGVS, NGFS and FDS per absolute magnitude range. Top: complete samples. Middle: bright dwarfs with $-17$ < M$_g$ < $-12$. Bottom: faint dwarfs with $-12$ $\leq$ M$_g$ < $-8$. Due to the large difference of statistics between the MATLAS distributions and the others, we display the counts in log scale to ensure a better visibility. The MATLAS and cluster dwarfs show similar ranges of structural properties.}
  \label{fig:vs}
\end{figure*}

To investigate the similarities of the structural properties of dwarf populations located in low to high density environments, we focus our study on the distribution of each individual structural property. We use the results of the modelling of the 1589 MATLAS dwarfs and compare the S\'ersic index, effective radius, and axis ratio to the ones measured for the Virgo (NGVS; \citealt{Ferrarese2020}) and Fornax (NGFS; \citealt{Eigenthaler2018}; FDS; \citealt{Venhola2018}) dwarfs with the same absolute magnitude. We divide the MATLAS sample in two absolute magnitude bins, from $-17$ to $-12$ and from $-12$ to $-8$, where $-17$ is the brightest common absolute magnitude bin between the samples and $-12$ is the median M$_g$ of the MATLAS sample (see Figure \ref{fig:vs}). 
One can see that the MATLAS and cluster dwarfs show similar structural properties, with the bright dwarfs showing a broader range of Sérsic index and effective radius than the faint ones. 

To test if the distributions of each property of the MATLAS and cluster dwarfs are drawn from a common distribution, we have performed two-sample Kolmogorov-Smirnov (KS) tests on the two absolute magnitude bins of bright (M$_g$ = $-17$ to $-12$) and faint (M$_g$ = $-12$ to $-8$) dwarfs. The null hypothesis is that the two samples come from the same distribution, and we set a significance level of $\alpha =$ 0.05. We report the obtained p-values in Table \ref{tab:KStests}. We split our analysis considering on one side the dwarfs located in the core of clusters (the NGFS and NGVS samples) and on the other side dwarfs in the whole cluster (the FDS sample). In the cluster core, based on the p-values, we cannot reject the hypothesis that the properties are drawn from the same population as the ones of the MATLAS dwarfs when considering: 1) the Sérsic index of all NGFS dwarfs, NGVS faint dwarfs, 2) the axis ratio of all dwarfs, and 3) the effective radius of NGFS faint dwarfs and NGVS bright dwarfs. But, we reject the null hypothesis for: 1) the Sérsic index of NGVS bright dwarfs, and 2) the effective radius of NGFS bright dwarfs, NGVS faint dwarfs. Thus, in the core of clusters, the p-values of most of the properties of the bright and faint dwarfs suggest that these galaxies are likely drawn from the same population as the MATLAS dwarfs, implying the possibility of a similar formation scenario for these galaxies.
However, we find a different result when looking at the whole Fornax cluster dwarf population, as we reject the null hypothesis for all the properties but the Sérsic index of the faint dwarfs. This suggests that even though the dwarfs in the core of clusters possibly share the same formation background as the MATLAS dwarfs, it may not be the case for all the dwarfs in the cluster environment.

\begin{table}
\begin{center}
\begin{tabular}{ccccccc} 
\hline
p-values & \multicolumn{2}{c}{Sérsic index} & \multicolumn{2}{c}{Axis ratio} & \multicolumn{2}{c}{Effective radius}\\
\hline
Sample & Bright & Faint & Bright & Faint & Bright & Faint \\
\hline
NGFS & 0.83 & 0.16 & 0.40 & 0.27 & 0.04 & 0.26\\ 
NGVS & 2.40e-3 & 0.36 & 0.20 & 0.55 & 0.08 & 2.75e-3\\ 
FDS & 0.05 & 0.02 & 0.02 & 0.03 & 0.02 & 1.07e-4\\ 
\hline
\end{tabular}
\end{center}
\caption{\label{tab:KStests}The two-sample KS tests p-values for the bright (M$_g$ = $-17$ to $-12$) and faint (M$_g$ = $-12$ to $-8$) dwarf samples as compared to the MATLAS dwarfs properties.}
\end{table}

\begin{figure*}
  \centering
  \includegraphics[scale=0.54]{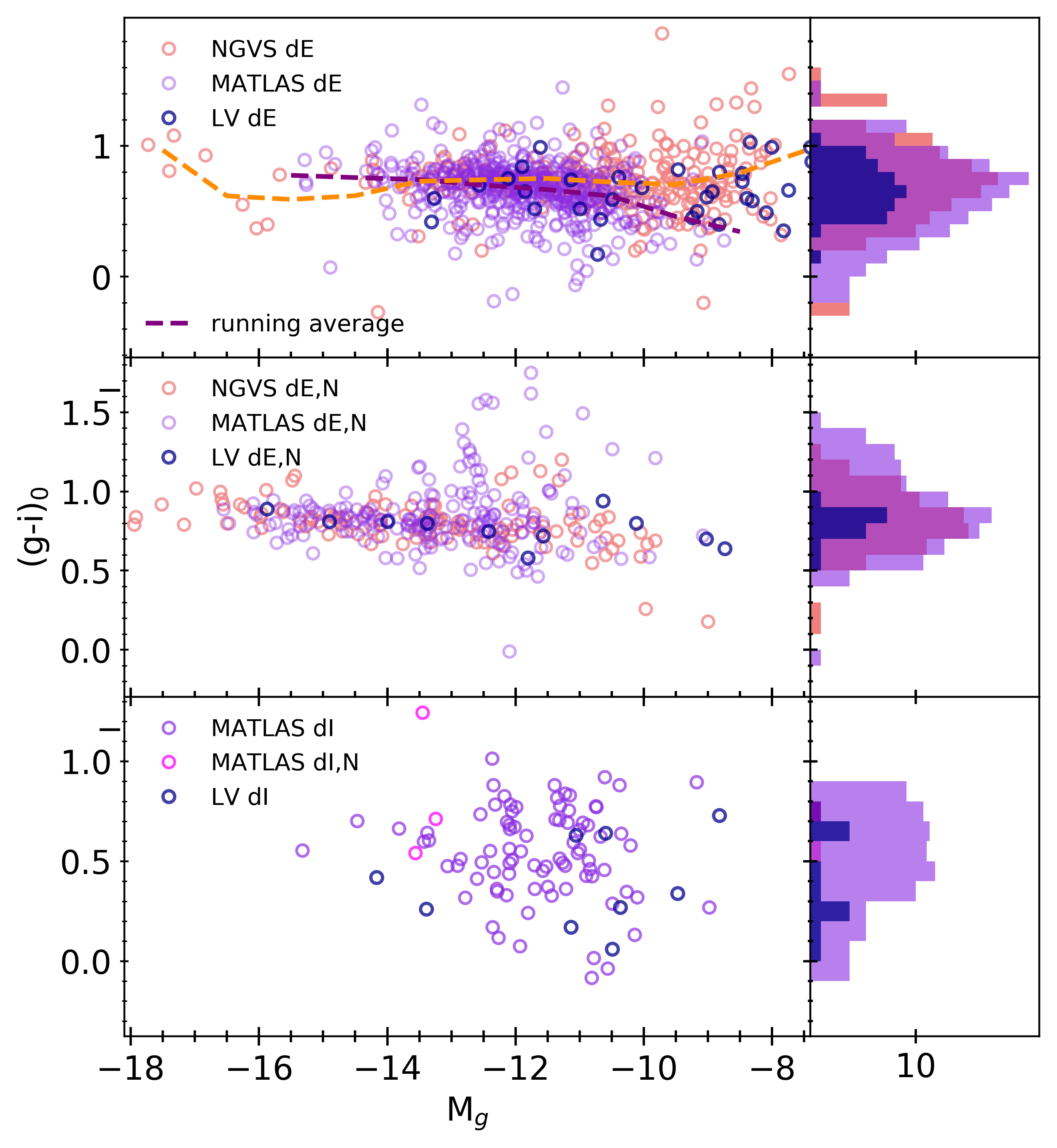}
  \includegraphics[scale=0.54]{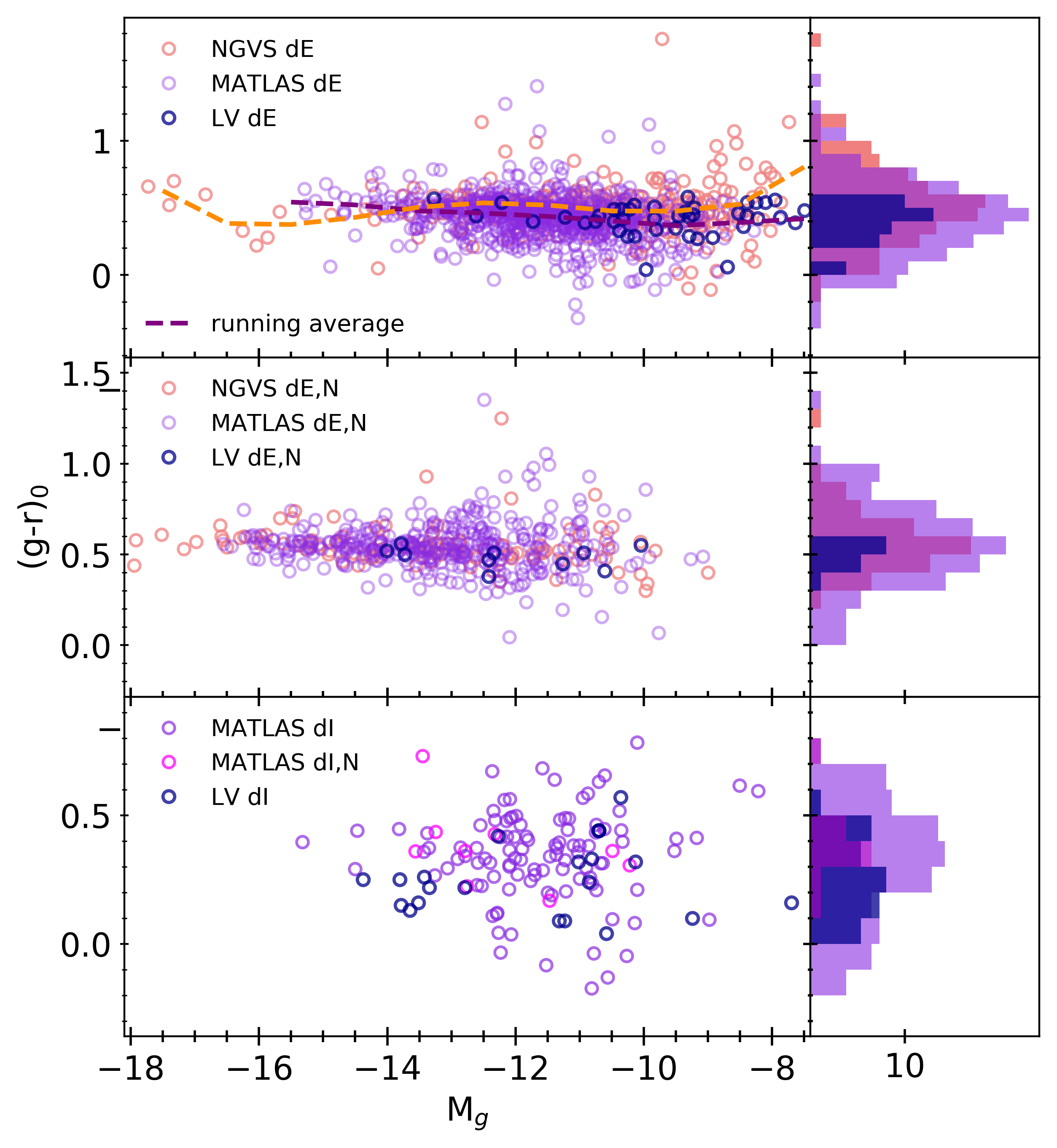}
  \caption{Colour-magnitude diagrams of the MATLAS dwarfs (purple) according to the different morphologies compared with the NGVS (orange) and LV (dark blue) dwarfs. Left: $(g-i)_0$ colour comparison. Right: $(g-r)_0$ colour comparison. Dashed lines: running average of the $(g-i)_0$ and $(g-r)_0$ colours for the dE population of MATLAS and NGVS. The colours are corrected for Galactic extinction. The dwarfs are showing similar colours, independently of their local environment.}
  \label{fig:colormag_vs}
\end{figure*}

\subsection{Colours of the MATLAS dwarfs}
\label{section:dwarfcolor}

To obtain the colours of the galaxies successfully modelled in the g-band, we ran the exact same model on the i and r band images, leaving only the magnitude and sky values free to change in the input parameters of \textsc{galfit}. Fewer fields have available observations in the i and r bands than in the g band, thus we obtained a \textsc{galfit} model for 782 and 1307 dwarfs in the i and r-band, respectively.\\

To look for the influence of the environment on the dwarfs colours and compare the stellar population according to their morphology of the MATLAS dwarfs, we compare our samples with dwarfs located in high and medium to low density environments. To perform a more robust comparison, we limit the selected samples to the ones observed with the filters of MegaCam on the CFHT. Therefore, we are comparing the MATLAS dwarfs to those from the NGVS \citep{Janssen2019} and to dwarfs in the LV situated around isolated galaxies or in groups \citep{Carlsten2020}. We present colour-magnitude relations (CMRs, \citealt{Bell2004}) for the Galactic extinction corrected\footnote{The extinction corrections are from \citet{Schlafly2011}.} $(g-r)_0$ and $(g-i)_0$ colours in Figure \ref{fig:colormag_vs}. To be consistent in our comparison, all the magnitudes are issued from single S\'{e}rsic modelling (i.e. for the nucleated dwarfs, the nuclei were masked or modelled by a PSF or King profile). The CMRs are divided according to morphological type with the dEs at the top, the dE,N in the middle and the dIs, dI,N at the bottom. To ensure a better visibility of the dE populations' colours of MATLAS and NGVS, we computed the running average of the colours, represented by dashed lines. We note that, in contrary to the nuclei (see Section \ref{section:nucprop}), we do not indicate the \textsc{galfit} statistical error on the colour of the dwarfs, as the error bars would appear smaller than the size of the marker. Focusing on the $(g-i)_0$ colour, the MATLAS dEs, dE,N and dIs have a median colour of 0.70$\pm$0.20, 0.82$\pm$0.22 and 0.55$\pm$0.23, respectively. While the MATLAS dEs, dE,N and dIs have a median $(g-r)_0$ colour of 0.46$\pm$0.16, 0.54$\pm$0.13 and 0.36$\pm$0.18, respectively. Looking at the different morphologies, we note that dE,N show redder colours than dEs and that dIs appear to be bluer than dEs. Comparing the MATLAS dwarfs to the ones from the Virgo cluster and the LV, one can see that for all morphological types, the dwarfs are showing similar colours, independently of their local environment, meaning that the MATLAS dEs are, on average, as red as clusters dwarfs. 
This result contradicts the findings of simulations \citep{Mistani2016} and observations \citep{Haines2008,Geha2012} that star-forming, blue, dwarf galaxies are found in majority in low-density environments.
We also note that the obtained colours are consistent with measurements from different instruments in groups \citep{Mueller2017,Mueller2018} and in the Fornax cluster \citep{Venhola2019}.

\begin{figure}
  \centering
    \includegraphics[scale=0.35]{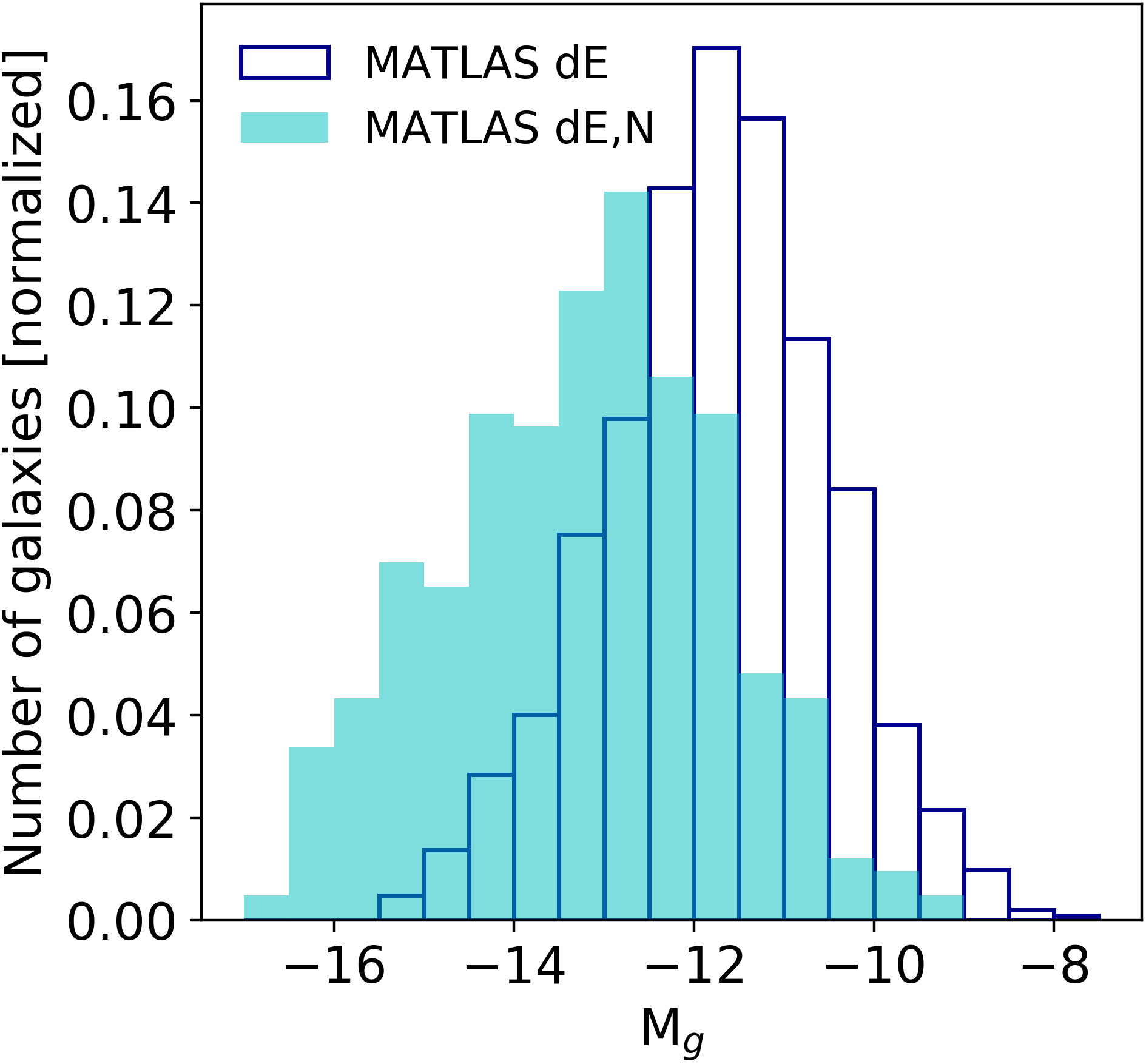}
    \includegraphics[scale=0.35]{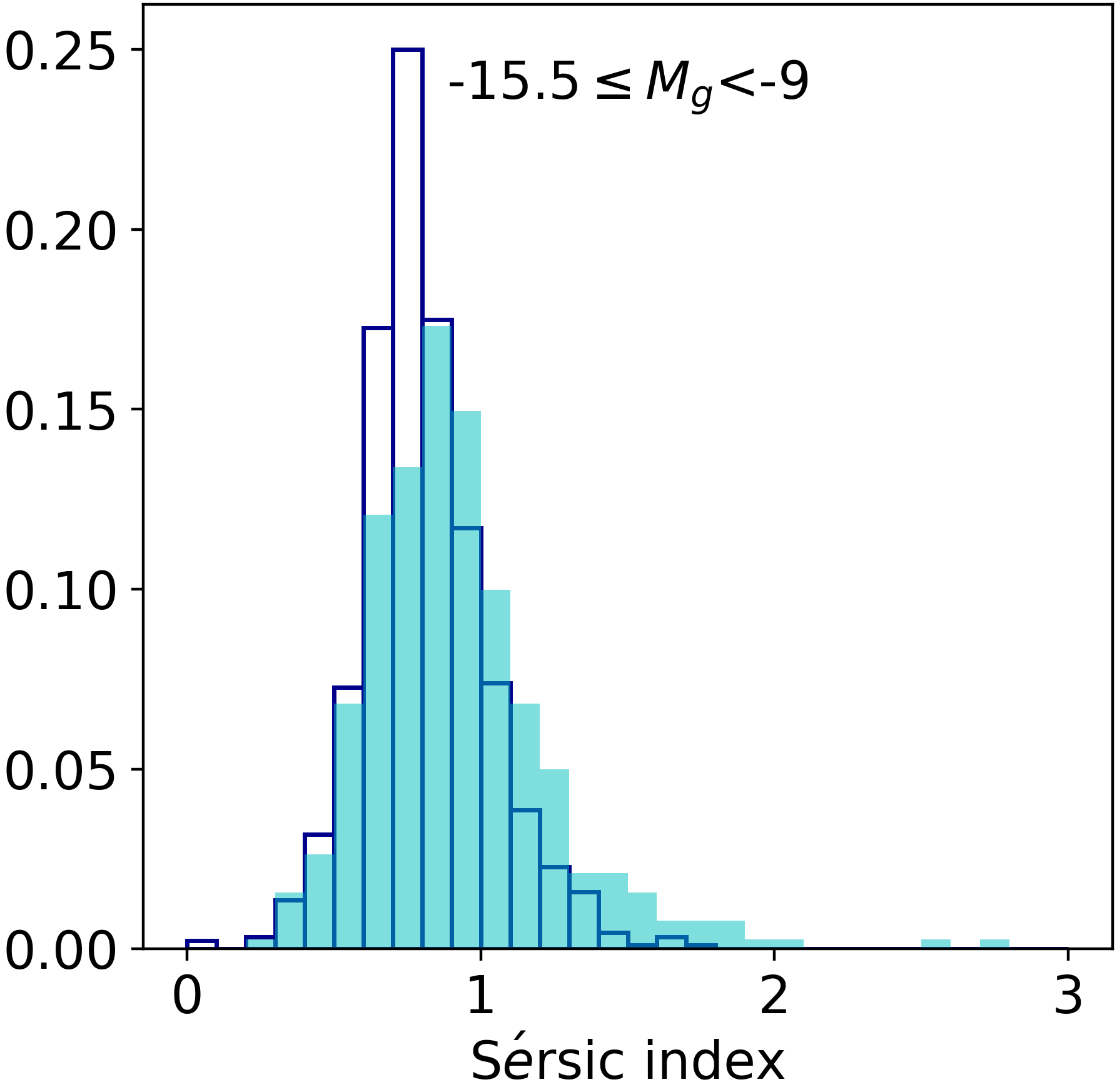}
    \includegraphics[scale=0.35]{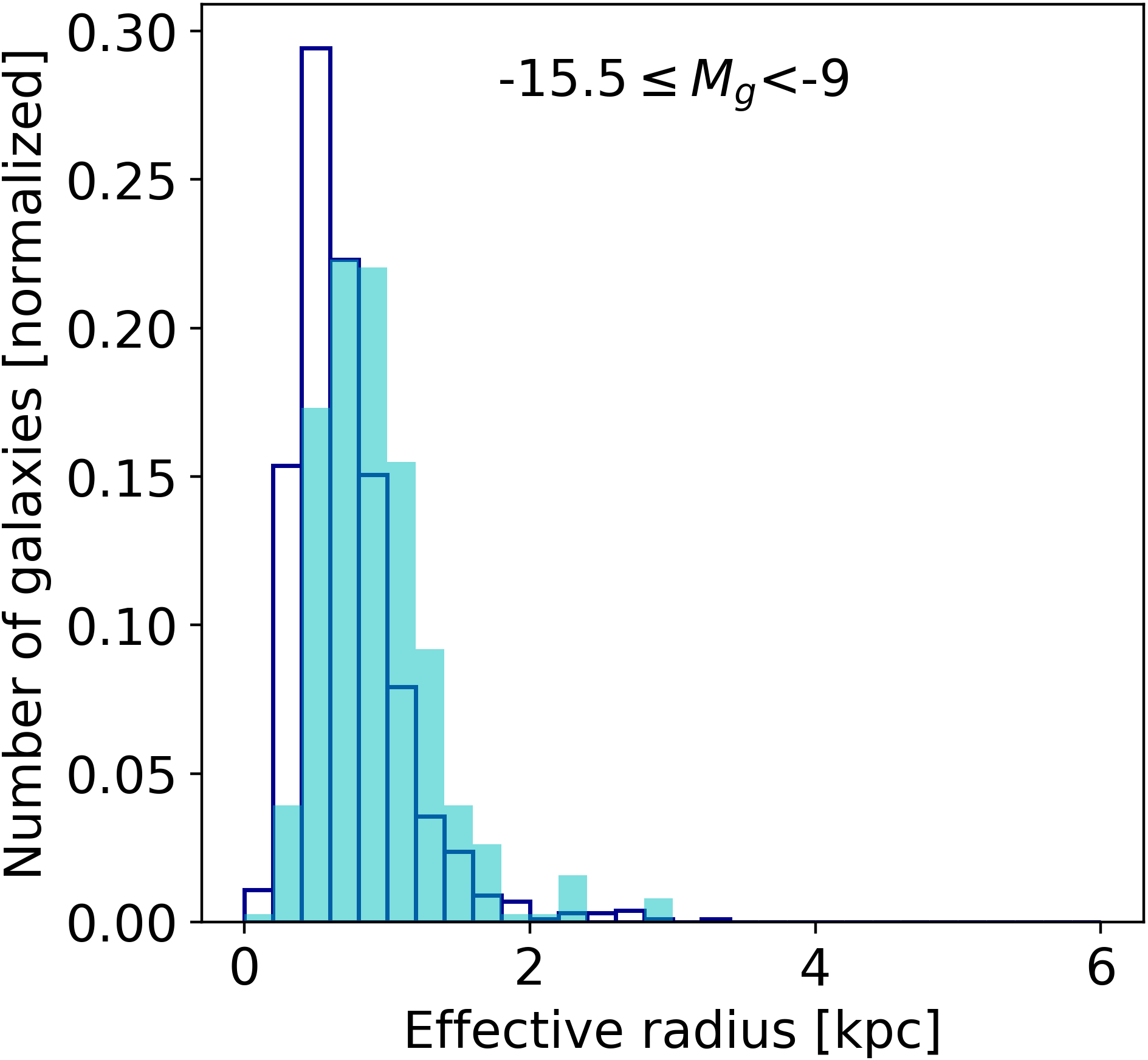}
    \includegraphics[scale=0.35]{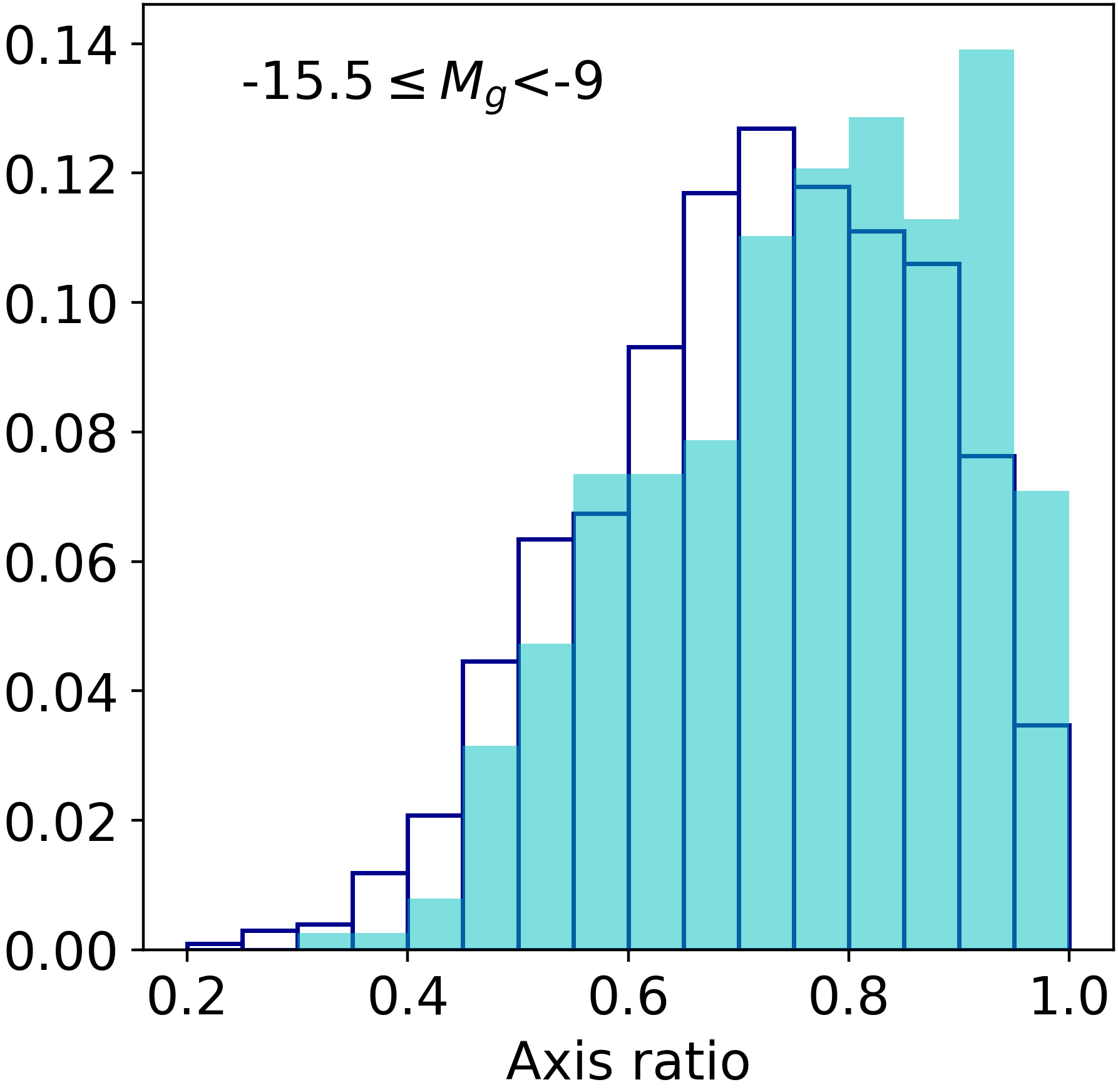}
  \caption{Comparison of distribution of structural properties between MATLAS nucleated and non-nucleated elliptical dwarf galaxies. The histograms are normalized to ensure a better visibility. A shift is visible between the distributions, with the nucleated dwarfs being rounder than non-nucleated dwarfs and less nuclei observed in dwarfs with a S{\'e}rsic index < 1 and a R$_e$ < 1 kpc.}
  \label{fig:properties}
\end{figure}

\subsection{The structural properties of the MATLAS nucleated dwarfs}

In the high density environment of the Virgo and Fornax clusters' core, populated by dEs, about 30\% of the dwarf galaxies show a central nucleus \citep{Janssen2019,Ordenes2018}. However, only 3 dwarf galaxies of the LG show a central nuclei: the M31 satellites NGC205 and M32 \citep{Kent1987,Lauer1998,Mateo1998,Butler2005,DeRijcke2006} and the MW satellite Sagittarius dSph \citep{Mateo1998,Monaco2005,Bellazzini2008}. In the MATLAS sample, about 23\% of the dwarfs are nucleated (Section \ref{section:dwarfsample}) with $\sim$10\% of the irregulars and $\sim$39\% of the ellipticals. The MATLAS dE sample shows a slightly larger fraction of nucleated than in Virgo and Fornax clusters, which is consistent given the less restrictive offset criteria chosen for the nucleus definition (see Section \ref{section:nuclei}).
A first study of the nucleated sample, looking at the fraction of nucleated of the MATLAS dwarfs as a function of M$_g$, has been done in \citet{Habas2020}. The results are in agreement with the findings in the Virgo and Fornax clusters, i.e., that the brighter dwarf galaxies tend to be more nucleated. In this section, we focus on four parameters of the nucleated dwarf population (absolute magnitude, S\'ersic index, effective radius and axis ratio) that we compare to the non-nucleated population.

In the scaling relation of Figure \ref{fig:scaling_relation}, the nucleated population of the MATLAS sample is represented in blue while the non-nucleated one is in red. We see a trend of the nucleated dwarfs to be larger and brighter than the non-nucleated ones. To look more in detail at this trend, we compare the distributions of structural parameters of these two dwarf populations. To avoid any possible biases from the dI population, we consider here only the dEs. In Figure \ref{fig:properties} are represented the distributions of the absolute magnitude, S\'ersic index, effective radius, and axis ratio. As seen in the top left panel, the non-nucleated and nucleated population have different ranges of M$_g$, with the nucleated dwarfs being brighter. The structural properties of the nucleated dwarfs are compared to those of the non-nucleated dwarfs within the same absolute magnitude range, i.e. $-15.5\leq$ M$_g$ < $-9$. We observe in all cases a shift of the distributions peaks between the nucleated (in cyan) and non-nucleated (in blue) dwarfs. These differences mean that nucleated are rounder than non-nucleated dwarfs and that we observe less nuclei in dwarfs with a faint centre (i.e. a S\'ersic index < 1) and a small size (i.e. a R$_e$ < 1 kpc). A similar trend has also been noted for Fornax and Virgo dwarfs \citep{Eigenthaler2018, Sanchez-Janssen2016}.

\section{The properties of the compact central nuclei}
\label{section:nuclei}
The nucleated dwarf galaxies and their nuclei have mainly been studied in the environment of clusters (Coma, Virgo and Fornax, \citealt{denBrok2014,Janssen2019,Ordenes2018}) and nearby groups \citep{Georgiev2009,Fahrion2020}. In this section, we investigate the properties of the nuclei of the MATLAS dwarfs, located in low-to-moderate density environments, as compared to clusters dwarf nuclei.
We define a nucleus as a compact source near the dwarf photocentre (within $\sim$0.5 R$_e$, Section \ref{section:nucprop}) that appears to be the brightest compact source within the galaxy's effective radius. When two to three point sources of similar magnitude were located close to the galaxy photocentre and appeared brighter than the surrounding point sources, we defined them as multiple nuclei (see Section \ref{section:multinuc}).
This selection gave us a sample of 508 nucleated dwarf candidates. As NSCs in dwarfs can have a similar magnitude to the brightest globular clusters, we have estimated the probability of a nuclei to actually be confused with a globular cluster projected near the centre of the dwarf. The Monte-Carlo simulation used for this purpose is described in Appendix \ref{Appendix_A1}. The probability that a bright GC is found near the centre of the dwarf depends on the distance of the dwarf which determines how many GCs will be above the detection limit. When the nucleated dwarfs are located at distances $\leq$ 20 Mpc [$>$ 20 Mpc], we compute a probability of contamination $\lesssim$ 10\% for galaxy photocentre separations below 0.5\arcsec\ [below 1.5\arcsec]. A contamination of our sample by foreground stars is also possible, since they can be confused with unresolved nuclei. Based on the sample of stars observed by the Gaia mission (see Appendix \ref{Appendix_A2} for details), we estimated a Galactic star contamination rate of only 0.2\% (1/508) in the nucleated sample. We removed this dwarf from the nucleated sample. We note that another source of contamination can come from background galaxies shining through in the centre.

In this section we describe how we have extracted the structural and photometric properties of the dwarf nuclei (single or multiple) and present the results.

\subsection{Nucleus modelling}
\label{section:doubleprofile}

\begin{figure}
  \centering
    \includegraphics[width=\linewidth]{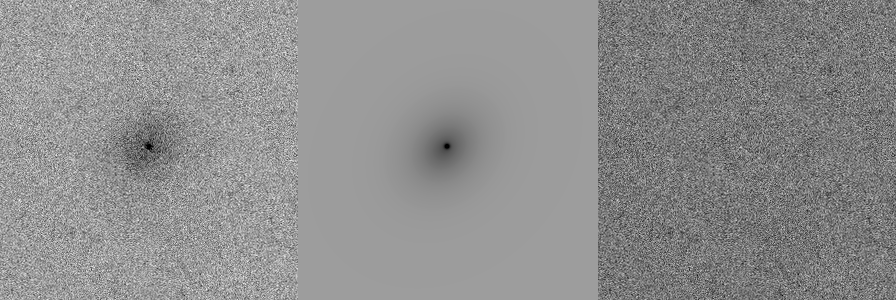}
    \includegraphics[width=\linewidth]{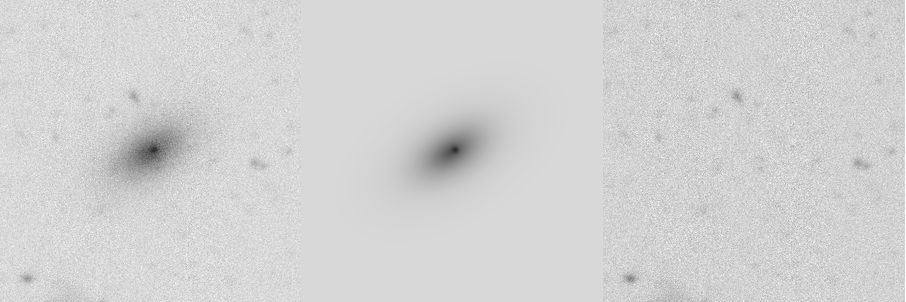}
    \includegraphics[width=\linewidth]{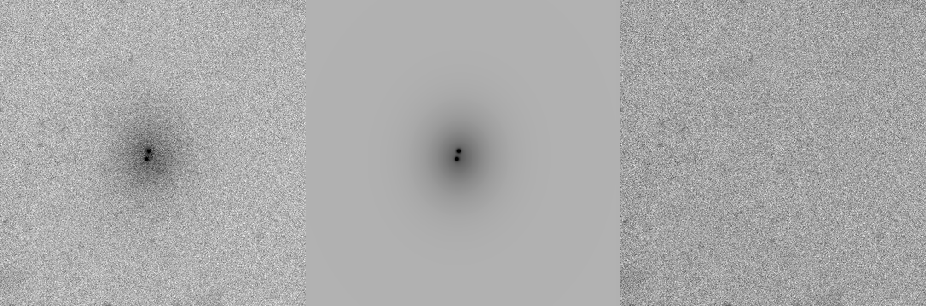}
  \caption{Examples of multiple component fitting. Top: S\'ersic + PSF profile. Centre: S\'ersic + King profile. Bottom: S\'ersic + double PSF profile. The two top rows cutouts are 56\arcsec\ $\times$ 56\arcsec\ while the bottom row cutouts are 67\arcsec\ $\times$ 67\arcsec. All images have North up and East left.}
  \label{fig:SersicPSF}
\end{figure}

To measure the properties of the dwarf nuclei, we used a double profile fit (or multiple profile fit for multiple nuclei) while running \textsc{galfit}. The profile fit is composed of a S\'{e}rsic fit for the diffuse component and a PSF (unresolved nucleus) or a King profile (slightly extended nucleus) for each of the nuclei (see Figure \ref{fig:SersicPSF}). If a \textsc{galfit} model for the diffuse component was available from the fits with the nucleus masked, we used its parameters as input for \textsc{galfit}.
When \textsc{galfit} encountered a problem finding the position of the nucleus while fitting a PSF or a King profile, we used \textsc{daophot} to detect it and establish its coordinates.

We visually inspected the results and determined the best models for the nuclei. However, when our visual inspection could not distinguish between the King and PSF models, we used the azimuthally averaged surface brightness profile of both cutout and model images to identify the best fit. This method, based on \citet{Jedrzejewski1987}, fits isophotes and returns the averaged intensity value at each interval of semimajor axis from the centre of the cleaned cutout image towards the edges. The obtained intensity for each image was then converted to surface brightness after subtracting the background.\\ 
Combining both PSF and King profile results, we have \textsc{galfit} models for $\sim$58\% of the nuclei.

\begin{figure*}
\centering
\includegraphics[scale=0.52]{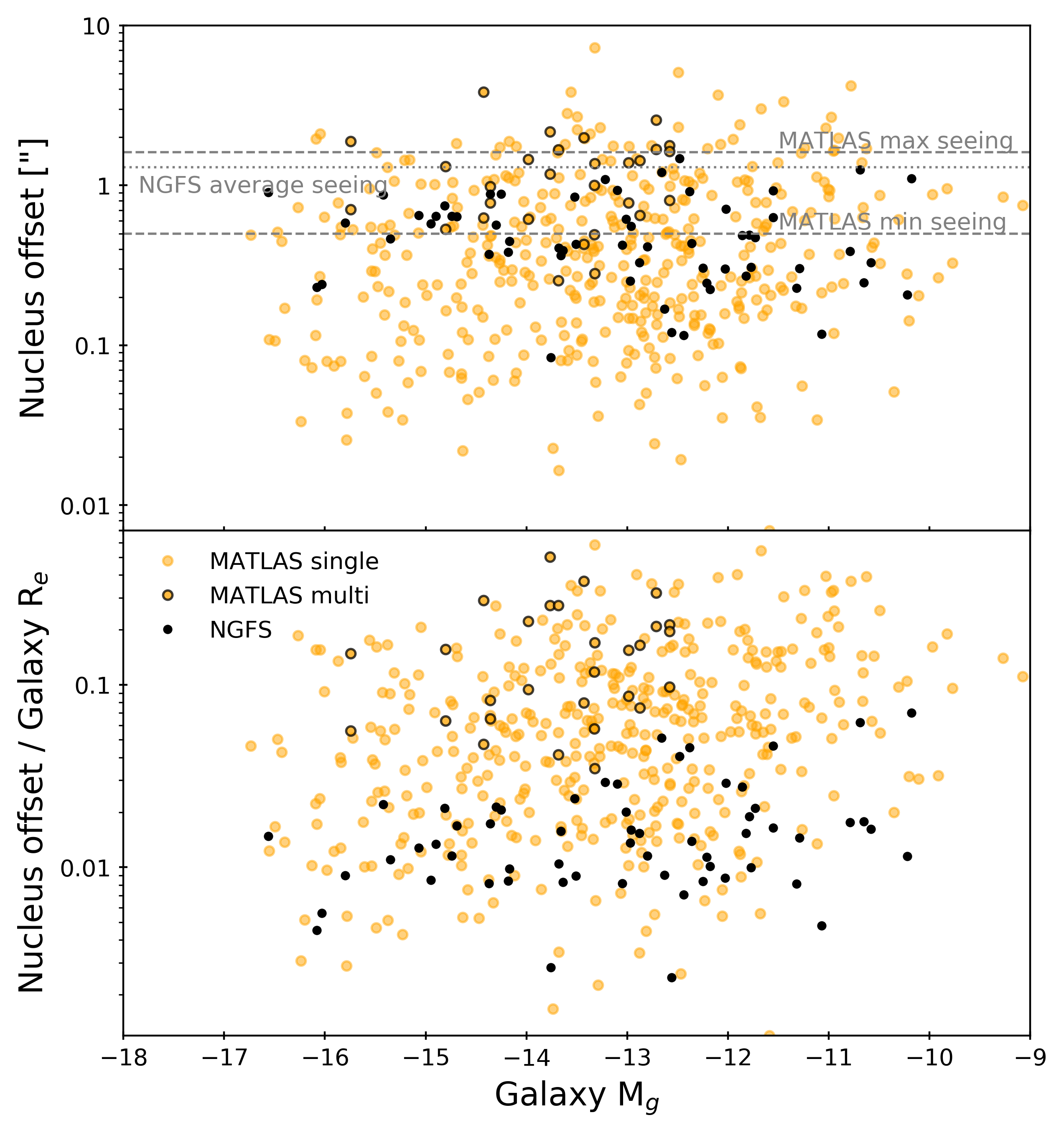}
\includegraphics[scale=0.52]{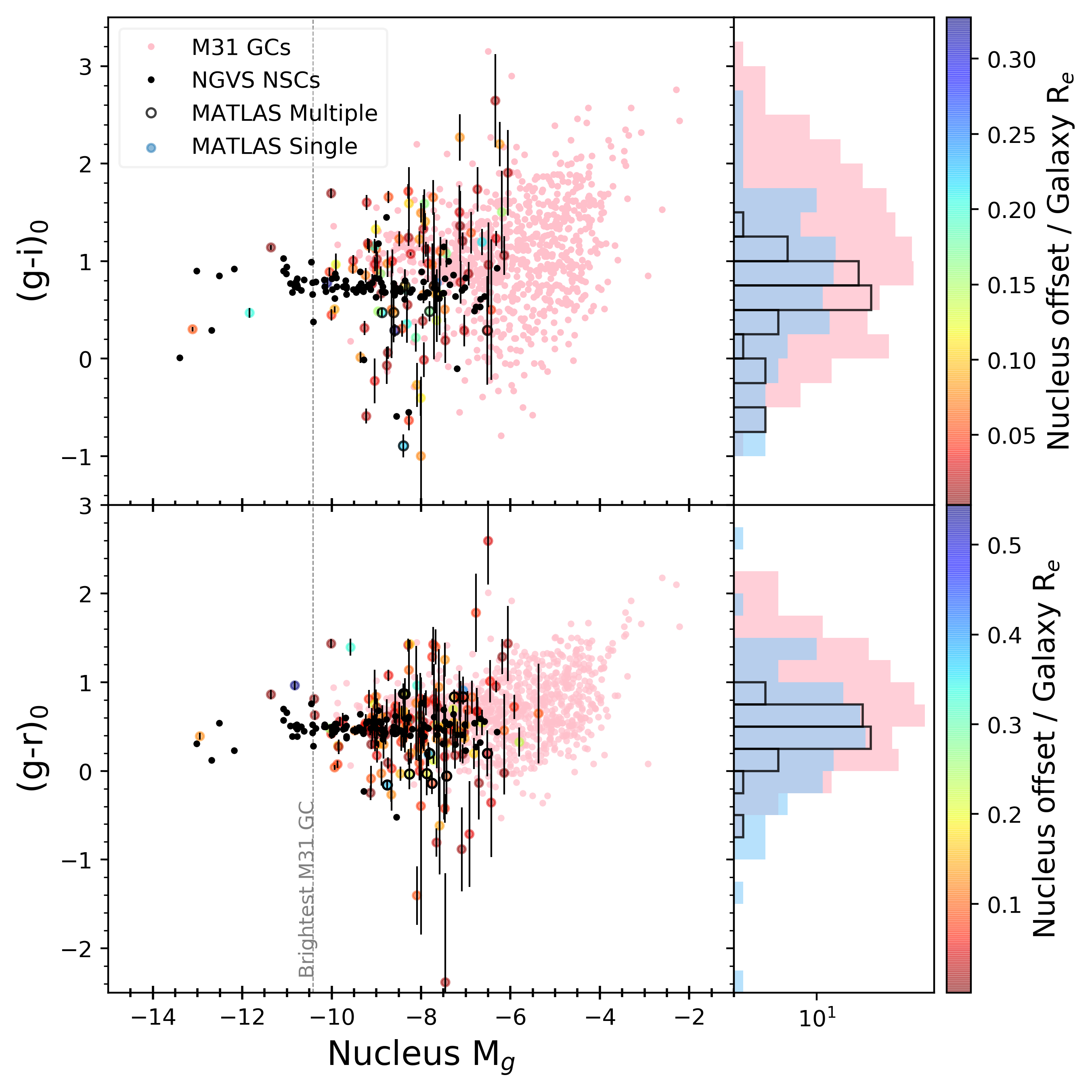}
\includegraphics[width=\linewidth]{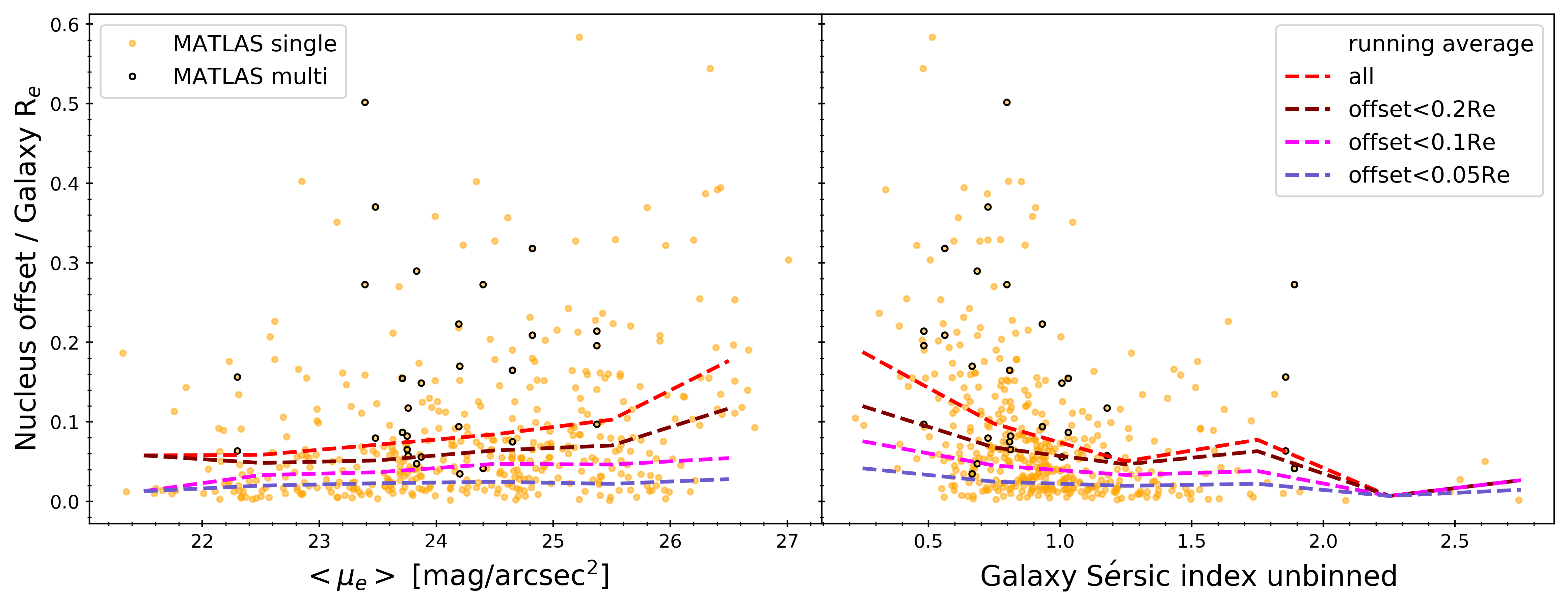}
\caption{Properties of the MATLAS nuclei. We defined a nucleus as a compact source within $\sim$ 0.5 R$_e$ of the dwarf that is the brightest one within one R$_e$. Top left panel: Estimated offset between the photocentre of the host galaxy and the nucleus in arcsecond (top) and fraction of the dwarf R$_e$ (bottom) as a function of the absolute magnitude M$_g$ of the dwarf. Yellow dots: MATLAS dwarfs with a single nucleus. Circled yellow dots: MATLAS dwarfs with mutiple nuclei. Black dots: nucleated dwarfs in NGFS. Dashed lines: minimum and maximum seeing of the MATLAS field g band images. Dotted line: average seeing of NGFS g-band images. Top right panel: Colour-magnitude diagrams of the nuclei compared to M31 globular clusters and NGVS dwarfs nuclei. Top: $(g-i)_0$ colour comparison. Bottom: $(g-r)_0$ colour comparison. Dashed line: M$_g$ of the brightest M31 GC. Colorbars: offset of the nucleus from the photocentre expressed in fraction of R$_e$ of the galaxy host. Error bars: \textsc{galfit} statistical errors on the colours. A similar range of M$_g$ is observed for the MATLAS nuclei and NGVS NSCs, while the MATLAS nuclei show similar ($g-i$) and ($g-r$) colours than M31 GCs. Bottom panel: Nucleus-photocentre separation expressed in fraction of the dwarf R$_e$ as a function of the average surface brightness within R$_e$ in the g-band (left) and the Sérsic index of the dwarf (right). Yellow dots: MATLAS dwarfs with a single nucleus. Circled yellow dots: MATLAS dwarfs with multiple nuclei. Dashed lines: running average of the offset for different offset ranges. A trend of the separation increasing in the fainter dwarfs with fainter centres is visible.} 
\label{fig:nucprop}
\end{figure*}

\subsection{Catalogues of NSCs}
\label{section:NSCcat}

We used three catalogues of NSCs from cluster dwarfs to compare against the MATLAS nuclei. All the nuclei were modelled by a PSF or a King model. We focus on the Virgo and Fornax clusters by using the nuclei catalogues from the NGVS \citep{Janssen2019} and NGFS \citep{Ordenes2018}, to which we add the catalogue of NSCs located in bright dwarfs and more massive galaxies of the Virgo cluster from \citet{Cote2006}. This last study, the ACS Virgo cluster survey (ACSVCS), makes use of the high-resolution data of the Advanced Camera for Surveys (ACS) of the Hubble Space telescope to resolve the NSCs of Virgo elliptical galaxies with M$_g \lesssim -15$. These three catalogues provide us with different information. The NGVS and NGFS both contain the photometric properties, such as M$_g$ or colours, of both the hosts and the nuclei for a sample of dwarfs of similar luminosity to the MATLAS ones. However, only NGFS has information concerning the position of the NSCs as compared to the dwarf photocentre, allowing us to extract an offset value. We note that, similarly to the colour comparison of the dwarfs (see Section \ref{section:dwarfcolor}), we restrict the use of the colour information of the NSC to the NGVS, computed using observations from the filters of MegaCam on the CFHT. The ACSVCS, providing the luminosity of both the galaxies and nuclei, allows us to compare the nuclei of faint dwarfs to the ones of bright dEs and more massive ellipticals as well as the photometry extracted from unresolved nuclei observed with a ground-based telescope to the one of resolved nuclei observed with a space telescope in galaxies of similar luminosity.

\subsection{Properties of the nuclei}
\label{section:nucprop}

\begin{figure*}
\centering
\includegraphics[width=0.49\linewidth]{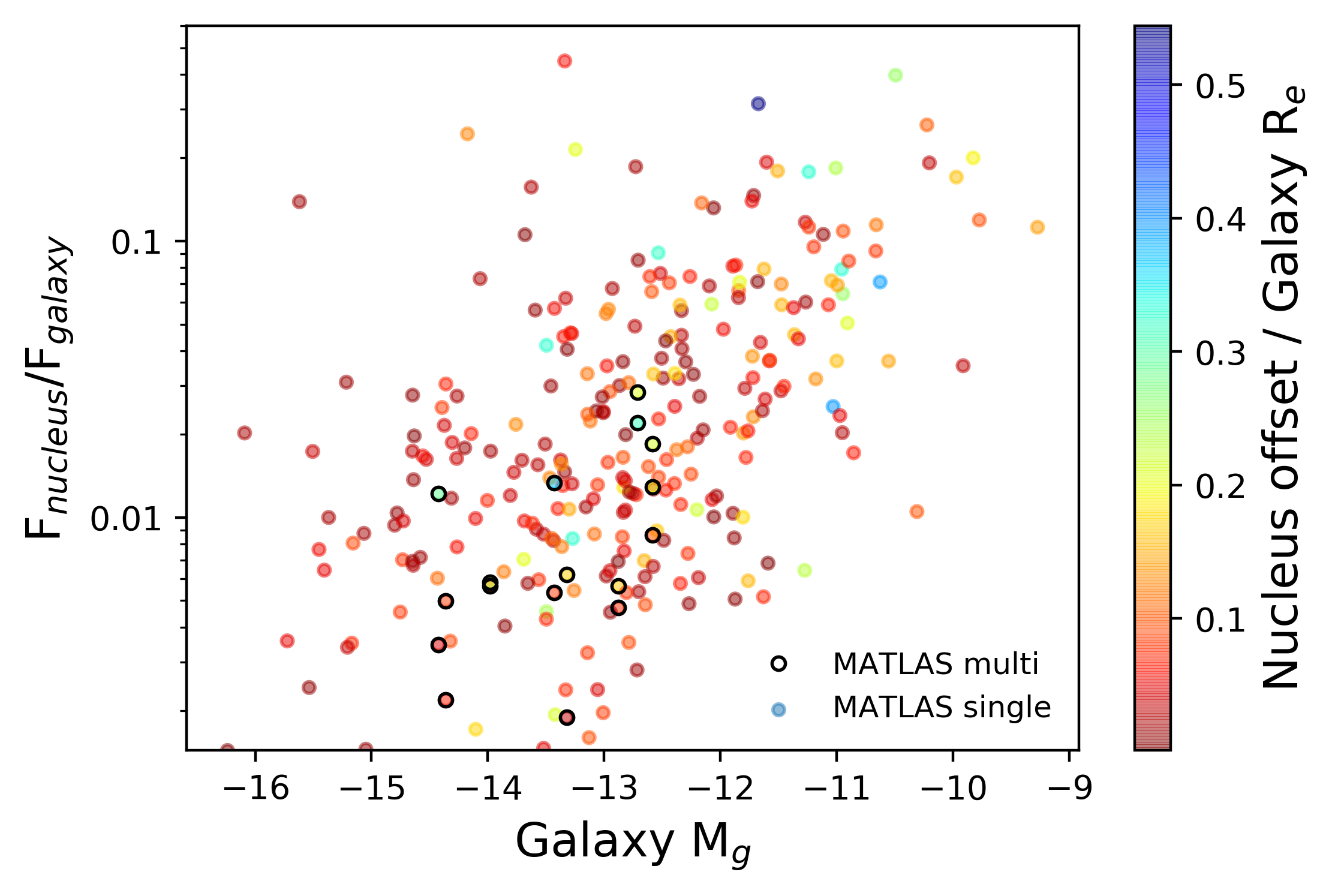}
\includegraphics[width=0.49\linewidth]{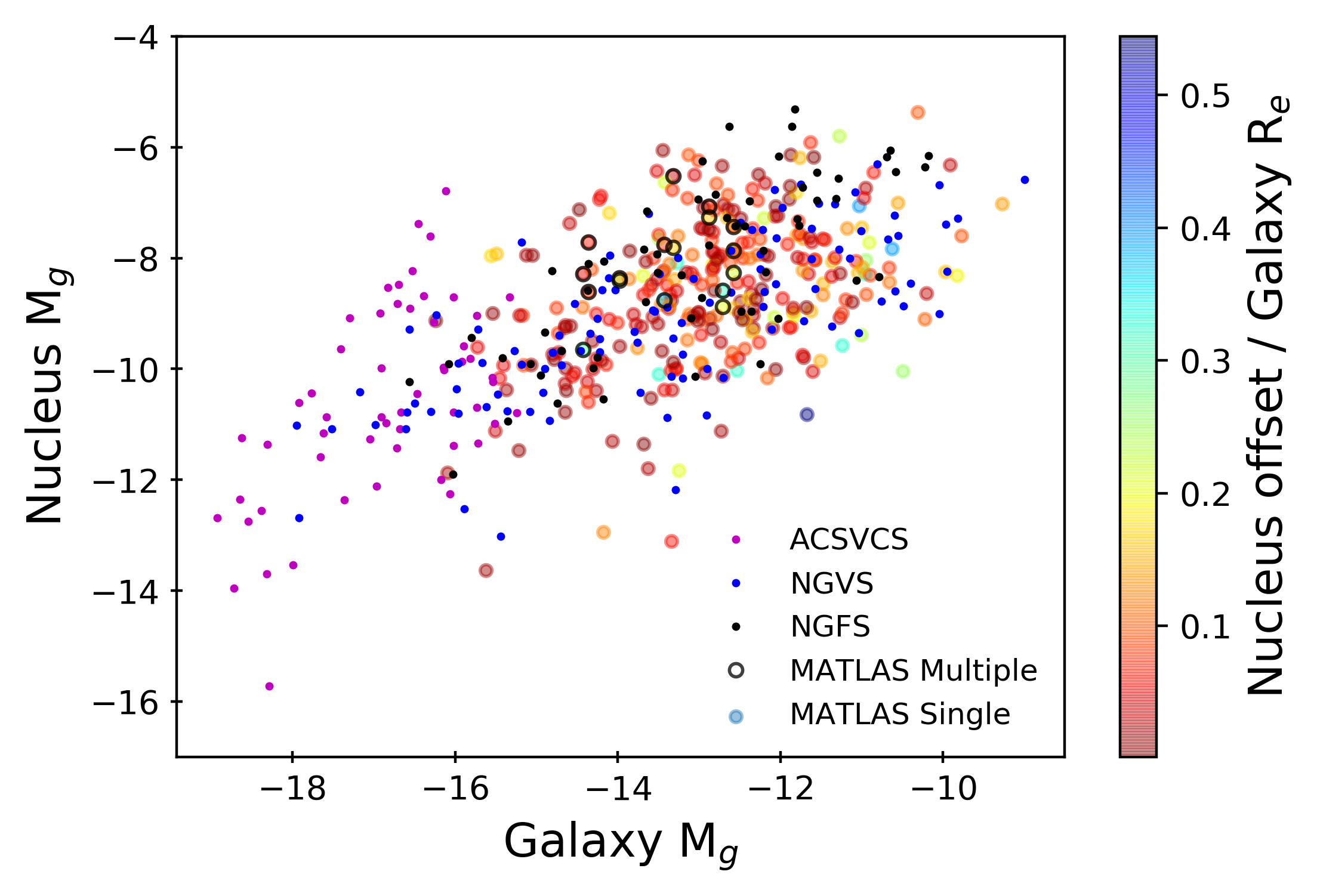}
\includegraphics[width=\linewidth]{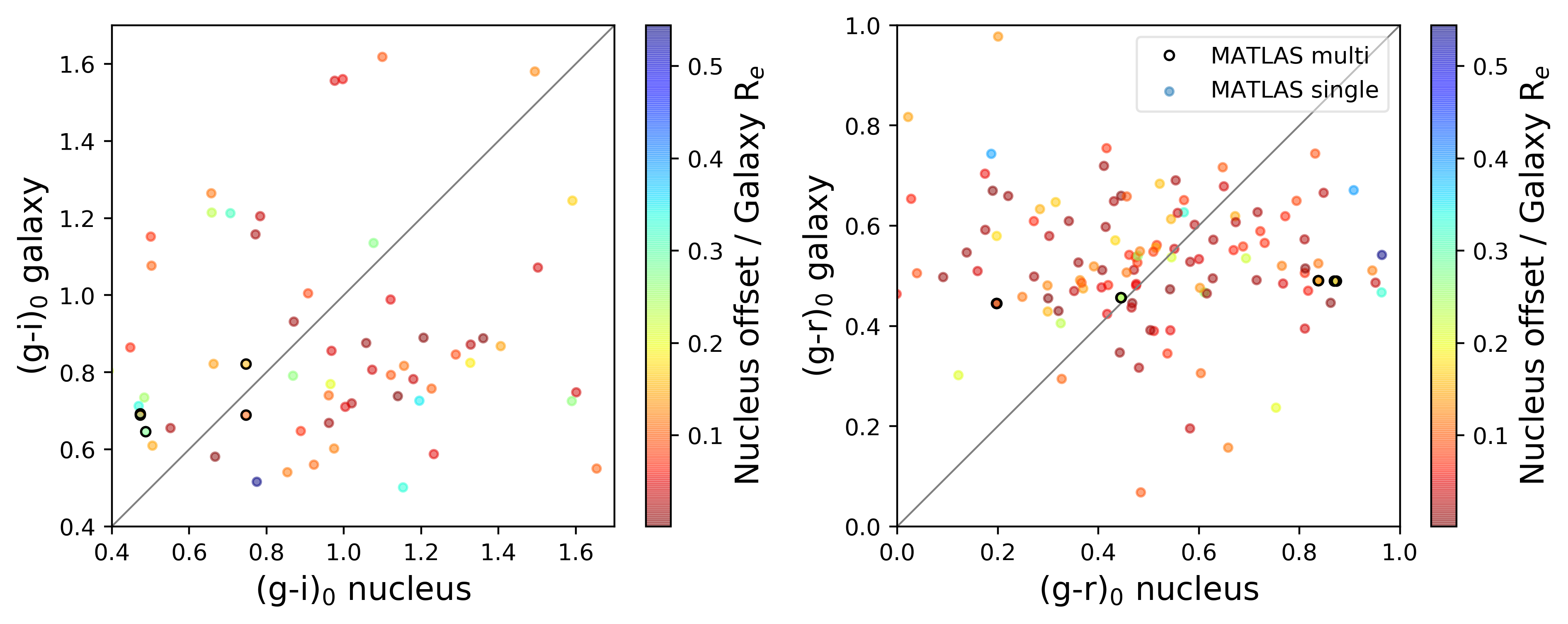}
\caption{Relations between the nucleus (as defined in Figure \ref{fig:nucprop}) and the dwarf host. Top left: Contribution of the nucleus to the total luminosity of the galaxy per M$_g$. Top right: The absolute magnitude of nucleus as a function of the absolute magnitude the host galaxy for NGFS (black dots) and NGVS (blue dots) dwarf galaxies, as well as ACSVCS elliptical galaxies (magenta dots) and the MATLAS dwarfs (open black dots: multiple nuclei, multicolour dots: single nucleus). The MATLAS, NGFS and NGVS dwarfs are showing similar absolute magnitudes for both the nuclei and galaxies. Moreover, the correlation between the magnitude of the nucleus and the magnitude of its host found for the elliptical galaxies appears to extend to the dwarfs. Bottom: Colour-colour diagrams of the galaxy as a function of the nucleus. Left: $(g-i)_0$ colour comparison. Right: $(g-r)_0$ colour comparison. No tendency for the nuclei to be bluer or redder than the galaxy is observed. Colorbars: offset of the nucleus from the photocentre expressed in fraction of R$_e$ of the galaxy host.}
\label{fig:hostnucrelation}
\end{figure*}

We discuss here several properties of the nuclei based on the \textsc{galfit} results. We note that neither the distance of the dwarf or its morphology affect the type of model that best fits the nuclei. 
All the properties are available in Table \ref{tab:catalog_nuclei}.\\

\begin{table*}
	\small
   \caption{\label{tab:catalog_nuclei}Properties of the MATLAS nuclei.}
  \begin{center}
   \begin{tabular}{ccccccccc}
    \toprule
     ID & \multicolumn{2}{c}{Distance} & RA & Dec & Offset & m$_g$ & $(g-i)_0$ & $(g-r)_0$\\
       & \multicolumn{2}{c}{(Mpc)} & (deg) & (deg) & (arcsec) &  &  & \\
    (1) & (2) & (3) & (4) & (5) & (6) & (7) & (8) & (9)\\
    \toprule
MATLAS-8 & 29.5 & -- & 18.7565 & -1.4806 & 0.62 & 24.41 & 0.66 & 0.42\\
MATLAS-9 & 29.5 & -- & 18.7779 & -1.2737 & 1.87 & -- & -- & -- \\
MATLAS-9 & 29.5 & -- & 18.7788 & -1.2745 & 2.60 & -- & -- & -- \\
MATLAS-10 & 29.5 & -- & 18.7948 & -1.4718 & 0.50 & -- & -- & -- \\
MATLAS-12 & 29.5 & -- & 18.8184 & -1.5825 & 0.14 & -- & -- & -- \\
MATLAS-20 & 30.9 & -- & 19.5842 & 3.4332 & 0.58 & 24.80 & 0.39 & 0.36\\
MATLAS-25 & 30.9 & -- & 19.8589 & 3.3598 & 0.40 & 24.74 & -- & 0.52\\
MATLAS-32 & 35.9 & -- & 20.3636 & 9.1801 & 0.56 & -- & -- & -- \\
MATLAS-43 & 35.9 & -- & 20.6259 & 8.7853 & 0.25 & -- & -- & -- \\
MATLAS-44 & 35.9 & -- & 20.6799 & 9.4184 & 0.12 & 24.83 & 1.33 & 0.55\\
... & ... & ... & ... & ... & ... & ... & ... \\
    \bottomrule
		\end{tabular}
	\end{center}
   \begin{tablenotes}
      \small
      \item \textbf{Notes.} The full table is available as supplementary material as well as at CDS. Columns meanings: (1) Dwarf ID; (2) Distance of the assumed host ETG; (3) Distance measurement from $^{(a)}$SDSS DR13 database, $^{(b)}$Poulain et al., in preparation, $^{(c)}$\citet{Ann2015}, $^{(d)}$\citet{Karachentsev2013}, $^{(e)}$\citet{Mueller2021}; (4) and (5) Right ascension and declination of the nucleus; (6) Offset distance between the nucleus and the dwarf photocentre; (7) Apparent magnitude in the g band; (8) $g-i$ colour corrected for Galactic extinction; (9) g$-$r colour corrected for Galactic extinction.
    \end{tablenotes}
\end{table*}

\subsubsection{Offset nuclei}

In Figure \ref{fig:nucprop}, we present the offset distance separating the nucleus from the photocentre of the dwarf, expressed in arcsecond and fraction of the dwarf R$_e$, as a function of the dwarf M$_g$. We indicate the minimum (0.5\arcsec) and maximum (1.61\arcsec) seeing of the MATLAS fields. We compare our nucleated dwarfs to the nucleated dwarfs from NGFS, studied using images of similar quality as MATLAS (average seeing of 1.3\arcsec\ in the g-band, \citealt{Eigenthaler2018}).
The NGFS nuclei have been selected with a maximum offset from the photocentre of 3\arcsec. However, due to the fact that not all nucleated dwarf have been modelled, the maximum offset shown in Figure \ref{fig:nucprop} is 1.5\arcsec\ [0.07 R$_e$].
The MATLAS nucleated sample show offsets up to 7\arcsec\ [0.58 R$_e$] with a median value of 0.4\arcsec\ [0.06 R$_e$]. Compared to NGFS dwarfs, the MATLAS sample shows similar range of offsets in arcseconds but show more nuclei with a larger offset in terms of fraction of R$_e$.
For both MATLAS and NGFS nucleated samples, the positions of the photocentre and the nucleus is determined by \textsc{galfit} while modelling both the galaxy and nucleus. Taking into account the seeing of the MATLAS field images, we can say that we observe offsets nuclei from the photocentre.
We have represented the offsets of the MATLAS sample by a colorbar in the top right panel of Figure \ref{fig:nucprop} as well as in Figure \ref{fig:hostnucrelation} to look for any effect of the size of the offset on the luminosities and colours of the nuclei. We find no evidence for the nucleus and galaxy to be brighter (or fainter) and bluer (or redder) with a larger offset.\\
Simulations have shown that dwarf nuclei can be off-centred by up to a few kpc when hosting a massive black hole due to dynamical perturbations \citep{Bellovary2019}. \citet{Binggeli2000} have found a typical displacement of 1\arcsec\ for a sample of 78 dE,N situated in the Virgo cluster. The offset of one of them has later been confirmed by \citet{Chung2019} using the better resolution data from NGVS. 
They also observed that the offset tends to increase with a decreasing effective surface brightness of the host. A similar result was found in \citet{Barazza2003} for a sample of 16 dE,N also in the Virgo cluster. In the bottom row of Figure \ref{fig:nucprop} are represented the separation of the MATLAS nuclei as a function of the average surface brightness within R$_e$, <$\mu_e$>, and the Sérsic index of the dwarf. We display the running average of the offset for different offset ranges. We see a similar trend of the offset increasing in the fainter dwarfs, and moreover, we note an increase of the offset for dwarfs a with fainter centre (smaller Sérsic index). Both \citet{Binggeli2000} and \citet{Barazza2003} suggest that this displacement is caused by the oscillation of the nucleus around the centre of the host galaxy due to a less strong gravitational potential (simulations of \citealt{Miller1992, Taga1998}). We note that this increase of offset could also be due to the fact that it is more difficult to determine the photocentre of objects with lower central surface brightness.\\

\subsubsection{Nuclei colours}

The colour of the nuclei can provide valuable insight on their nature and formation scenario. CMRs of the nuclei are visible in Figure \ref{fig:nucprop} for the $(g-i)_0$ 95 nuclei) and $(g-r)_0$ (162 nuclei) colours. We compare our nuclei to NSCs from NGVS and to M31's population of GCs \citep{Peacock2010}. We indicate \textsc{galfit} statistical error on the colours. The error is getting larger for an apparent magnitude of the nucleus m$_g \gtrsim 25$ and greater values of the field image seeing. The MATLAS nuclei show a similar range of M$_g$ than NGVS NSCs. When compared to the magnitudes of the GCs around M31, the brightest nuclei in both surveys are $\sim$ 4 magnitudes brighter than the brightest GC. Concerning the colours, MATLAS nuclei show a similar distribution of colours to the GC population of similar M$_g$ and a broader range of colours than NGVS NSCs. As already mentioned is the above paragraph, the large range of colour observed for MATLAS is likely not due to the large offsets of the nuclei measured.

When comparing the colours of the dwarf nuclei, we note that we do not distinguish between NSCs and AGNs (or AGNs embedded in NSCs). The low-redshift quasars can show $(g-r)$ colour in the range $0-1.5$ \citep{Richards2002}. The identification of the main emission source(s) for the MATLAS nuclei will be investigated in a future publication.\\

\subsection{The nuclei and their dwarf hosts}

We now focus our study on the relations between the photometric properties of the nuclei and their dwarf host.

Based on the sample of nuclei successfully modelled, we find a low contribution of the nucleus to the total luminosity of the galaxy (including the nucleus) with a median of 1.7\%. We note a contribution above 20\% for 6 nuclei, with a maximum contribution of 44.8\%. The contribution for each modelled nucleus is visible in Figure \ref{fig:hostnucrelation}. The nuclei of the bright Virgo dwarfs from \citet{Binggeli2000} have a contribution $\lesssim$ 10\% which is consistent with our findings. Despite the large scatter, we see a trend for the contribution to increase towards faint dwarfs, similarly to the nucleus-to-galaxy mass ratio increase towards low-mass dwarfs observed in NGFS \citep{Ordenes2018}.

In the Figure \ref{fig:hostnucrelation}, we also show the scaling relation between the magnitude of the nucleus and the magnitude of the host galaxy. We represent in magenta the sample of Virgo elliptical galaxies from ACSVCS, in blue the NGVS nucleated dwarfs and in black the NGFS nucleated dwarfs. The multicolour dots correspond to the MATLAS nucleated dwarfs, whose nuclei have been successfully modelled by a PSF or a King profile. Their colours indicate the offset of the nucleus, with the largest in blue and the smallest in dark red. We can see that the MATLAS dwarfs have similar absolute magnitudes to the NGVS and NGFS dwarfs for both the nuclei and the hosts. A correlation between the magnitude of the nucleus and the magnitude of its host has been found for the ACSVCS elliptical galaxies, and this correlation appears to extend to the NGVS, NGFS and MATLAS dwarfs. 

We plot at the bottom of Figure \ref{fig:hostnucrelation} the colours of the galaxy as a function of the colours of the nuclei. One can see that there is no clear correlation, as the nuclei are showing both bluer and redder colours than the galaxies.

\subsection{Dwarfs with multiple nuclei}
\label{section:multinuc}

The presence of multiple nuclei has been observed in both dwarf and more massive galaxies. In the case of more massive galaxies, the presence of multiple nuclei can be observed during the late stages of the process of galaxy merging. They can occur under the following processes: merging between two galaxies and merging between a galaxy and one of its satellites. The coexistence of AGNs or SMBH coupled to a stellar source (NSC or stellar disc) have been reported in the litterature (examples of dual AGNs in \citealt{Koss2012}; NSC or stellar disc -- M31 in \citealt{Lauer1993}). However, the presence of multiple nuclei in dwarf galaxies has so far been poorly studied. One example is the work of \citet{Debattista2006}. They focus on a double nucleated dwarf elliptical galaxy situated in the Virgo cluster and suggest that the nature of this double nucleus is a nuclear disc surrounding a central massive black hole, as seen in M31 or in the low-luminosity galaxy NGC4486B \citep{Lauer1996}. However, they do not reject the hypothesis of two merging GCs, although they argue it is unlikely. Another example is the study of \citet{Pak2016}. They present a double nucleated dwarf lenticular galaxy located in the Ursa Major cluster. They report that this galaxy shows a boxy shape that, in addition to the double nuclei, could be the result of a merger between two galaxies.

Within our sample of nucleated dwarf galaxies, we classified a dwarf galaxy as multinucleated when two to three-point sources of similar magnitude were located close to the galaxy photocentre and appeared brighter than the surrounding point sources. The resulted sample consists of 4 dwarfs with two bright nuclei (our best candidates), 11 with a two faint nuclei and one with a three faint nuclei. As discussed above in Section \ref{section:doubleprofile}, for those galaxies, a multiple component profile (a PSF or King model for each nucleus in addition to a S\'ersic model) was used for \textsc{galfit}. In Figure \ref{fig:SersicPSF}, we show a \textsc{galfit} modelling result for a double nucleated dwarf using a S\'ersic profile coupled to a PSF for both nuclei. We obtained a \textsc{galfit} model of both the galaxy and its nuclei for 8 of the 15 double nucleated as well as for the triple nucleated and a \textsc{galfit} model of the galaxy alone for 3 double nucleated. The multiple nucleated successfully modelled are represented with black edged dots in Figures \ref{fig:nucprop} and \ref{fig:hostnucrelation}.
As for the single nucleated, the magnitude of the nuclei seems to be correlated with the one of the host galaxies. The nuclei off-centre distances also spans the same range of amplitudes and galaxy magnitudes as for the single nucleated.

As tidal features and isophotal distortions (boxy) are associated with merger remnants, we have looked for these features in our multinucleated dwarf sample using their corresponding field images. None of our multinucleated dwarfs have boxy isophotes. However, we find tidal features for one candidate. This galaxy shows a tidal tail that indicates a late stage merger between two dwarf galaxies and could provide as well an explanation for the presence of the two bright nuclei.

\section{Environmental role on the formation scenarios}

We have presented the structural and photometric properties of the MATLAS dwarfs as well as of their nuclei. We have compared these results with the findings in different density environments. We now discuss the implications in terms of formation scenarios and evolution of the dwarf galaxies and their central compact nuclei.

\subsection{Dwarfs formation scenario, evolution and environment}

In this section, we analyse the different findings regarding the MATLAS dwarfs population to estimate the effect of the environment on the formation and evolution of the dwarf galaxies.
We know from \citet{Habas2020} that we observe a morphology-density relation for the MATLAS dwarfs with a tendency for the dIs to be located in lower local density environments than dEs and for the nucleated population to be situated in higher local density environments than non-nucleated dwarfs. This relation is also observed in higher density environments. The dIs are preferentially found in the outskirts of clusters, the LG and nearby groups in the LV \citep{Ferguson1990,Sillman2003,Cote2009,Mcconachie2012} and the nucleated are in majority situated close to the centre of the clusters \citep{Ferguson1989}. This morphology-density relation, coupled to the fact that we find a difference of structural properties between the dE and dE,N populations, implies that the local environment plays a role in shaping dwarf galaxies.

Several formation scenarios of the dEs, based on external processes, were put forward according to the environment. 
In the LG, simulations show that tidal stripping is needed to recover the observed properties of the dwarfs \citep{Mayer2001,Sawala2012}.
In clusters, the harassment and ram-pressure stripping of late-type galaxies during their infall are favored to explain the red colours of the elliptical dwarfs \citep{Boselli2008,Mastropietro2005,Steyrleithner2020}.
However, as this work shows, the similarities observed between the photometric and structural properties of the dwarfs located in vastly different environments imply that they are likely drawn from the same population and thus have a similar formation scenario.
\citet{Murali2000} shows that ram-pressure stripping is not efficient to form the Milky Way satellites.
The finding of isolated rotating dEs in the field (with a distance to a massive neighbour > 1 Mpc) at distances between 27 and 83 Mpc by \citet{Janz2017} indicate that they are not necessarily the result of harassment and ram-pressure stripping in clusters. 
In the low-density environment of MATLAS, we find a dE population with, on average, colours as red as the dEs in clusters, which also implies that dEs are not uniquely the product of morphological transformation due to ram-pressure stripping and galaxy harassment in high-density environments. This result also suggests that some, if not all, dEs in the cluster environment may have formed prior to infall.

\subsection{Formation scenario of the compact central nuclei}

In the following, we discuss the results concerning the MATLAS nucleated dwarfs sample and their nuclei in the context of formation scenario of the nucleus. Currently, two main formation scenarios are under debate: the in-situ formation by gas infall and the migration followed by the merging of GCs towards the centre. After this discussion, we will also examine the hypothesis of the nuclei as being ultra compact dwarfs (UCDs) progenitors.

\begin{figure}
\centering
\includegraphics[width=\linewidth]{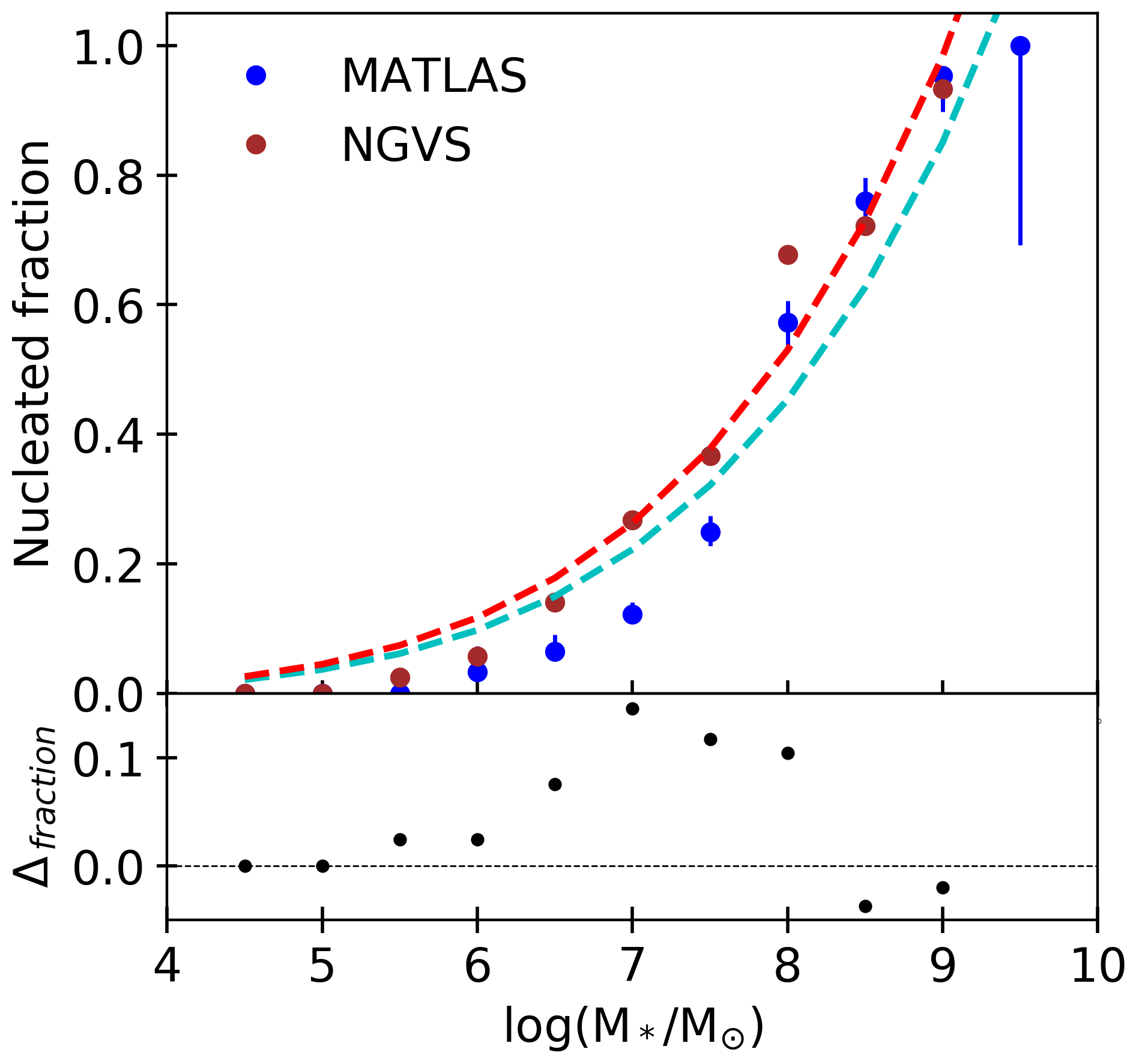}
\caption{Nucleated fraction as a function of the stellar mass. Blue dots: MATLAS dwarfs. Red dots: NGVS dwarfs. Dashed lines: best power-law fit. Error bars: 1$\sigma$ binomial confidence intervals. The nucleated fraction of the MATLAS sample falls systematically below the NGVS sample.}
\label{fig:nucfraction_Mstar}
\end{figure}

\subsubsection{Nucleated fraction mass dependence}

We investigate the nucleated dwarf fraction as a function of the stellar mass as in \citet{Janssen2019}. We estimated the stellar masses of our MATLAS dwarfs using the stellar mass-to-light ratios from \citet{Bell2003}. We chose to compute the stellar masses from the $g-r$ colour, as we have a larger number of modelled galaxies than with the $g-i$ colour. The nucleated fraction was calculated per bin of 0.5 in stellar mass. The result can be seen in Figure \ref{fig:nucfraction_Mstar}. The MATLAS sample is shown in blue and the NGVS sample is in red. We have fitted a power-law function for both samples and the best fits are shown with the dashed lines. \citet{Janssen2019} found that the nucleated fraction increases with stellar mass, which we also observe in the low-to-moderate density environments of the MATLAS sample.
As well as this mass dependence, \citet{Janssen2019} report a possible effect of the environment on the nucleated fraction. They compared their fraction to the ones of the LG, the Coma cluster, and Fornax cluster, considering the clusters core populations (within $\sim$ 0.25 R$_{vir}$). Comparing to the Virgo and Fornax clusters that show an almost identical fraction for a similar mass bin, they found that the fraction for Coma is systematically larger while the fraction for the LG is smaller. Similarly, for the MATLAS sample, we observe that the fraction of nucleated is lower than the one of NGVS at a given stellar mass. Moreover, while we observe dwarfs with stellar masses as low as log($M_*/M_{\odot}$) = 5.17 and 4.8 for the MATLAS and NGVS, respectively, the value of the stellar mass from which the fraction becomes non-null seems to be higher towards lower density environments, with a value of M$_* \sim 5 \times 10^5 M_{\odot}$ for NGVS and M$_* \sim 1 \times 10^6 M_{\odot}$ for MATLAS. \citet{Zanatta2021} and \citet{Carlsten2021} report a similar effect of the environment on the nucleated fraction of dwarfs located in groups from the LV. As mentioned in \citet{Janssen2019}, these results suggest that nuclei in low-to-moderate density environments tend to form in lower quantities than in higher density environments.\\

One possible explanation for this environmental difference could be the effect of tidal perturbations on the dwarfs. \citet{Oh2000} made use of numerical simulations to test the migration scenario in cluster dwarfs. They found that dwarfs located in the outskirts of clusters experience tidal disruption that extends GCs orbits and lengthens the dynamical friction time-scales, making the formation of nuclei difficult. While in the core cluster, the tidal perturbations compress and protect the dwarf integrity, leading to the formation of nuclei by the migration and merging of the GCs.

To test this hypothesis, we compare the nucleated fraction of dwarfs with M$_*$ in the range $10^7-10^9$M$_{\odot}$ located in different environments: cluster core, cluster outskirt, group, group outskirt and field. \citet{Janssen2019} found a nucleated fraction of 77\%, 53\% and 56\% in the cluster core of Coma, Virgo and Fornax, respectively, and a fraction of 29\% in the LG. We compute a fraction of 51\% for the MATLAS dwarfs located within the projected virial radii of a group\footnote{Considering the galaxy groups located at distances < 50 Mpc in the catalogue from \citet{Kourkchi2017}, with virial radii adapted using H$_0$ = 70 km s$^{-1}$ Mpc$^{-1}$.}, and 39\% for the MATLAS dwarfs located beyond. We use the FDS sample to estimate the fraction of nucleated in and outside the core ($\sim$ 0.25 R$_{vir}$) of the Fornax cluster. We find a nucleated fraction of 56\% in the core, as \citet{Janssen2019}, and 19\% outside. The results are consistent with what we would expect from the effects of tidal perturbations. The dwarfs in the outskirts of Fornax and in the LG suffer from tidal disruption, leading to a fraction smaller than in cluster core. The MATLAS dwarfs are found in groups with masses in the range $\sim3\times10^{8}-9\times10^{13}$M$_{\odot}$, reaching masses similar to the Fornax cluster and explaining the obtained fraction, similar to the Fornax core but still slightly smaller due to the tidally disrupted dwarfs found in less massive groups. In fact, we observe a decreasing nucleated fraction towards less massive groups, with 48\% for dwarfs in groups with a mass below $5\times10^{12}$, 51\% for a mass range $5\times10^{12}-10^{13}$ and 52\% above $10^{13}$. While the MATLAS dwarfs located in lower density environments, especially in groups outskirts, are more likely to undergo tidal disruption, leading to a smaller nucleated fraction.

\subsubsection{Relations between the nuclei and their host}

Some studies have put forward a mass dependence for the formation scenario of the nuclei with a turnover at a stellar mass of M$_* \sim 10^9$ M$_{\odot}$. They suggest that the in-situ process dominates at the high galaxy mass regime while nuclei of low-mass galaxies are more likely to be formed from GCs migration and merging \citep{Turner2012,Fahrion2021}.
The observed correlation between the nucleus and galaxy magnitude for both dwarfs and more massive ellipticals is consistent with both formation scenarios \citep{Lotz2004}. Coupled to the wide range of relative colours of nuclei and the host galaxies, this argues that the galaxy and the nucleus have formed together and then followed a different evolution \citep{Grant2005}. We can link this result to the mass dependence of the nucleated fraction and suggest that, similarly to \citet{Ordenes2018} and \citet{Janssen2019}, the NSCs can form together with the galaxy but get more easily disrupted in lower mass dwarfs.

In the context of the migration scenario being dominant in low-mass galaxies, observing dwarfs nuclei with colours similar to GCs as well as dwarfs showing multiple nuclei suggest that these nuclei may form from GCs migration and merging. However we cannot reject a possible contamination of a small fraction of the nuclei sample by GC, and we note that the presence of multiple nuclei can also be explained by nucleated dwarf galaxies merging, when coupled to signs of merging process such as tidal tails.

Another formation scenario of the dwarfs nuclei, combining both the in-situ and migration ones, is the wet migration scenario. The presence or absence of clusters merger in this scenario affects the star formation activity of the NSC, with its quenching after a merger. This scenario can explain the presence in the MATLAS sample of blue nuclei, possibly off-centred, as well as the observation of double nuclei. We note that \citet{Paudel2020} found off-centred nuclei-like star-forming regions in red dEs whose properties are consistent with the wet migration scenario.

Our observations also are consistent with the findings of spectroscopic studies. While NSC and galaxy photometry give an estimate of the age and metallicity of their stellar population, the spectroscopic studies allow one to measure these quantities and compute the star formation history. Comparative studies between dE nuclei and their host galactic main body have revealed mixed results. \citet{Paudel2011} show that the nuclei are significantly younger and metal-rich than the host galaxies, while \citet{Fahrion2020} and \citet{Johnston2020} show that nuclei have lower metallicities than their host galaxies. On the other hand, \citet{Spengler2017} show no clear differences in age or metallicity of the dE nuclei with their host galaxies. 
Due to its large size, low surface brightness and numerous GCs, the dwarf MATLAS-2019 was particularly focused in recent studies. \citet{Forbes2019} suggest that one of the central bright GCs might be an NSC. Considering this assumption and the results of the spectroscopic study of \citet{Mueller2020}, these bright star clusters (named GC5 and GC6 in that study) show no significant difference in metallicity as compared to the host galaxy, which would be in agreement with the results of \citet{Spengler2017}, such as the absence of relation between the nuclei and hosts colours found in the MATLAS sample.

\subsubsection{MATLAS nuclei as UCD progenitors}

\begin{figure}
\centering
\includegraphics[width=\linewidth]{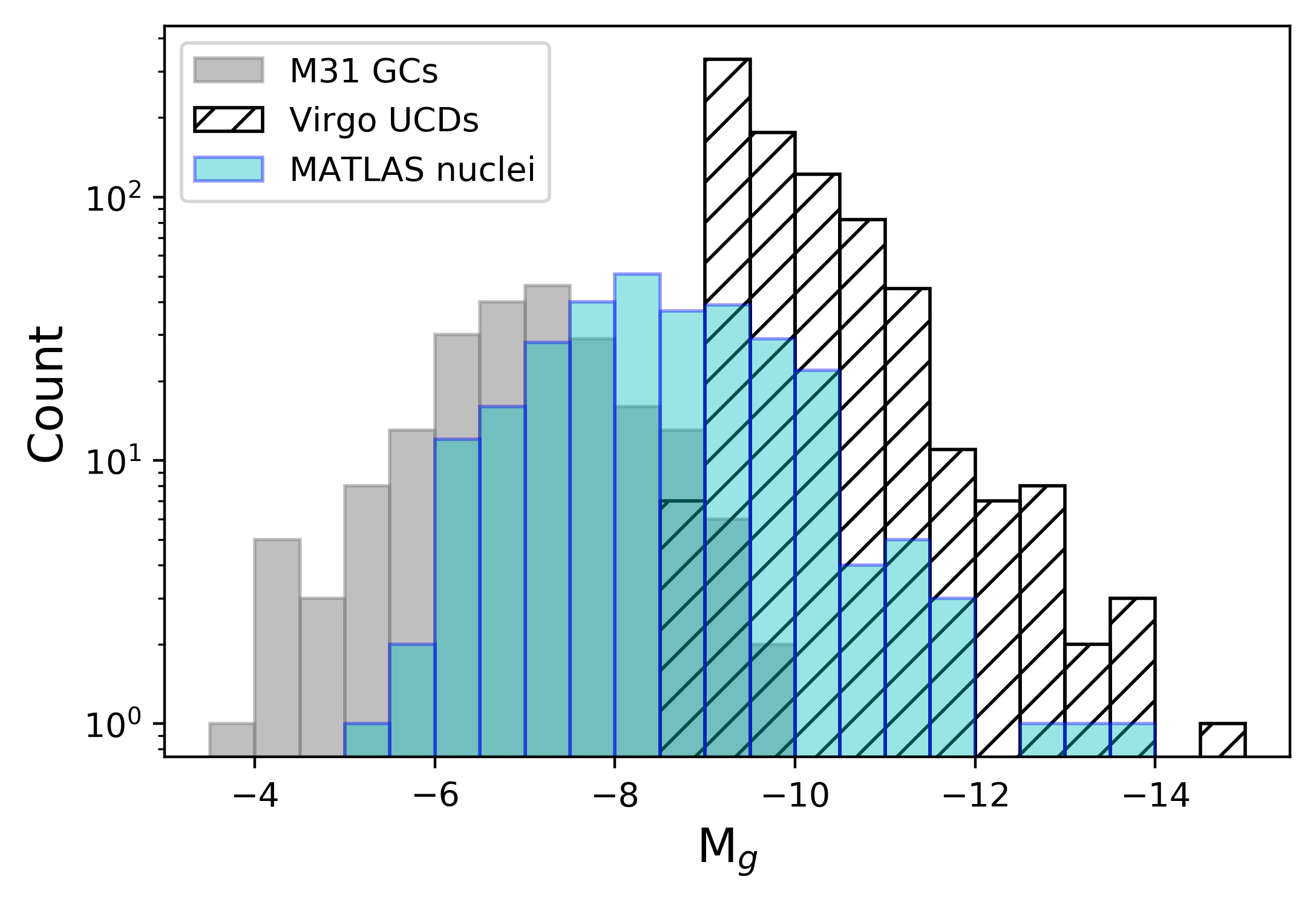}
\caption{Comparison of M$_g$ between MATLAS nuclei, M31 GCs and UCDs from the Virgo cluster. The overlapping of UCDs absolute magnitudes with some of the dwarfs nuclei suggests that some of the MATLAS nuclei may be progenitors of UCDs. The counts are displayed in log scale to ensure a better visibility.}
\label{fig:nuclei_vs_GC_UCD}
\end{figure}

Having discussed the formation scenario of the nuclei, we now focus on one of the possible formation scenarios of UCDs that suggests that they are the remnants of tidally disrupted nucleated dwarf galaxies \citep{Bassino1994,Bekki2001,Voggel2016}. In Figure \ref{fig:nuclei_vs_GC_UCD} we show the distribution of M$g$ of the nuclei modelled by both a King profile or a PSF that we compare to UCDs located in the Virgo cluster \citep{Liu2020} and to GCs from M31 \citep{Peacock2010}. We can see that the distribution of MATLAS nuclei magnitudes falls in between the UCDs and GCs. As UCDs have absolute magnitudes that overlap the distribution of the bright dwarfs nuclei (as well as the brightest GCs), this result suggests that some of the MATLAS nuclei may be progenitors of UCDs. However, due to the lack of information concerning the size of the MATLAS nuclei, we cannot test the robustness of this result.\\

\section{Conclusion}

We have studied the structure and morphology of a sample of 2210 dwarf galaxies located in the low-to-moderate density environments of the MATLAS fields. We visually identified a sample of 508 nucleated galaxies that includes a sub-sample of 16 dwarfs showing possible multiple nuclei. 
With the use of the software \textsc{galfit}, we have modelled the dwarf galaxies with a S\'ersic profile and their compact central nuclei with a PSF or a King profile and obtain a reliable modelling for 1589 galaxies (1022 dEs, 142 dIs, 415 dE,N and 10 dI,N), among them 292 nuclei. We provided the catalogues of structural and photometric properties of the MATLAS dwarfs as well as the nuclei properties. We have compared the structural and photometric properties of the dwarfs with galaxies from the LG, LV as well as from the Virgo and Fornax clusters. We report the following results:

\begin{itemize}
    \item Considering dwarfs in the same range of M$_g$, the MATLAS and clusters dwarfs are showing similar range of R$_e$, axis-ratio and Sérsic index while the LV and LG dwarfs are showing a similar range of R$_e$ to the MATLAS dwarfs.
    \item Two-sample KS tests, using a significance level $\alpha=0.05$, have shown that we cannot reject the hypothesis, for most of the structural properties of bright and faint MATLAS and clusters core dwarfs, that they are drawn from the same population.
    \item The MATLAS dwarfs show similar $(g-i)_0$ and $(g-r)_0$ colours as compared to the dwarfs located in the Virgo cluster and in the LV, meaning that we find, on average, that dEs are as red in low-to-moderate density environments as they are in the cluster environment.
    \item As observed in clusters, the MATLAS nucleated dwarfs are brighter and, for similar luminosities, are rounder than the non-nucleated dwarfs. We also observe fewer nuclei in dwarfs with a faint centre (i.e. a Sérsic index < 1) and a small size (i.e. a R$_e$ < 1 kpc).
\end{itemize}

We defined a nucleus as a compact source within $\sim$ 0.5 R$_e$ of the dwarf that is the brightest one within one R$_e$. Focusing on the properties of the nuclei in the nucleated sample, we find the following results:

\begin{itemize}
    \item We observe a systematic displacement of the nuclei that tend to increase towards dwarfs with low <$\mu_e$> and low Sérsic index. This can be caused by the oscillation of the nucleus around the centre of the host galaxy due to a less strong gravitational potential or by an increase of the uncertainty in the determination of the photocentre.
    \item The MATLAS, NGFS and NGVS nuclei show a similar range of M$_g$. The MATLAS nuclei have ranges of $g-i$ and $g-r$ colours similar to M31 GCs.
    \item The larger range of colours observed in the MATLAS sample, as compared to the NGVS one, is likely not due to the large offsets of the MATLAS nuclei measured.
    \item We find a low contribution of the nucleus to the total luminosity of the galaxy with a median of 1.7\%, consistent to the observed contribution of bright dwarf galaxies in the Virgo cluster, and a trend for the contribution of the nucleus to increase towards faint dwarfs.
    \item Our findings of the correlation between the nucleus and galaxy magnitude and the absence of relation between the colours of nuclei and their host galaxies is consistent with the scenario of galaxies and nuclei forming together but followed by a different evolution.
    \item We find an increasing nucleated fraction towards high stellar masses. For a similar mass, this fraction is systematically smaller than the one of the NGVS sample, suggesting that nuclei in low-to-moderate density environments tend to form in lower quantities than in higher density environments. This difference could be explained by tidal perturbations encountered in moderate density environments.
    \item The nuclei colours, offsets, and the observations of possible multiple nuclei are consistent with the migration and wet migration formation scenarios.
    \item As the bright MATLAS nuclei (M$_g$ < -8.5) show similar M$_g$ to UCDs from the Virgo cluster, they may be progenitors of UCDs.
\end{itemize}

Multiband ground-based imaging under good seeing conditions are a powerful probe to simultaneously study nuclei and their host galaxies. Ultimately, we need spectroscopy to confirm their association and get the information of age and metallicity to study their formation paths.

\section*{Acknowledgements}

We thank the referee for the constructive report that helped to improve this manuscript. Based on observations obtained with MegaPrime/MegaCam, a joint project of CFHT and CEA/IRFU, at the Canada-France-Hawaii Telescope (CFHT) which is operated by the National Research Council (NRC) of Canada, the Institut National des Science de l'Univers of the Centre National de la Recherche Scientifique (CNRS) of France, and the University of Hawaii. This work is based in part on data products produced at Terapix available at the Canadian Astronomy Data Centre as part of the Canada-France-Hawaii Telescope Legacy Survey, a collaborative project of NRC and CNRS. MP acknowledges the Vice Rector for Research of the University of Innsbruck for the granted scholarship. SP acknowledges support from the New Researcher Program (Shinjin grant No. 2019R1C1C1009600) through the National Research Foundation of Korea. SL acknowledges the support from the Sejong Science Fellowship Program through the National Research Foundation of Korea (NRF-2021R1C1C2006790). MB acknowledges the support from the Polish National Science Centre under the grant 2017/26/D/ST9/00449. This work has made use of data from the European Space Agency (ESA) mission {\it Gaia} (\url{https://www.cosmos.esa.int/gaia}), processed by the {\it Gaia} Data Processing and Analysis Consortium (DPAC,
\url{https://www.cosmos.esa.int/web/gaia/dpac/consortium}). Funding for the DPAC
has been provided by national institutions, in particular the institutions
participating in the {\it Gaia} Multilateral Agreement.

\section*{Data availability}
The data underlying this article are available at the CDS.

%%%%%%%%%%%%%%%%%%%%%%%%%%%%%%%%%%%%%%%%%%%%%%%%%%



\bibliographystyle{mnras}
\bibliography{MATLAS_dwarfs2v1.bib} 

\begin{thebibliography}{}
\makeatletter
\relax
\def\mn@urlcharsother{\let\do\@makeother \do\$\do\&\do\#\do\^\do\_\do\%\do\~}
\def\mn@doi{\begingroup\mn@urlcharsother \@ifnextchar [ {\mn@doi@}
  {\mn@doi@[]}}
\def\mn@doi@[#1]#2{\def\@tempa{#1}\ifx\@tempa\@empty \href
  {http://dx.doi.org/#2} {doi:#2}\else \href {http://dx.doi.org/#2} {#1}\fi
  \endgroup}
\def\mn@eprint#1#2{\mn@eprint@#1:#2::\@nil}
\def\mn@eprint@arXiv#1{\href {http://arxiv.org/abs/#1} {{\tt arXiv:#1}}}
\def\mn@eprint@dblp#1{\href {http://dblp.uni-trier.de/rec/bibtex/#1.xml}
  {dblp:#1}}
\def\mn@eprint@#1:#2:#3:#4\@nil{\def\@tempa {#1}\def\@tempb {#2}\def\@tempc
  {#3}\ifx \@tempc \@empty \let \@tempc \@tempb \let \@tempb \@tempa \fi \ifx
  \@tempb \@empty \def\@tempb {arXiv}\fi \@ifundefined
  {mn@eprint@\@tempb}{\@tempb:\@tempc}{\expandafter \expandafter \csname
  mn@eprint@\@tempb\endcsname \expandafter{\@tempc}}}

\bibitem[\protect\citeauthoryear{{Adami} et~al.,}{{Adami}
  et~al.}{2006}]{Adami2006}
{Adami} C.,  et~al., 2006, \mn@doi [\aap] {10.1051/0004-6361:20053758}, \href
  {https://ui.adsabs.harvard.edu/abs/2006A&A...459..679A} {459, 679}

\bibitem[\protect\citeauthoryear{{Aguerri} \&
  {Gonz{\'a}lez-Garc{\'\i}a}}{{Aguerri} \&
  {Gonz{\'a}lez-Garc{\'\i}a}}{2009}]{Aguerri2009}
{Aguerri} J.~A.~L.,  {Gonz{\'a}lez-Garc{\'\i}a} A.~C.,  2009, \mn@doi [\aap]
  {10.1051/0004-6361:200810339}, \href
  {https://ui.adsabs.harvard.edu/abs/2009A&A...494..891A} {494, 891}

\bibitem[\protect\citeauthoryear{{Ann}, {Seo}  \& {Ha}}{{Ann}
  et~al.}{2015}]{Ann2015}
{Ann} H.~B.,  {Seo} M.,   {Ha} D.~K.,  2015, \mn@doi [\apjs]
  {10.1088/0067-0049/217/2/27}, \href
  {https://ui.adsabs.harvard.edu/abs/2015ApJS..217...27A} {217, 27}

\bibitem[\protect\citeauthoryear{{Antonini}, {Capuzzo-Dolcetta},
  {Mastrobuono-Battisti}  \& {Merritt}}{{Antonini} et~al.}{2012}]{Antonini2012}
{Antonini} F.,  {Capuzzo-Dolcetta} R.,  {Mastrobuono-Battisti} A.,   {Merritt}
  D.,  2012, \mn@doi [\apj] {10.1088/0004-637X/750/2/111}, \href
  {https://ui.adsabs.harvard.edu/abs/2012ApJ...750..111A} {750, 111}

\bibitem[\protect\citeauthoryear{{Arenou} et~al.,}{{Arenou}
  et~al.}{2018}]{Gaia2018}
{Arenou} F.,  et~al., 2018, \mn@doi [\aap] {10.1051/0004-6361/201833234}, \href
  {https://ui.adsabs.harvard.edu/abs/2018A&A...616A..17A} {616, A17}

\bibitem[\protect\citeauthoryear{{Barazza}, {Binggeli}  \& {Jerjen}}{{Barazza}
  et~al.}{2003}]{Barazza2003}
{Barazza} F.~D.,  {Binggeli} B.,   {Jerjen} H.,  2003, \mn@doi [\aap]
  {10.1051/0004-6361:20030872}, 407, 121

\bibitem[\protect\citeauthoryear{{Bassino}, {Muzzio}  \& {Rabolli}}{{Bassino}
  et~al.}{1994}]{Bassino1994}
{Bassino} L.~P.,  {Muzzio} J.~C.,   {Rabolli} M.,  1994, \mn@doi [\apj]
  {10.1086/174514}, \href
  {https://ui.adsabs.harvard.edu/abs/1994ApJ...431..634B} {431, 634}

\bibitem[\protect\citeauthoryear{{Bekki}}{{Bekki}}{2007}]{Bekki2007}
{Bekki} K.,  2007, \mn@doi [\pasa] {10.1071/AS07008}, \href
  {https://ui.adsabs.harvard.edu/abs/2007PASA...24...77B} {24, 77}

\bibitem[\protect\citeauthoryear{{Bekki}, {Couch}  \& {Drinkwater}}{{Bekki}
  et~al.}{2001}]{Bekki2001}
{Bekki} K.,  {Couch} W.~J.,   {Drinkwater} M.~J.,  2001, \mn@doi [\apjl]
  {10.1086/320339}, \href
  {https://ui.adsabs.harvard.edu/abs/2001ApJ...552L.105B} {552, L105}

\bibitem[\protect\citeauthoryear{{Bell}, {McIntosh}, {Katz}  \&
  {Weinberg}}{{Bell} et~al.}{2003}]{Bell2003}
{Bell} E.~F.,  {McIntosh} D.~H.,  {Katz} N.,   {Weinberg} M.~D.,  2003, \mn@doi
  [\apjs] {10.1086/378847}, \href
  {https://ui.adsabs.harvard.edu/abs/2003ApJS..149..289B} {149, 289}

\bibitem[\protect\citeauthoryear{{Bell} et~al.,}{{Bell}
  et~al.}{2004}]{Bell2004}
{Bell} E.~F.,  et~al., 2004, \mn@doi [\apj] {10.1086/420778}, \href
  {https://ui.adsabs.harvard.edu/abs/2004ApJ...608..752B} {608, 752}

\bibitem[\protect\citeauthoryear{{Bellazzini} et~al.,}{{Bellazzini}
  et~al.}{2008}]{Bellazzini2008}
{Bellazzini} M.,  et~al., 2008, \mn@doi [\aj] {10.1088/0004-6256/136/3/1147},
  \href {https://ui.adsabs.harvard.edu/abs/2008AJ....136.1147B} {136, 1147}

\bibitem[\protect\citeauthoryear{{Bellovary}, {Cleary}, {Munshi}, {Tremmel},
  {Christensen}, {Brooks}  \& {Quinn}}{{Bellovary}
  et~al.}{2019}]{Bellovary2019}
{Bellovary} J.~M.,  {Cleary} C.~E.,  {Munshi} F.,  {Tremmel} M.,  {Christensen}
  C.~R.,  {Brooks} A.,   {Quinn} T.~R.,  2019, \mn@doi [\mnras]
  {10.1093/mnras/sty2842}, \href
  {https://ui.adsabs.harvard.edu/abs/2019MNRAS.482.2913B} {482, 2913}

\bibitem[\protect\citeauthoryear{{Bertin}}{{Bertin}}{2011}]{Bertin2011}
{Bertin} E.,  2011, in {Evans} I.~N.,  {Accomazzi} A.,  {Mink} D.~J.,   {Rots}
  A.~H.,  eds,  Astronomical Society of the Pacific Conference Series Vol. 442,
  Astronomical Data Analysis Software and Systems XX. p.~435

\bibitem[\protect\citeauthoryear{{Bertin} \& {Arnouts}}{{Bertin} \&
  {Arnouts}}{1996}]{Bertin1996}
{Bertin} E.,  {Arnouts} S.,  1996, \mn@doi [\aaps] {10.1051/aas:1996164}, 117,
  393

\bibitem[\protect\citeauthoryear{{Binggeli}, {Barazza}  \& {Jerjen}}{{Binggeli}
  et~al.}{2000}]{Binggeli2000}
{Binggeli} B.,  {Barazza} F.,   {Jerjen} H.,  2000, \aap, 359, 447

\bibitem[\protect\citeauthoryear{{Boselli}, {Boissier}, {Cortese}  \&
  {Gavazzi}}{{Boselli} et~al.}{2008}]{Boselli2008}
{Boselli} A.,  {Boissier} S.,  {Cortese} L.,   {Gavazzi} G.,  2008, \mn@doi
  [\apj] {10.1086/525513}, \href
  {https://ui.adsabs.harvard.edu/abs/2008ApJ...674..742B} {674, 742}

\bibitem[\protect\citeauthoryear{{Butler} \& {Mart{\'\i}nez-Delgado}}{{Butler}
  \& {Mart{\'\i}nez-Delgado}}{2005}]{Butler2005}
{Butler} D.~J.,  {Mart{\'\i}nez-Delgado} D.,  2005, \mn@doi [\aj]
  {10.1086/429524}, \href
  {https://ui.adsabs.harvard.edu/abs/2005AJ....129.2217B} {129, 2217}

\bibitem[\protect\citeauthoryear{Cappellari et~al.,}{Cappellari
  et~al.}{2011}]{Capellari2011}
Cappellari M.,  et~al., 2011, \mn@doi [\mnras]
  {10.1111/j.1365-2966.2010.18174.x}, 413, 813

\bibitem[\protect\citeauthoryear{{Carlsten}, {Greco}, {Beaton}  \&
  {Greene}}{{Carlsten} et~al.}{2020}]{Carlsten2020}
{Carlsten} S.~G.,  {Greco} J.~P.,  {Beaton} R.~L.,   {Greene} J.~E.,  2020,
  \mn@doi [\apj] {10.3847/1538-4357/ab7758}, \href
  {https://ui.adsabs.harvard.edu/abs/2020ApJ...891..144C} {891, 144}

\bibitem[\protect\citeauthoryear{{Carlsten}, {Greene}, {Beaton}  \&
  {Greco}}{{Carlsten} et~al.}{2021}]{Carlsten2021}
{Carlsten} S.~G.,  {Greene} J.~E.,  {Beaton} R.~L.,   {Greco} J.~P.,  2021,
  arXiv e-prints, \href {https://ui.adsabs.harvard.edu/abs/2021arXiv210503440C}
  {p. arXiv:2105.03440}

\bibitem[\protect\citeauthoryear{Chung, Rey, Sung, Kim, Lee  \& Lee}{Chung
  et~al.}{2019}]{Chung2019}
Chung J.,  Rey S.-C.,  Sung E.-C.,  Kim S.,  Lee Y.,   Lee W.,  2019, \mn@doi
  [\apj] {10.3847/1538-4357/ab25e8}, 879, 97

\bibitem[\protect\citeauthoryear{C{\^{o}}t{\'{e}} et~al.,}{C{\^{o}}t{\'{e}}
  et~al.}{2006}]{Cote2006}
C{\^{o}}t{\'{e}} P.,  et~al., 2006, \mn@doi [\apjs] {10.1086/504042}, 165, 57

\bibitem[\protect\citeauthoryear{{C{\^o}t{\'e}}, {Draginda}, {Skillman}  \&
  {Miller}}{{C{\^o}t{\'e}} et~al.}{2009}]{Cote2009}
{C{\^o}t{\'e}} S.,  {Draginda} A.,  {Skillman} E.~D.,   {Miller} B.~W.,  2009,
  \mn@doi [\aj] {10.1088/0004-6256/138/4/1037}, \href
  {https://ui.adsabs.harvard.edu/abs/2009AJ....138.1037C} {138, 1037}

\bibitem[\protect\citeauthoryear{{De Rijcke}, {Prugniel}, {Simien}  \&
  {Dejonghe}}{{De Rijcke} et~al.}{2006}]{DeRijcke2006}
{De Rijcke} S.,  {Prugniel} P.,  {Simien} F.,   {Dejonghe} H.,  2006, \mn@doi
  [\mnras] {10.1111/j.1365-2966.2006.10377.x}, \href
  {https://ui.adsabs.harvard.edu/abs/2006MNRAS.369.1321D} {369, 1321}

\bibitem[\protect\citeauthoryear{Debattista, Ferreras, Pasquali, Seth, Rijcke
  \& Morelli}{Debattista et~al.}{2006}]{Debattista2006}
Debattista V.~P.,  Ferreras I.,  Pasquali A.,  Seth A.,  Rijcke S.~D.,
  Morelli L.,  2006, \mn@doi [\apj] {10.1086/509783}, 651, L97

\bibitem[\protect\citeauthoryear{{Dekel} \& {Silk}}{{Dekel} \&
  {Silk}}{1986}]{Dekel1986}
{Dekel} A.,  {Silk} J.,  1986, \mn@doi [\apj] {10.1086/164050}, \href
  {https://ui.adsabs.harvard.edu/abs/1986ApJ...303...39D} {303, 39}

\bibitem[\protect\citeauthoryear{{Draper}, {Gray}, {Berry}  \&
  {Taylor}}{{Draper} et~al.}{2014}]{Draper2014}
{Draper} P.~W.,  {Gray} N.,  {Berry} D.~S.,   {Taylor} M.,  2014, {GAIA:
  Graphical Astronomy and Image Analysis Tool} (\mn@eprint {ascl} {1403.024})

\bibitem[\protect\citeauthoryear{Duc et~al.,}{Duc et~al.}{2014}]{Duc2014}
Duc P.-A.,  et~al., 2014, \mn@doi [\mnras] {10.1093/mnras/stu2019}, 446, 120

\bibitem[\protect\citeauthoryear{{Dunn}}{{Dunn}}{2015}]{Dunn2015}
{Dunn} J.~M.,  2015, \mn@doi [\mnras] {10.1093/mnras/stv1629}, \href
  {https://ui.adsabs.harvard.edu/abs/2015MNRAS.453.1799D} {453, 1799}

\bibitem[\protect\citeauthoryear{Eigenthaler et~al.,}{Eigenthaler
  et~al.}{2018}]{Eigenthaler2018}
Eigenthaler P.,  et~al., 2018, \mn@doi [\apj] {10.3847/1538-4357/aaab60}, 855,
  142

\bibitem[\protect\citeauthoryear{{Fahrion} et~al.,}{{Fahrion}
  et~al.}{2020}]{Fahrion2020}
{Fahrion} K.,  et~al., 2020, \mn@doi [\aap] {10.1051/0004-6361/201937120},
  \href {https://ui.adsabs.harvard.edu/abs/2020A&A...634A..53F} {634, A53}

\bibitem[\protect\citeauthoryear{{Fahrion} et~al.,}{{Fahrion}
  et~al.}{2021}]{Fahrion2021}
{Fahrion} K.,  et~al., 2021, arXiv e-prints, \href
  {https://ui.adsabs.harvard.edu/abs/2021arXiv210406412F} {p. arXiv:2104.06412}

\bibitem[\protect\citeauthoryear{{Ferguson} \& {Binggeli}}{{Ferguson} \&
  {Binggeli}}{1994}]{Ferguson1994}
{Ferguson} H.~C.,  {Binggeli} B.,  1994, \mn@doi [\aapr] {10.1007/BF01208252},
  \href {https://ui.adsabs.harvard.edu/abs/1994A&ARv...6...67F} {6, 67}

\bibitem[\protect\citeauthoryear{{Ferguson} \& {Sandage}}{{Ferguson} \&
  {Sandage}}{1989}]{Ferguson1989}
{Ferguson} H.~C.,  {Sandage} A.,  1989, \mn@doi [\apjl] {10.1086/185577}, 346,
  L53

\bibitem[\protect\citeauthoryear{{Ferguson} \& {Sandage}}{{Ferguson} \&
  {Sandage}}{1990}]{Ferguson1990}
{Ferguson} H.~C.,  {Sandage} A.,  1990, in {Hollenbach} D.~J.,  {Thronson}
  Harley~A. J.,  eds,  NASA Conference Publication Vol. 3084, NASA Conference
  Publication. p.~281

\bibitem[\protect\citeauthoryear{{Ferguson} \& {Sandage}}{{Ferguson} \&
  {Sandage}}{1991}]{Ferguson1991}
{Ferguson} H.~C.,  {Sandage} A.,  1991, \mn@doi [\aj] {10.1086/115721}, \href
  {https://ui.adsabs.harvard.edu/abs/1991AJ....101..765F} {101, 765}

\bibitem[\protect\citeauthoryear{Ferrarese et~al.,}{Ferrarese
  et~al.}{2012}]{Ferrarese2012}
Ferrarese L.,  et~al., 2012, \mn@doi [\apjs] {10.1088/0067-0049/200/1/4}, 200,
  4

\bibitem[\protect\citeauthoryear{{Ferrarese} et~al.,}{{Ferrarese}
  et~al.}{2020}]{Ferrarese2020}
{Ferrarese} L.,  et~al., 2020, \mn@doi [\apj] {10.3847/1538-4357/ab339f}, \href
  {https://ui.adsabs.harvard.edu/abs/2020ApJ...890..128F} {890, 128}

\bibitem[\protect\citeauthoryear{{Forbes}, {Gannon}, {Couch}, {Iodice},
  {Spavone}, {Cantiello}, {Napolitano}  \& {Schipani}}{{Forbes}
  et~al.}{2019}]{Forbes2019}
{Forbes} D.~A.,  {Gannon} J.,  {Couch} W.~J.,  {Iodice} E.,  {Spavone} M.,
  {Cantiello} M.,  {Napolitano} N.,   {Schipani} P.,  2019, \mn@doi [\aap]
  {10.1051/0004-6361/201935499}, \href
  {https://ui.adsabs.harvard.edu/abs/2019A&A...626A..66F} {626, A66}

\bibitem[\protect\citeauthoryear{{Gaia Collaboration} et~al.,}{{Gaia
  Collaboration} et~al.}{2016}]{Gaiamission}
{Gaia Collaboration} et~al., 2016, \mn@doi [\aap]
  {10.1051/0004-6361/201629272}, \href
  {https://ui.adsabs.harvard.edu/abs/2016A&A...595A...1G} {595, A1}

\bibitem[\protect\citeauthoryear{{Gaia Collaboration} et~al.,}{{Gaia
  Collaboration} et~al.}{2018}]{GaiaDR2}
{Gaia Collaboration} et~al., 2018, \mn@doi [\aap]
  {10.1051/0004-6361/201833051}, \href
  {https://ui.adsabs.harvard.edu/abs/2018A&A...616A...1G} {616, A1}

\bibitem[\protect\citeauthoryear{{Geha}, {Blanton}, {Yan}  \& {Tinker}}{{Geha}
  et~al.}{2012}]{Geha2012}
{Geha} M.,  {Blanton} M.~R.,  {Yan} R.,   {Tinker} J.~L.,  2012, \mn@doi [\apj]
  {10.1088/0004-637X/757/1/85}, \href
  {https://ui.adsabs.harvard.edu/abs/2012ApJ...757...85G} {757, 85}

\bibitem[\protect\citeauthoryear{{Georgiev}, {Hilker}, {Puzia}, {Goudfrooij}
  \& {Baumgardt}}{{Georgiev} et~al.}{2009}]{Georgiev2009}
{Georgiev} I.~Y.,  {Hilker} M.,  {Puzia} T.~H.,  {Goudfrooij} P.,   {Baumgardt}
  H.,  2009, \mn@doi [\mnras] {10.1111/j.1365-2966.2009.14776.x}, \href
  {https://ui.adsabs.harvard.edu/abs/2009MNRAS.396.1075G} {396, 1075}

\bibitem[\protect\citeauthoryear{{Georgiev}, {Puzia}, {Goudfrooij}  \&
  {Hilker}}{{Georgiev} et~al.}{2010}]{Georgiev2010}
{Georgiev} I.~Y.,  {Puzia} T.~H.,  {Goudfrooij} P.,   {Hilker} M.,  2010,
  \mn@doi [\mnras] {10.1111/j.1365-2966.2010.16802.x}, \href
  {https://ui.adsabs.harvard.edu/abs/2010MNRAS.406.1967G} {406, 1967}

\bibitem[\protect\citeauthoryear{{Graham} \& {Driver}}{{Graham} \&
  {Driver}}{2005}]{Graham2005}
{Graham} A.~W.,  {Driver} S.~P.,  2005, \mn@doi [\pasa] {10.1071/AS05001},
  \href {https://ui.adsabs.harvard.edu/abs/2005PASA...22..118G} {22, 118}

\bibitem[\protect\citeauthoryear{Grant, Kuipers  \& Phillipps}{Grant
  et~al.}{2005}]{Grant2005}
Grant N.~I.,  Kuipers J.~A.,   Phillipps S.,  2005, \mn@doi [\mnras]
  {10.1111/j.1365-2966.2005.09518.x}, 363, 1019

\bibitem[\protect\citeauthoryear{{Grebel}}{{Grebel}}{2001}]{Grebel2001}
{Grebel} E.~K.,  2001, in {de Boer} K.~S.,  {Dettmar} R.-J.,   {Klein} U.,
  eds, ~ Vol. 40, Dwarf galaxies and their environment. p.~45 (\mn@eprint
  {arXiv} {astro-ph/0107208})

\bibitem[\protect\citeauthoryear{{Grebel}}{{Grebel}}{2004}]{Grebel2004}
{Grebel} E.~K.,  2004, in {McWilliam} A.,  {Rauch} M.,  eds, Origin and
  Evolution of the Elements. p.~234 (\mn@eprint {arXiv} {astro-ph/0403222})

\bibitem[\protect\citeauthoryear{{Guillard}, {Emsellem}  \&
  {Renaud}}{{Guillard} et~al.}{2016}]{Guillard2016}
{Guillard} N.,  {Emsellem} E.,   {Renaud} F.,  2016, \mn@doi [\mnras]
  {10.1093/mnras/stw1570}, \href
  {https://ui.adsabs.harvard.edu/abs/2016MNRAS.461.3620G} {461, 3620}

\bibitem[\protect\citeauthoryear{{Habas} et~al.,}{{Habas}
  et~al.}{2020}]{Habas2020}
{Habas} R.,  et~al., 2020, \mn@doi [\mnras] {10.1093/mnras/stz3045}, \href
  {https://ui.adsabs.harvard.edu/abs/2020MNRAS.491.1901H} {491, 1901}

\bibitem[\protect\citeauthoryear{{Haines}, {Gargiulo}  \& {Merluzzi}}{{Haines}
  et~al.}{2008}]{Haines2008}
{Haines} C.~P.,  {Gargiulo} A.,   {Merluzzi} P.,  2008, \mn@doi [\mnras]
  {10.1111/j.1365-2966.2008.12954.x}, \href
  {https://ui.adsabs.harvard.edu/abs/2008MNRAS.385.1201H} {385, 1201}

\bibitem[\protect\citeauthoryear{{Harris}}{{Harris}}{2010}]{Harris2010}
{Harris} W.~E.,  2010, arXiv e-prints, \href
  {https://ui.adsabs.harvard.edu/abs/2010arXiv1012.3224H} {p. arXiv:1012.3224}

\bibitem[\protect\citeauthoryear{{Harris} \& {van den Bergh}}{{Harris} \& {van
  den Bergh}}{1981}]{Harris1981}
{Harris} W.~E.,  {van den Bergh} S.,  1981, \mn@doi [\aj] {10.1086/113047},
  \href {https://ui.adsabs.harvard.edu/abs/1981AJ.....86.1627H} {86, 1627}

\bibitem[\protect\citeauthoryear{Harris, Harris  \& Alessi}{Harris
  et~al.}{2013}]{Harris2013}
Harris W.~E.,  Harris G. L.~H.,   Alessi M.,  2013, \mn@doi [\apj]
  {10.1088/0004-637x/772/2/82}, 772, 82

\bibitem[\protect\citeauthoryear{{Ho}, {Li}, {Barth}, {Seigar}  \& {Peng}}{{Ho}
  et~al.}{2011}]{Ho2011}
{Ho} L.~C.,  {Li} Z.-Y.,  {Barth} A.~J.,  {Seigar} M.~S.,   {Peng} C.~Y.,
  2011, \mn@doi [\apjs] {10.1088/0067-0049/197/2/21}, \href
  {https://ui.adsabs.harvard.edu/abs/2011ApJS..197...21H} {197, 21}

\bibitem[\protect\citeauthoryear{{Janz}, {Penny}, {Graham}, {Forbes}  \&
  {Davies}}{{Janz} et~al.}{2017}]{Janz2017}
{Janz} J.,  {Penny} S.~J.,  {Graham} A.~W.,  {Forbes} D.~A.,   {Davies} R.~L.,
  2017, \mn@doi [\mnras] {10.1093/mnras/stx634}, \href
  {https://ui.adsabs.harvard.edu/abs/2017MNRAS.468.2850J} {468, 2850}

\bibitem[\protect\citeauthoryear{{Jedrzejewski}}{{Jedrzejewski}}{1987}]{Jedrzejewski1987}
{Jedrzejewski} R.~I.,  1987, \mn@doi [\mnras] {10.1093/mnras/226.4.747}, 226,
  747

\bibitem[\protect\citeauthoryear{{Johnston} et~al.,}{{Johnston}
  et~al.}{2020}]{Johnston2020}
{Johnston} E.~J.,  et~al., 2020, \mn@doi [\mnras] {10.1093/mnras/staa1261},
  \href {https://ui.adsabs.harvard.edu/abs/2020MNRAS.495.2247J} {495, 2247}

\bibitem[\protect\citeauthoryear{Jord{\'{a}}n et~al.,}{Jord{\'{a}}n
  et~al.}{2006}]{Jordan2006}
Jord{\'{a}}n A.,  et~al., 2006, \mn@doi [\apj] {10.1086/509119}, 651, L25

\bibitem[\protect\citeauthoryear{{Karachentsev}, {Makarov}  \&
  {Kaisina}}{{Karachentsev} et~al.}{2013}]{Karachentsev2013}
{Karachentsev} I.~D.,  {Makarov} D.~I.,   {Kaisina} E.~I.,  2013, \mn@doi [\aj]
  {10.1088/0004-6256/145/4/101}, \href
  {https://ui.adsabs.harvard.edu/abs/2013AJ....145..101K} {145, 101}

\bibitem[\protect\citeauthoryear{{Kaviraj}, {Martin}  \& {Silk}}{{Kaviraj}
  et~al.}{2019}]{Sugata2019}
{Kaviraj} S.,  {Martin} G.,   {Silk} J.,  2019, \mn@doi [\mnras]
  {10.1093/mnrasl/slz102}, \href
  {https://ui.adsabs.harvard.edu/abs/2019MNRAS.489L..12K} {489, L12}

\bibitem[\protect\citeauthoryear{{Kent}}{{Kent}}{1987}]{Kent1987}
{Kent} S.~M.,  1987, \mn@doi [\aj] {10.1086/114472}, \href
  {https://ui.adsabs.harvard.edu/abs/1987AJ.....94..306K} {94, 306}

\bibitem[\protect\citeauthoryear{{Koss}, {Mushotzky}, {Treister}, {Veilleux},
  {Vasudevan}  \& {Trippe}}{{Koss} et~al.}{2012}]{Koss2012}
{Koss} M.,  {Mushotzky} R.,  {Treister} E.,  {Veilleux} S.,  {Vasudevan} R.,
  {Trippe} M.,  2012, \mn@doi [\apjl] {10.1088/2041-8205/746/2/L22}, \href
  {https://ui.adsabs.harvard.edu/abs/2012ApJ...746L..22K} {746, L22}

\bibitem[\protect\citeauthoryear{{Kourkchi} \& {Tully}}{{Kourkchi} \&
  {Tully}}{2017}]{Kourkchi2017}
{Kourkchi} E.,  {Tully} R.~B.,  2017, \mn@doi [\apj]
  {10.3847/1538-4357/aa76db}, \href
  {https://ui.adsabs.harvard.edu/abs/2017ApJ...843...16K} {843, 16}

\bibitem[\protect\citeauthoryear{{Lauer} et~al.,}{{Lauer}
  et~al.}{1993}]{Lauer1993}
{Lauer} T.~R.,  et~al., 1993, \mn@doi [\aj] {10.1086/116737}, \href
  {https://ui.adsabs.harvard.edu/abs/1993AJ....106.1436L} {106, 1436}

\bibitem[\protect\citeauthoryear{{Lauer} et~al.,}{{Lauer}
  et~al.}{1996}]{Lauer1996}
{Lauer} T.~R.,  et~al., 1996, \mn@doi [\apjl] {10.1086/310344}, \href
  {https://ui.adsabs.harvard.edu/abs/1996ApJ...471L..79L} {471, L79}

\bibitem[\protect\citeauthoryear{{Lauer}, {Faber}, {Ajhar}, {Grillmair}  \&
  {Scowen}}{{Lauer} et~al.}{1998}]{Lauer1998}
{Lauer} T.~R.,  {Faber} S.~M.,  {Ajhar} E.~A.,  {Grillmair} C.~J.,   {Scowen}
  P.~A.,  1998, \mn@doi [\aj] {10.1086/300617}, \href
  {https://ui.adsabs.harvard.edu/abs/1998AJ....116.2263L} {116, 2263}

\bibitem[\protect\citeauthoryear{{Lim} et~al.,}{{Lim} et~al.}{2020}]{Lim2020}
{Lim} S.,  et~al., 2020, \mn@doi [\apj] {10.3847/1538-4357/aba433}, \href
  {https://ui.adsabs.harvard.edu/abs/2020ApJ...899...69L} {899, 69}

\bibitem[\protect\citeauthoryear{{Lindegren} et~al.,}{{Lindegren}
  et~al.}{2018}]{Gaiaastro}
{Lindegren} L.,  et~al., 2018, \mn@doi [\aap] {10.1051/0004-6361/201832727},
  \href {https://ui.adsabs.harvard.edu/abs/2018A&A...616A...2L} {616, A2}

\bibitem[\protect\citeauthoryear{{Liu} et~al.,}{{Liu} et~al.}{2020}]{Liu2020}
{Liu} C.,  et~al., 2020, \mn@doi [\apjs] {10.3847/1538-4365/abad91}, \href
  {https://ui.adsabs.harvard.edu/abs/2020ApJS..250...17L} {250, 17}

\bibitem[\protect\citeauthoryear{{Loose}, {Kruegel}  \& {Tutukov}}{{Loose}
  et~al.}{1982}]{Loose1982}
{Loose} H.~H.,  {Kruegel} E.,   {Tutukov} A.,  1982, \aap, \href
  {https://ui.adsabs.harvard.edu/abs/1982A&A...105..342L} {105, 342}

\bibitem[\protect\citeauthoryear{Lotz, Miller  \& Ferguson}{Lotz
  et~al.}{2004}]{Lotz2004}
Lotz J.~M.,  Miller B.~W.,   Ferguson H.~C.,  2004, \mn@doi [\apj]
  {10.1086/422871}, 613, 262

\bibitem[\protect\citeauthoryear{{Marleau}, {Clancy}  \& {Bianconi}}{{Marleau}
  et~al.}{2013}]{Marleau2013}
{Marleau} F.~R.,  {Clancy} D.,   {Bianconi} M.,  2013, \mn@doi [\mnras]
  {10.1093/mnras/stt1503}, \href
  {https://ui.adsabs.harvard.edu/abs/2013MNRAS.435.3085M} {435, 3085}

\bibitem[\protect\citeauthoryear{{Marleau}, {Clancy}, {Habas}  \&
  {Bianconi}}{{Marleau} et~al.}{2017}]{Marleau2017}
{Marleau} F.~R.,  {Clancy} D.,  {Habas} R.,   {Bianconi} M.,  2017, \mn@doi
  [\aap] {10.1051/0004-6361/201629832}, \href
  {https://ui.adsabs.harvard.edu/abs/2017A&A...602A..28M} {602, A28}

\bibitem[\protect\citeauthoryear{{Mastropietro}, {Moore}, {Mayer},
  {Debattista}, {Piffaretti}  \& {Stadel}}{{Mastropietro}
  et~al.}{2005}]{Mastropietro2005}
{Mastropietro} C.,  {Moore} B.,  {Mayer} L.,  {Debattista} V.~P.,  {Piffaretti}
  R.,   {Stadel} J.,  2005, \mn@doi [\mnras]
  {10.1111/j.1365-2966.2005.09579.x}, \href
  {https://ui.adsabs.harvard.edu/abs/2005MNRAS.364..607M} {364, 607}

\bibitem[\protect\citeauthoryear{{Mateo}}{{Mateo}}{1998}]{Mateo1998}
{Mateo} M.~L.,  1998, \mn@doi [\araa] {10.1146/annurev.astro.36.1.435}, \href
  {https://ui.adsabs.harvard.edu/abs/1998ARA&A..36..435M} {36, 435}

\bibitem[\protect\citeauthoryear{{Mayer}, {Governato}, {Colpi}, {Moore},
  {Quinn}, {Wadsley}, {Stadel}  \& {Lake}}{{Mayer} et~al.}{2001}]{Mayer2001}
{Mayer} L.,  {Governato} F.,  {Colpi} M.,  {Moore} B.,  {Quinn} T.,  {Wadsley}
  J.,  {Stadel} J.,   {Lake} G.,  2001, \mn@doi [\apjl] {10.1086/318898}, \href
  {https://ui.adsabs.harvard.edu/abs/2001ApJ...547L.123M} {547, L123}

\bibitem[\protect\citeauthoryear{{McConnachie}}{{McConnachie}}{2012}]{Mcconachie2012}
{McConnachie} A.~W.,  2012, \mn@doi [\aj] {10.1088/0004-6256/144/1/4}, \href
  {https://ui.adsabs.harvard.edu/abs/2012AJ....144....4M} {144, 4}

\bibitem[\protect\citeauthoryear{{Mezcua} \& {Dom{\'\i}nguez
  S{\'a}nchez}}{{Mezcua} \& {Dom{\'\i}nguez S{\'a}nchez}}{2020}]{Mezcua2020}
{Mezcua} M.,  {Dom{\'\i}nguez S{\'a}nchez} H.,  2020, \mn@doi [\apjl]
  {10.3847/2041-8213/aba199}, \href
  {https://ui.adsabs.harvard.edu/abs/2020ApJ...898L..30M} {898, L30}

\bibitem[\protect\citeauthoryear{{Miller} \& {Smith}}{{Miller} \&
  {Smith}}{1992}]{Miller1992}
{Miller} R.~H.,  {Smith} B.~F.,  1992, \mn@doi [\apj] {10.1086/171523}, 393,
  508

\bibitem[\protect\citeauthoryear{{Misgeld}, {Mieske}  \& {Hilker}}{{Misgeld}
  et~al.}{2008}]{Misgeld2008}
{Misgeld} I.,  {Mieske} S.,   {Hilker} M.,  2008, \mn@doi [\aap]
  {10.1051/0004-6361:200810014}, \href
  {https://ui.adsabs.harvard.edu/abs/2008A&A...486..697M} {486, 697}

\bibitem[\protect\citeauthoryear{{Mistani} et~al.,}{{Mistani}
  et~al.}{2016}]{Mistani2016}
{Mistani} P.~A.,  et~al., 2016, \mn@doi [\mnras] {10.1093/mnras/stv2435}, \href
  {https://ui.adsabs.harvard.edu/abs/2016MNRAS.455.2323M} {455, 2323}

\bibitem[\protect\citeauthoryear{{Molina}, {Reines}, {Greene}, {Darling}  \&
  {Condon}}{{Molina} et~al.}{2021}]{Molina2021}
{Molina} M.,  {Reines} A.~E.,  {Greene} J.~E.,  {Darling} J.,   {Condon} J.~J.,
   2021, \mn@doi [\apj] {10.3847/1538-4357/abe120}, \href
  {https://ui.adsabs.harvard.edu/abs/2021ApJ...910....5M} {910, 5}

\bibitem[\protect\citeauthoryear{{Monaco}, {Bellazzini}, {Ferraro}  \&
  {Pancino}}{{Monaco} et~al.}{2005}]{Monaco2005}
{Monaco} L.,  {Bellazzini} M.,  {Ferraro} F.~R.,   {Pancino} E.,  2005, \mn@doi
  [\mnras] {10.1111/j.1365-2966.2004.08579.x}, \href
  {https://ui.adsabs.harvard.edu/abs/2005MNRAS.356.1396M} {356, 1396}

\bibitem[\protect\citeauthoryear{{Moore}, {Katz}, {Lake}, {Dressler}  \&
  {Oemler}}{{Moore} et~al.}{1996}]{Moore1996}
{Moore} B.,  {Katz} N.,  {Lake} G.,  {Dressler} A.,   {Oemler} A.,  1996,
  \mn@doi [\nat] {10.1038/379613a0}, \href
  {https://ui.adsabs.harvard.edu/abs/1996Natur.379..613M} {379, 613}

\bibitem[\protect\citeauthoryear{{M{\"u}ller}, {Jerjen}  \&
  {Binggeli}}{{M{\"u}ller} et~al.}{2017}]{Mueller2017}
{M{\"u}ller} O.,  {Jerjen} H.,   {Binggeli} B.,  2017, \mn@doi [\aap]
  {10.1051/0004-6361/201628921}, \href
  {https://ui.adsabs.harvard.edu/abs/2017A&A...597A...7M} {597, A7}

\bibitem[\protect\citeauthoryear{{M{\"u}ller}, {Jerjen}  \&
  {Binggeli}}{{M{\"u}ller} et~al.}{2018}]{Mueller2018}
{M{\"u}ller} O.,  {Jerjen} H.,   {Binggeli} B.,  2018, \mn@doi [\aap]
  {10.1051/0004-6361/201832897}, \href
  {https://ui.adsabs.harvard.edu/abs/2018A&A...615A.105M} {615, A105}

\bibitem[\protect\citeauthoryear{{M{\"u}ller} et~al.,}{{M{\"u}ller}
  et~al.}{2020}]{Mueller2020}
{M{\"u}ller} O.,  et~al., 2020, \mn@doi [\aap] {10.1051/0004-6361/202038351},
  \href {https://ui.adsabs.harvard.edu/abs/2020A&A...640A.106M} {640, A106}

\bibitem[\protect\citeauthoryear{{M{\"u}ller} et~al.,}{{M{\"u}ller}
  et~al.}{2021}]{Mueller2021}
{M{\"u}ller} O.,  et~al., 2021, arXiv e-prints, \href
  {https://ui.adsabs.harvard.edu/abs/2021arXiv210110659M} {p. arXiv:2101.10659}

\bibitem[\protect\citeauthoryear{{Murali}}{{Murali}}{2000}]{Murali2000}
{Murali} C.,  2000, \mn@doi [\apjl] {10.1086/312462}, \href
  {https://ui.adsabs.harvard.edu/abs/2000ApJ...529L..81M} {529, L81}

\bibitem[\protect\citeauthoryear{{Neumayer}, {Seth}  \& {Boeker}}{{Neumayer}
  et~al.}{2020}]{Neumayer2020}
{Neumayer} N.,  {Seth} A.,   {Boeker} T.,  2020, arXiv e-prints, \href
  {https://ui.adsabs.harvard.edu/abs/2020arXiv200103626N} {p. arXiv:2001.03626}

\bibitem[\protect\citeauthoryear{{Oh} \& {Lin}}{{Oh} \& {Lin}}{2000}]{Oh2000}
{Oh} K.~S.,  {Lin} D.~N.~C.,  2000, \mn@doi [\apj] {10.1086/317118}, \href
  {https://ui.adsabs.harvard.edu/abs/2000ApJ...543..620O} {543, 620}

\bibitem[\protect\citeauthoryear{Ordenes-Brice{\~{n}}o
  et~al.,}{Ordenes-Brice{\~{n}}o et~al.}{2018}]{Ordenes2018}
Ordenes-Brice{\~{n}}o Y.,  et~al., 2018, \mn@doi [\apj]
  {10.3847/1538-4357/aac1b8}, 860, 4

\bibitem[\protect\citeauthoryear{{Pak}, {Paudel}, {Lee}  \& {Kim}}{{Pak}
  et~al.}{2016}]{Pak2016}
{Pak} M.,  {Paudel} S.,  {Lee} Y.,   {Kim} S.~C.,  2016, \mn@doi [\aj]
  {10.3847/0004-6256/151/6/141}, \href
  {https://ui.adsabs.harvard.edu/abs/2016AJ....151..141P} {151, 141}

\bibitem[\protect\citeauthoryear{{Pasetto}, {Chiosi}  \& {Carraro}}{{Pasetto}
  et~al.}{2003}]{Pasetto2003}
{Pasetto} S.,  {Chiosi} C.,   {Carraro} G.,  2003, \mn@doi [\aap]
  {10.1051/0004-6361:20030673}, \href
  {https://ui.adsabs.harvard.edu/abs/2003A&A...405..931P} {405, 931}

\bibitem[\protect\citeauthoryear{{Paudel} \& {Yoon}}{{Paudel} \&
  {Yoon}}{2020}]{Paudel2020}
{Paudel} S.,  {Yoon} S.-J.,  2020, \mn@doi [\apjl] {10.3847/2041-8213/aba6ed},
  \href {https://ui.adsabs.harvard.edu/abs/2020ApJ...898L..47P} {898, L47}

\bibitem[\protect\citeauthoryear{{Paudel}, {Lisker}  \& {Kuntschner}}{{Paudel}
  et~al.}{2011}]{Paudel2011}
{Paudel} S.,  {Lisker} T.,   {Kuntschner} H.,  2011, \mn@doi [\mnras]
  {10.1111/j.1365-2966.2011.18256.x}, \href
  {https://ui.adsabs.harvard.edu/abs/2011MNRAS.413.1764P} {413, 1764}

\bibitem[\protect\citeauthoryear{{Peacock}, {Maccarone}, {Knigge}, {Kundu},
  {Waters}, {Zepf}  \& {Zurek}}{{Peacock} et~al.}{2010}]{Peacock2010}
{Peacock} M.~B.,  {Maccarone} T.~J.,  {Knigge} C.,  {Kundu} A.,  {Waters}
  C.~Z.,  {Zepf} S.~E.,   {Zurek} D.~R.,  2010, \mn@doi [\mnras]
  {10.1111/j.1365-2966.2009.15952.x}, \href
  {https://ui.adsabs.harvard.edu/abs/2010MNRAS.402..803P} {402, 803}

\bibitem[\protect\citeauthoryear{{Peng}, {Ho}, {Impey}  \& {Rix}}{{Peng}
  et~al.}{2010}]{Peng2010}
{Peng} C.~Y.,  {Ho} L.~C.,  {Impey} C.~D.,   {Rix} H.-W.,  2010, \mn@doi [\aj]
  {10.1088/0004-6256/139/6/2097}, \href
  {https://ui.adsabs.harvard.edu/abs/2010AJ....139.2097P} {139, 2097}

\bibitem[\protect\citeauthoryear{{Richards} et~al.,}{{Richards}
  et~al.}{2002}]{Richards2002}
{Richards} G.~T.,  et~al., 2002, \mn@doi [\aj] {10.1086/340187}, \href
  {https://ui.adsabs.harvard.edu/abs/2002AJ....123.2945R} {123, 2945}

\bibitem[\protect\citeauthoryear{S{\'{a}}nchez-Janssen
  et~al.,}{S{\'{a}}nchez-Janssen et~al.}{2016}]{Sanchez-Janssen2016}
S{\'{a}}nchez-Janssen R.,  et~al., 2016, \mn@doi [\apj]
  {10.3847/0004-637x/820/1/69}, 820, 69

\bibitem[\protect\citeauthoryear{{S{\'a}nchez-Janssen}
  et~al.,}{{S{\'a}nchez-Janssen} et~al.}{2019}]{Janssen2019}
{S{\'a}nchez-Janssen} R.,  et~al., 2019, \mn@doi [\apj]
  {10.3847/1538-4357/aaf4fd}, \href
  {https://ui.adsabs.harvard.edu/abs/2019ApJ...878...18S} {878, 18}

\bibitem[\protect\citeauthoryear{{Sawala}, {Scannapieco}, {Maio}  \&
  {White}}{{Sawala} et~al.}{2010}]{Sawala2010}
{Sawala} T.,  {Scannapieco} C.,  {Maio} U.,   {White} S.,  2010, \mn@doi
  [\mnras] {10.1111/j.1365-2966.2009.16035.x}, \href
  {https://ui.adsabs.harvard.edu/abs/2010MNRAS.402.1599S} {402, 1599}

\bibitem[\protect\citeauthoryear{{Sawala}, {Scannapieco}  \& {White}}{{Sawala}
  et~al.}{2012}]{Sawala2012}
{Sawala} T.,  {Scannapieco} C.,   {White} S.,  2012, \mn@doi [\mnras]
  {10.1111/j.1365-2966.2011.20181.x}, \href
  {https://ui.adsabs.harvard.edu/abs/2012MNRAS.420.1714S} {420, 1714}

\bibitem[\protect\citeauthoryear{{Schlafly} \& {Finkbeiner}}{{Schlafly} \&
  {Finkbeiner}}{2011}]{Schlafly2011}
{Schlafly} E.~F.,  {Finkbeiner} D.~P.,  2011, \mn@doi [\apj]
  {10.1088/0004-637X/737/2/103}, \href
  {https://ui.adsabs.harvard.edu/abs/2011ApJ...737..103S} {737, 103}

\bibitem[\protect\citeauthoryear{{Secker}}{{Secker}}{1995}]{Secker1995}
{Secker} J.,  1995, \mn@doi [\pasp] {10.1086/133580}, \href
  {https://ui.adsabs.harvard.edu/abs/1995PASP..107..496S} {107, 496}

\bibitem[\protect\citeauthoryear{{S{\'e}rsic}}{{S{\'e}rsic}}{1963}]{Sersic1963}
{S{\'e}rsic} J.~L.,  1963, BAAA, \href
  {https://ui.adsabs.harvard.edu/abs/1963BAAA....6...41S} {6, 41}

\bibitem[\protect\citeauthoryear{{Sharina} et~al.,}{{Sharina}
  et~al.}{2008}]{Sharina2008}
{Sharina} M.~E.,  et~al., 2008, \mn@doi [\mnras]
  {10.1111/j.1365-2966.2007.12814.x}, \href
  {https://ui.adsabs.harvard.edu/abs/2008MNRAS.384.1544S} {384, 1544}

\bibitem[\protect\citeauthoryear{{Sills}, {Dalessandro}, {Cadelano},
  {Alfaro-Cuello}  \& {Kruijssen}}{{Sills} et~al.}{2019}]{Sills2019}
{Sills} A.,  {Dalessandro} E.,  {Cadelano} M.,  {Alfaro-Cuello} M.,
  {Kruijssen} J.~M.~D.,  2019, \mn@doi [\mnras] {10.1093/mnrasl/slz149}, \href
  {https://ui.adsabs.harvard.edu/abs/2019MNRAS.490L..67S} {490, L67}

\bibitem[\protect\citeauthoryear{{Simon}}{{Simon}}{2019}]{Simon2019}
{Simon} J.~D.,  2019, \mn@doi [\araa] {10.1146/annurev-astro-091918-104453},
  \href {https://ui.adsabs.harvard.edu/abs/2019ARA&A..57..375S} {57, 375}

\bibitem[\protect\citeauthoryear{{Skillman}, {C{\^o}t{\'e}}  \&
  {Miller}}{{Skillman} et~al.}{2003}]{Sillman2003}
{Skillman} E.~D.,  {C{\^o}t{\'e}} S.,   {Miller} B.~W.,  2003, \mn@doi [\aj]
  {10.1086/345964}, \href
  {https://ui.adsabs.harvard.edu/abs/2003AJ....125..593S} {125, 593}

\bibitem[\protect\citeauthoryear{{Spengler} et~al.,}{{Spengler}
  et~al.}{2017}]{Spengler2017}
{Spengler} C.,  et~al., 2017, \mn@doi [\apj] {10.3847/1538-4357/aa8a78}, \href
  {https://ui.adsabs.harvard.edu/abs/2017ApJ...849...55S} {849, 55}

\bibitem[\protect\citeauthoryear{Stetson}{Stetson}{1987}]{Stetson1987}
Stetson P.~B.,  1987, \mn@doi [Publications of the Astronomical Society of the
  Pacific] {10.1086/131977}, 99, 191

\bibitem[\protect\citeauthoryear{{Steyrleithner}, {Hensler}  \&
  {Boselli}}{{Steyrleithner} et~al.}{2020}]{Steyrleithner2020}
{Steyrleithner} P.,  {Hensler} G.,   {Boselli} A.,  2020, \mn@doi [\mnras]
  {10.1093/mnras/staa775}, \href
  {https://ui.adsabs.harvard.edu/abs/2020MNRAS.494.1114S} {494, 1114}

\bibitem[\protect\citeauthoryear{{Taga} \& {Iye}}{{Taga} \&
  {Iye}}{1998}]{Taga1998}
{Taga} M.,  {Iye} M.,  1998, \mn@doi [\mnras]
  {10.1046/j.1365-8711.1998.01753.x}, 299, 111

\bibitem[\protect\citeauthoryear{{Tremaine}, {Ostriker}  \&
  {Spitzer}}{{Tremaine} et~al.}{1975}]{Tremaine1975}
{Tremaine} S.~D.,  {Ostriker} J.~P.,   {Spitzer} L. J.,  1975, \mn@doi [\apj]
  {10.1086/153422}, \href
  {https://ui.adsabs.harvard.edu/abs/1975ApJ...196..407T} {196, 407}

\bibitem[\protect\citeauthoryear{{Turner}, {C{\^o}t{\'e}}, {Ferrarese},
  {Jord{\'a}n}, {Blakeslee}, {Mei}, {Peng}  \& {West}}{{Turner}
  et~al.}{2012}]{Turner2012}
{Turner} M.~L.,  {C{\^o}t{\'e}} P.,  {Ferrarese} L.,  {Jord{\'a}n} A.,
  {Blakeslee} J.~P.,  {Mei} S.,  {Peng} E.~W.,   {West} M.~J.,  2012, \mn@doi
  [\apjs] {10.1088/0067-0049/203/1/5}, \href
  {https://ui.adsabs.harvard.edu/abs/2012ApJS..203....5T} {203, 5}

\bibitem[\protect\citeauthoryear{{Valcke}, {de Rijcke}  \& {Dejonghe}}{{Valcke}
  et~al.}{2008}]{Valcke2008}
{Valcke} S.,  {de Rijcke} S.,   {Dejonghe} H.,  2008, \mn@doi [\mnras]
  {10.1111/j.1365-2966.2008.13654.x}, \href
  {https://ui.adsabs.harvard.edu/abs/2008MNRAS.389.1111V} {389, 1111}

\bibitem[\protect\citeauthoryear{{Venhola} et~al.,}{{Venhola}
  et~al.}{2018}]{Venhola2018}
{Venhola} A.,  et~al., 2018, \mn@doi [\aap] {10.1051/0004-6361/201833933},
  \href {https://ui.adsabs.harvard.edu/abs/2018A&A...620A.165V} {620, A165}

\bibitem[\protect\citeauthoryear{{Venhola} et~al.,}{{Venhola}
  et~al.}{2019}]{Venhola2019}
{Venhola} A.,  et~al., 2019, \mn@doi [\aap] {10.1051/0004-6361/201935231},
  \href {https://ui.adsabs.harvard.edu/abs/2019A&A...625A.143V} {625, A143}

\bibitem[\protect\citeauthoryear{Villegas et~al.,}{Villegas
  et~al.}{2010}]{Villegas2010}
Villegas D.,  et~al., 2010, \mn@doi [\apj] {10.1088/0004-637x/717/2/603}, 717,
  603

\bibitem[\protect\citeauthoryear{{Voggel}, {Hilker}  \& {Richtler}}{{Voggel}
  et~al.}{2016}]{Voggel2016}
{Voggel} K.,  {Hilker} M.,   {Richtler} T.,  2016, \mn@doi [\aap]
  {10.1051/0004-6361/201527070}, \href
  {https://ui.adsabs.harvard.edu/abs/2016A&A...586A.102V} {586, A102}

\bibitem[\protect\citeauthoryear{{Weisz} et~al.,}{{Weisz}
  et~al.}{2011}]{Weisz2011}
{Weisz} D.~R.,  et~al., 2011, \mn@doi [\apj] {10.1088/0004-637X/743/1/8}, \href
  {https://ui.adsabs.harvard.edu/abs/2011ApJ...743....8W} {743, 8}

\bibitem[\protect\citeauthoryear{{Young}, {Jerjen}, {L{\'o}pez-S{\'a}nchez}  \&
  {Koribalski}}{{Young} et~al.}{2014}]{Young2014}
{Young} T.,  {Jerjen} H.,  {L{\'o}pez-S{\'a}nchez} {\'A}.~R.,   {Koribalski}
  B.~S.,  2014, \mn@doi [\mnras] {10.1093/mnras/stu1646}, \href
  {https://ui.adsabs.harvard.edu/abs/2014MNRAS.444.3052Y} {444, 3052}

\bibitem[\protect\citeauthoryear{{Zanatta}, {S{\'a}nchez-Janssen},
  {Chies-Santos}, {de Souza}  \& {Blakeslee}}{{Zanatta}
  et~al.}{2021}]{Zanatta2021}
{Zanatta} E. J.~B.,  {S{\'a}nchez-Janssen} R.,  {Chies-Santos} A.~L.,  {de
  Souza} R.~S.,   {Blakeslee} J.~P.,  2021, arXiv e-prints, \href
  {https://ui.adsabs.harvard.edu/abs/2021arXiv210302123Z} {p. arXiv:2103.02123}

\bibitem[\protect\citeauthoryear{{den Brok} et~al.,}{{den Brok}
  et~al.}{2014}]{denBrok2014}
{den Brok} M.,  et~al., 2014, \mn@doi [\mnras] {10.1093/mnras/stu1906}, \href
  {https://ui.adsabs.harvard.edu/abs/2014MNRAS.445.2385D} {445, 2385}

\bibitem[\protect\citeauthoryear{{den Brok} et~al.,}{{den Brok}
  et~al.}{2015}]{denBrock2015}
{den Brok} M.,  et~al., 2015, \mn@doi [\apj] {10.1088/0004-637X/809/1/101},
  \href {https://ui.adsabs.harvard.edu/abs/2015ApJ...809..101D} {809, 101}

\bibitem[\protect\citeauthoryear{{van Dokkum}, {Abraham}, {Merritt}, {Zhang},
  {Geha}  \& {Conroy}}{{van Dokkum} et~al.}{2015}]{vanDokkum2015}
{van Dokkum} P.~G.,  {Abraham} R.,  {Merritt} A.,  {Zhang} J.,  {Geha} M.,
  {Conroy} C.,  2015, \mn@doi [\apjl] {10.1088/2041-8205/798/2/L45}, \href
  {https://ui.adsabs.harvard.edu/abs/2015ApJ...798L..45V} {798, L45}

\bibitem[\protect\citeauthoryear{{van den Bergh}}{{van den
  Bergh}}{1960}]{vandenBergh1960}
{van den Bergh} S.,  1960, \mn@doi [\apj] {10.1086/146821}, \href
  {https://ui.adsabs.harvard.edu/abs/1960ApJ...131..215V} {131, 215}

\makeatother
\end{thebibliography}

%%%%%%%%%%%%%%%%%%%%%%%%%%%%%%%%%%%%%%%%%%%%%%%%%%

%%%%%%%%%%%%%%%%% APPENDICES %%%%%%%%%%%%%%%%%%%%%
%\newpage
\appendix
\section{Contamination of the nuclei by globular clusters and foreground stars}

\subsection{Globular clusters simulations}
\label{Appendix_A1}
\begin{figure}
\centering
\includegraphics[width=\linewidth]{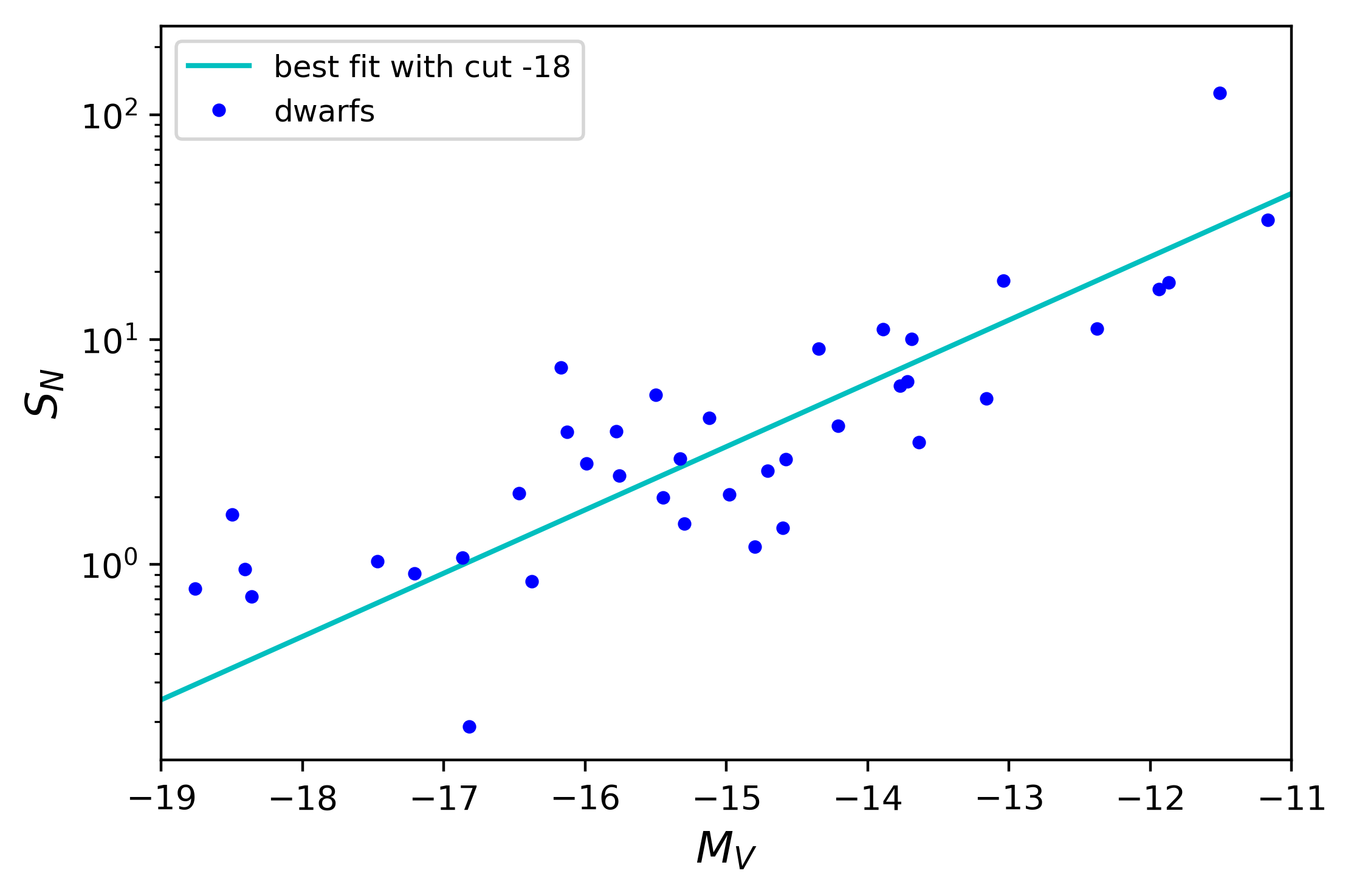}
\caption{Relation between the specific frequency and the absolute magnitude in the V band for the sample of dwarfs from \citet{Georgiev2010}. Blue dots represent the dwarfs. The formula for the best fit (cyan) is $log(S_N)=0.28M_V+4.74$.}
\label{fig:GeorgievFig3}
\end{figure}

As mentioned in Section \ref{section:nuclei}, we have estimated the probability for a globular cluster to be located at the centre of a dwarf and thus be visually confused with a compact central nucleus. For our calculation, we considered the sample of 425 nucleated dwarf galaxies with a diffuse part (masked nucleus) successfully modelled by a single S\'ersic profile. Our estimation was done by running 1000 Monte-Carlo simulations for each of the dwarfs. The simulation randomly distributes GCs around dwarfs and look for the presence of at least one of them at the centre, within a defined radius, which in addition would be the brightest source at maximum distance of one effective radius.

As a first step, we need to estimate the specific frequency (S$_N$) and total number of GCs (N$_{GC}$) for each dwarf. We do not have a robust catalogue of GCs corrected from contamination for each dwarf, therefore we used the linear relation log(S$_N$) versus M$_V$ found by \citet{Georgiev2010} for a sample of 73 dwarf galaxies (55 dIs, 5 dSphs, 3 dEs and 5 Sm) in the field environment observed with the Advanced Camera for Surveys. In Figure \ref{fig:GeorgievFig3}, we have plotted this relation and performed a linear regression on the data with a cut M$_V -18$ to be consistent with the MATLAS dwarfs sample. We used the best fit, $log(S_N)=0.28M_V+4.74$, to estimate the specific frequency of the MATLAS dwarfs coupled to the formula \(S_N \equiv N_{GC} 10^{0.4(M_V+15)}\) from \citet{Harris1981} to obtain the total number of GCs as a function of magnitude.

As a second step, now that we have an estimate of the total number of GCs for each dwarfs, given its absolute magnitude, we need to estimate the contamination by the GCs population from the host, i.e. the closest ETG.
To do this, we have used catalogs of GCs for the MATLAS fields. These catalogues were available for the field of 86\% of the nucleated sample. We have computed the density of GCs in a ring area defined by two circles centreed on the ETG with radii equal to the distance dwarf-ETG $\pm$ 3R$_e$, i.e., the area of projection of the GCs on the dwarf. As a result, 7\% of the MATLAS nucleated dwarfs show a contamination by at least 1 GCs, with a maximum of 4 GCs. As the median contamination value is 0, we assume that the contamination is null for the dwarfs having no available catalogue.

The third step is to create the sample of GCs. To define the globular clusters luminosity function (GCLF), we assume a Gaussian distribution with \(\mu = -7.4 + 0.04(M_V + 21.3)\) and \(\sigma = 1.2 - 0.10(M_V + 21.3)\) \citep{Harris2013,Jordan2006,Villegas2010}. We pick randomly the GCs following the GCLF and remove the GCs with m$_g$ < 24.5, the limit of detection of GCs in the MATLAS fields.

Our final step in this simulation is to project the GCs on the dwarf. We assume a spherical distribution and project the GCs within 3R$_e$ around the dwarfs. We pick random spherical coordinates that we convert first to Cartesian 3D and then 2D coordinates. We consider a GC as a possible nucleus if it is situated at a certain distance (0.5\arcsec\ and 1.5\arcsec, see discussion below) to the photocentre of the galaxy and if it is the brightest source GC within one effective radius, to be consistent with our visual classification of our nucleated sample.

\begin{figure}
\centering
\includegraphics[width=\linewidth]{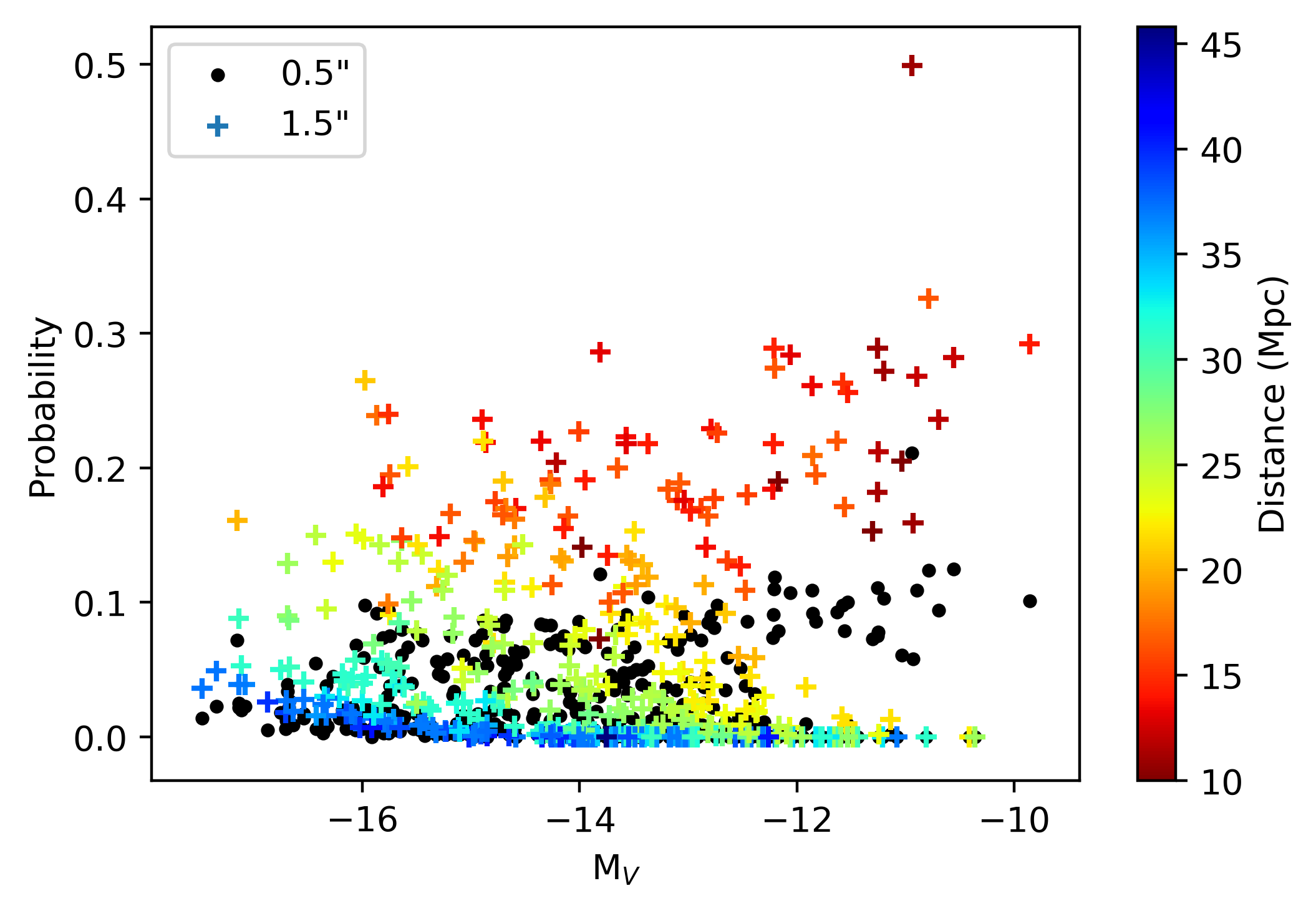}
\caption{Probability to find a GC within a radius of 0.5\arcsec\ (black dots) and 1.5\arcsec\ (coloured crosses) as a function of the dwarf M$_V$. The colorbar shows the distance to the dwarf.}
\label{fig:GCsimresults}
\end{figure}

We present the results for the nucleated dwarfs considering separations of 0.5\arcsec\ and 1.5\arcsec\ in Figure \ref{fig:GCsimresults}. In the case of a separation of 1.5\arcsec, we display the distance of the galaxy. The probability appears linked to the distance of the dwarf, with a higher probability for the closest dwarfs. This can be explained by the fact that the closer the dwarf, the larger the number of GCs visible. Thus, a robust sample of nucleated dwarfs with a probability of contamination $\lesssim$ 10\% can be defined by considering a maximum separation of 0.5\arcsec\ for the galaxies closer than 20 Mpc and a maximum separation of 1.5\arcsec\ for the more distant dwarfs.

\subsection{Galactic stars contamination}
\label{Appendix_A2}
In our sample of nucleated dwarfs, the unresolved nuclei can easily be confused with Galactic stars. Therefore, we want to estimate the contamination of our nucleated sample by foreground (Galactic) stars. The Gaia mission of the European Space Agency \citep{Gaiamission}, with its observations of more than a billion stars located in and around the Milky Way, provide us with the data to estimate the level of contamination. We have cross-matched our sample of nucleated with the most recent catalogue, Gaia DR2 \citep{GaiaDR2,Gaia2018,Gaiaastro}, with a maximum separation from the photocentre of 7\arcsec, the maximum separation observed for the MATLAS nuclei (Figure \ref{fig:nucprop}). Of the 508 nucleated dwarfs, we obtained 58 matches (11\%). To ensure the stellar nature of the matched sources, we selected only the sources that have a proper motion value larger than the proper motion error. We find one object matching the position of a nucleus, allowing us to estimate a contamination of 0.2\% of our nucleated sample by foreground stars.

%%%%%%%%%%%%%%%%%%%%%%%%%%%%%%%%%%%%%%%%%%%%%%%%%%

% Don't change these lines
\bsp	% typesetting comment
\label{lastpage}
\end{document}